\documentclass[6pt]{article}
\usepackage[dvips]{graphicx}
\usepackage[tight]{subfigure}
\usepackage{amssymb}
\usepackage{amsmath}
 
\usepackage{blindtext}
\usepackage{geometry}
\def\be{\begin{equation}}
\def\ee{\end{equation}}
\def\bea{\begin{eqnarray}}
\def\eea{\end{eqnarray}}
\def\beaN{\begin{eqnarray*}}
\def\eeaN{\end{eqnarray*}}
\def\ed{\end{document}}
\def\bit{\begin{itemize}}
\def\eit{\end{itemize}}

\def\~{\tilde}

\def\m{\label}
\def\l{\left}
\def\r{\right}

\def\goto{\rightarrow}

\usepackage{authblk}

\usepackage[dvipsnames]{xcolor}
\usepackage{graphicx}

\def\be{\begin{equation}}
\def\ee{\end{equation}}
\def\bea{\begin{eqnarray}}
\def\eea{\end{eqnarray}}

\begin{document}

\title{ \bf Suppressed Intrinsic Curvature Gravity}
\author[1]{Ido Ben-Dayan\thanks{Electronic address: \texttt{idobd@ariel.ac.il}}}
\author[1,2]{Elena Emtsova\thanks{Electronic address: \texttt{elenaem@ariel.ac.il}}}

\affil[1]{Physics Department, Ariel University, Ariel 40700, Israel}

\affil[2]{Department of Physics, Bar Ilan University, Ramat Gan 5290002, Israel}

\date{\small \today}

\date{\small \today}
\maketitle
\begin{abstract}
{We consider various mechanisms of modifying the effect of intrinsic curvature in gravity with respect to General Relativity. 
 Two primary approaches for suppressing intrinsic curvature are studied. First, by considering a Lagrange multiplier or an auxiliary field. Second, by non-minimal coupling between a scalar field and the intrinsic curvature scalar. We promote the foliation to a dynamical field, getting a fully covariant, and foliation independent theory.
We reproduce the basic solutions of FLRW cosmology, black hole solutions, Lense-Thirring effect, and gravitational waves. The speed of gravitational waves is modified in comparison to the speed of light. A certain limit of our theory corresponds to the lowest-order Carroll gravity. Hence, our theory is a different UV completion of Carroll gravity, compared to the usual expansion of a small speed of light. Carroll gravity limit also has an enhanced symmetry, making the reduced or vanishing intrinsic curvature technically natural. Finally, our construction defines a one parameter family of theories, that sets the relative strength of the intrinsic and extrinsic curvature, and General Relativity corresponds to a specific value of this parameter.}
\end{abstract}
\section{Introduction}
Curvature is a central concept of gravity, since Einstein's theory of General Relativity (GR) states that energy curves spacetime. The Einstein-Hilbert action then consists of a single function, the Ricci curvature scalar. Matter is then minimally coupled to gravity. Considering GR in the Hamiltonian formalism, one has to foliate spacetime, and given a foliation, one has two curvature tensors, conveniently dubbed the intrinsic and extrinsic curvature. The extrinsic curvature exists also in flat spacetime and is natural in the sense that it depends on the variation of the normal to the curve along the curve. On the other hand, the intrinsic curvature is special to curved spacetime.  Curiously, in the most immediate and applicative solutions - black holes and Cosmology, the intrinsic curvature vanishes. In black holes, this is simply due to the black hole being the vacuum solution. In Cosmology, the intrinsic curvature is the spatial curvature term of the Friedmann equations and could be large. Moreover, given a "reasonably small" initial spatial curvature, it should have been a rather dominant observed energy component by now. Thus, its observational absence is an existing puzzle, and inflation or its alternatives are generally invoked to explain it \cite{Lehners:2010fy,Battefeld:2014uga,Ben-Dayan:2008fhy,Ben-Dayan:2009fyj,Ben-Dayan:2010vsj,Ben-Dayan:2013fva, Ben-Dayan:2016iks, Ben-Dayan:2018ksd, Artymowski:2019cdg,Artymowski:2019jlh, Artymowski:2020pci, Ben-Dayan:2023rlj}.

Indeed, during inflation, the spatial curvature becomes negligible due to the quasiexponential expansion of the universe. The curvature term \( \frac{k}{a^2} \) in the Friedmann equation rapidly decreases as the scale factor \( a(t) \) grows exponentially. This solves the flatness problem by driving \( \Omega_k \), the relative curvature energy density close to zero, regardless of the initial curvature. 

Note that such explanations do not alter the global curvature of the universe, but simply state that the observable universe is a rather small part of the whole universe, which is why its curvature has not been observed.
It is then interesting to ask whether one can have a mechanism that dynamically reduces the intrinsic curvature, something like the axion and the strong CP problem \cite{Peccei:1977hh}. Such a mechanism perhaps based on symmetry could provide a more fundamental solution to the flatness problem and could lead to interesting model building.

Recent observations have reignited the interest in the question of spatial curvature, particularly concerning the possibility that the universe may not be perfectly flat. A notable study 
 \cite{DiValentino:2019qzk}, argues that data from the Planck satellite \cite{Planck:2018vyg} could be interpreted as favoring a closed universe with positive curvature. This result directly challenges the prevailing assumption of spatial flatness, which has been a cornerstone of the standard cosmological model. The implications of these findings are profound, as they raise the possibility of a “crisis for cosmology,” wherein current theoretical frameworks and observational interpretations may need to be reconsidered. 

If the evidence for a closed universe is confirmed, it would necessitate a significant reevaluation of inflationary models, which typically predict an extremely flat universe due to the rapid exponential expansion during the inflationary epoch. Additionally, it would require revisiting our understanding of the global topology of the universe, which could have far-reaching consequences for fundamental physics and cosmological observations. 
 The possibility of a non-flat universe also has implications for the interpretation of cosmic microwave background (CMB) anisotropies, large-scale structure (LSS) surveys, and Supernovae observations  {\cite{Planck:2018vyg,DES:2024jxu,DES:2025xii,DESI:2025zgx,DESI:2025zpo,BOSS:2016wmc,Pan-STARRS1:2017jku,SPT-3G:2025bzu,ACT:2025fju,Brout:2022vxf}.} {Recent work has further emphasized that the observational signatures of spatial curvature in FLRW models can vary depending on the method of inference and the underlying assumptions~\cite{Shimon:2024mbm}.} This potential discrepancy underscores the need for further observational precision and theoretical exploration to fully understand the geometry of the universe.

The Hubble tension refers to the significant discrepancy between early-universe estimates of the Hubble constant \( H_0 \), derived from CMB observations assuming the spatially flat \(\Lambda\)CDM model, and higher values obtained from local measurements using Cepheids and Type Ia supernovae \cite{Planck:2018vyg, Verde:2019ivm, Freedman:2024eph, Li:2025ife,Li:2025lfp, CosmoVerse:2025txj}. Since spatial curvature affects the relation between distances and redshift, even small deviations from flatness can shift cosmological parameter estimates and potentially alleviate this tension. Consequently, theoretical efforts to resolve the tension also explore modifications to dark energy, interactions with dark matter, additional relativistic species, and early dark energy \cite{Artymowski:2020zwy, DiValentino:2021solutions, Artymowski:2021fkw, Ben-Dayan:2023rgt, Ben-Dayan:2023htq, Dainotti:2023future, Khalife:2023hubble, Ben-Dayan:2024uvx}. Clarifying the source of the Hubble tension remains essential for testing the consistency of the standard cosmological model and identifying possible new physics.
 
The role of modifications to General Relativity (GR) is central to ongoing efforts in understanding cosmological dynamics. A wide range of extensions to GR, such as scalar-tensor theories, higher-dimensional models, and emergent gravity scenarios provide powerful frameworks for exploring the interplay between curvature and the evolution of the universe \cite{Clifton:2011jh, Sahlu:2024pxq, deRham:2014zqa}.  In a somewhat tangential manner, the \textit{Carroll limit} of GR has emerged as a particularly intriguing direction from the theoretical perspective. In this limit, the speed of light \( c \to 0 \), and the gravitational action simplifies in such a way that the intrinsic curvature is suppressed, while extrinsic curvature dominates the spacetime geometry. This leads to a natural mechanism for achieving a nearly flat universe. 
Studies such as \cite{Hansen:2021fxi, deBoer:2023fnj} show that the second-order term in the Carroll expansion of the action—the leading order in \( c \)—contains the only extrinsic curvature, effectively removing the intrinsic curvature contribution. Taking $c\rightarrow 0$ cannot describe phenomenology, but can give insight to various theoretical aspects of (quantum) gravity.

This article aims to explore the mechanism of modifying the intrinsic curvature by coupling it to a scalar field and considering the VEV of this field. By analyzing models with an additional scalar field, 
we investigate how these couplings can suppress curvature and their implications for gravity. Through this approach, we hope to contribute to the broader understanding of curvature's role in shaping gravity, the universe’s evolution and its potential connection to fundamental physics.

There are several ways to engineer theories that suppress the intrinsic curvature of a chosen hypersurface. The most direct (and admittedly naive) option is to add a Lagrange--multiplier term of the form $\lambda\,{}^{(3)}\!R$ to the action: the $\lambda$--equation then enforces ${}^{(3)}\!R=0$. The drawback is conceptual as well as technical: since ${}^{(3)}\!R$ is defined only after selecting a foliation, such a construction is not manifestly covariant and, in practice, becomes tied to a preferred slicing (or even to a preferred class of foliations). A more satisfactory route is to embed the hypersurface data covariantly into the dynamics by introducing a scalar field $A$ and defining the relevant leaves as level sets $A=\mathrm{const}$. One can then build a covariant scalar--tensor mechanism that drives ${}^{(3)}\!R\to 0$ on these leaves, while keeping the fundamental formulation coordinate, diffeomorphism and foliation invariant: the notion of foliation becomes secondary, emerging from the scalar dynamics rather than being imposed by hand. This perspective, however, raises a complementary question—under what conditions do the level sets of $A$ realize the physically desired hypersurfaces, and how can this be ensured dynamically? In this paper we start from the naive non-covariant $\lambda\,{}^{(3)}\!R$ idea and develop progressively more elaborate covariant scalar--tensor completions capable of vanishing the intrinsic curvature on the hypersurface of interest.

The rest of this article is organized as follows: Section \ref{sec:basics} provides geometric preliminaries, focusing on foliations and their role in decomposing spacetime. Section \ref{sec:lagmult} introduces theories with a Lagrange multiplier or an auxiliary field, the corresponding scalar-tensor reformulations and explore their implications for the intrinsic curvature. Section \ref{sec:nonmincoupl} examines an alternative approach of non-minimal coupling to gravity, while Section \ref{sec:applications} investigates applications of these theories to astrophysical and cosmological scenarios. In Section 6, we extend the analysis to rotating black holes using a modified Newman-Janis algorithm. Finally, Section \ref{sec:concl} summarizes our findings and outlines future research directions.
 \section{Geometric basics}\setcounter{equation}{0} \label{sec:basics}
 
Let us introduce useful geometrical quantities, based on \cite{Jha:2022svf}.
\emph{Conventions.}
We adopt the metric signature $(-,+,+,+)$ and set $c=1$.
Greek letters are avoided; we use Latin indices $a,b,c,\dots=0,1,2,3$.
Round and square brackets denote symmetrization and antisymmetrization as:
$X_{(ab)}\equiv \tfrac12(X_{ab}+X_{ba})$, $X_{[ab]}\equiv \tfrac12(X_{ab}-X_{ba})$.
Covariant derivatives are written $\nabla_a$ and use metric-compatible Levi-Civita connection, so, $\nabla_a g_{bc}=0$.
 First, we need to introduce a foliation. Foliation is a decomposition of the four-dimensional spacetime into a family of non-overlapping three-dimensional hypersurfaces labeled by some physical parameter. In the Arnowitt-Deser-Misner (ADM) formalism each slice is a hypersurface of constant time $t$.
 \begin{figure}[h]
    \centering
    \includegraphics[width=0.5\textwidth]{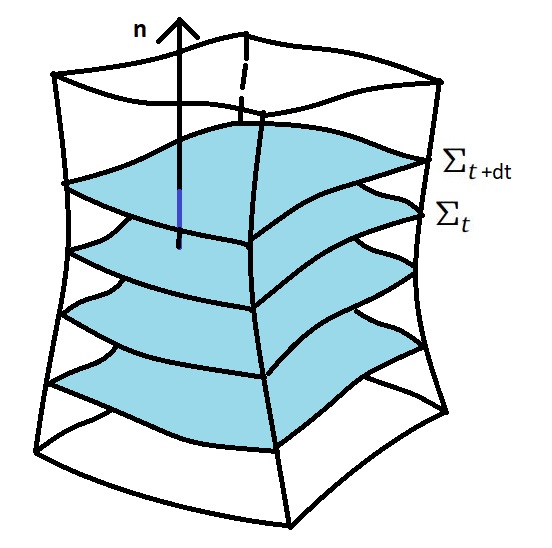} 
    \caption{Foliation of the spacetime in ADM formalism}
    \label{fig:foliation}
\end{figure}
 The ADM metric decomposition expresses the spacetime metric as:
\begin{equation}
    ds^2 = -N^2 dt^2 + \gamma_{ij} (dx^i + N^i dt)(dx^j + N^j dt),
\end{equation}
where
 \(N\) is the lapse function, determining the rate of proper time flow normal to the hypersurface; \(N^i\) is the shift vector, encoding the change of spatial coordinates across time slices; and \(\gamma_{ij}\) is the spatial metric induced on each slice (induced metric).
\noindent\emph{Note.} The induced metric, which is intrinsically three–dimensional, will be treated as its canonical 4D extension by padding zeros in the time rows/columns, i.e. $\gamma^{0a}=0=\gamma^{a0}$ and $\gamma^{ij}$ is the spatial block; the same convention applies to all 3D tensors below, which are regarded as 4D objects obtained by projection and extended with vanishing $0$-index-components.
The lapse function \(N\) is given by:
\begin{equation}\label{lapse}
    N=\frac{1}{\sqrt{-g^{00}}}.
\end{equation}
 This decomposition allows one to study the dynamics of spacetime as the evolution of these hypersurfaces. The unit timelike normal vector 
$n_m$ ($n_m n^m = -1$) is defined on each spacelike hypersurface 
$\Sigma$
  in the foliation of spacetime. It represents the four-velocity of a normal observer whose worldline is always orthogonal to $\Sigma$. In practice, by introducing a normal vector one can define a foliation. An example of such a foliation is depicted in figure \ref{fig:foliation}.
In the ADM formalism, the following expressions for the normal vector in (-+++) signature were derived in \cite{Jha:2022svf}:
\begin{equation}\label{normal}
    \begin{array}{cccc}
        n_m=-\frac{\delta^0_m}{\sqrt{-g^{00}}},\\
         n^m=-\frac{g^{0m}}{\sqrt{-g^{00}}},
    \end{array}
\end{equation}
where $g_{m n}$ is the metric of the 4-dimensional spacetime and $g^{m n}$ is its inverse.
The induced metric \(\gamma_{mn}\) projects the four-dimensional space-time geometry onto the three-dimensional hypersurface. It encodes the intrinsic geometry of the hypersurface, as viewed from within. The induced metric and its inverse are defined in general as:
\begin{equation}\label{gammagen}
    \begin{array}{cccc}
      \gamma^m_n = \delta^m_n + n^m n_n, \\
       \gamma_{mn}= g_{mn} +n_m n_n, \\
          \gamma^{mn}= g^{mn} +n^m n^n.
    \end{array}
\end{equation}
Substituting (\ref{normal}) we get the expressions of the intrinsic metric in the ADM formalism:
\begin{equation}\label{gamma}
    \begin{array}{cccc}
      \gamma^m_n  =  \delta^m_n- \frac{\delta^0_m g^{0n}}{g^{00}}, \\
       \gamma_{mn}= g_{mn}-\frac{\delta^0_m \delta^0_n}{g^{00}}, \\
          \gamma^{mn}= g^{mn}-\frac{g^{0m}g^{0n}}{g^{00}} .
    \end{array}
\end{equation}
Then one can get the useful expressions 
\begin{equation}\label{dergamma}
\begin{array}{cccc}
         \frac{\partial \gamma^{ij}}{\partial g^{ab} }= \delta^i_a  \delta^j_b - \delta^0_a  \delta^i_b \frac{g^{0j}}{g^{00}}- \delta^0_a  \delta^j_b \frac{g^{0i}}{g^{00}} + \delta^0_a  \delta^0_b \frac{g^{0i}g^{0j}}{(g^{00})^2}, \\
      \frac{\partial \gamma_{ij}}{\partial g_{ab} }=\delta^a_i  \delta^b_j+\frac{\delta^0_i  \delta^0_j}{(g^{00})^2} \frac{\partial g^{00}}{\partial g_{ab}}  =\delta^a_i  \delta^b_j-\frac{\delta^0_i  \delta^0_j}{(g^{00})^2} g^{a0} g^{b0}.
\end{array}
\end{equation}
The Riemann curvature tensor describes the curvature of a manifold and is given by:
\begin{equation}
    R^{r}{}_{smn} = \partial_{m} \Gamma^{r}{}_{ns} - \partial_{n} \Gamma^{r}{}_{ms} 
    + \Gamma^{r}{}_{mq} \Gamma^{q}{}_{ns} - \Gamma^{r}{}_{nl} \Gamma^{l}{}_{ms},
\end{equation}
where \( \Gamma^{r}_{mn} \) are the components of the Levi-Civita connection:
\begin{equation}
   \Gamma^{l}{}_{mn} = \frac{1}{2} g^{ls} \left( \frac{\partial g_{sm}}{\partial x^n} + \frac{\partial g_{sn}}{\partial x^m} - \frac{\partial g_{mn}}{\partial x^s} \right).
\end{equation}
The Ricci tensor is obtained by contracting the Riemann tensor:
\begin{equation}
    R_{mn} = R^{r}{}_{mrn}.
\end{equation}
The Ricci scalar is the trace of the Ricci tensor:
\begin{equation}
    R = g^{mn} R_{mn}.
\end{equation}
The curvature of the hypersurface can be characterized in two ways: intrinsic curvature, which describes the geometry within the hypersurface, and extrinsic curvature, which describes how the hypersurface is embedded in the larger spacetime.

The intrinsic curvature is captured by the three-dimensional Riemann tensor \({}^{(3)}R^m {}_{nab}\), Ricci tensor \({}^{(3)}R_{ab}\), and 3-curvature scalar \({}^{(3)}R\) which depend only on the induced metric \(\gamma_{mn}\) and its first and second derivatives. 
The intrinsic Riemann curvature tensor of each hypersurface induced by the 3-metric \( \gamma_{mn} \) is given by the following equations: (1) the 3-geometry  Christoffel symbols, defined as:
\begin{equation}
{}^{(3)}\Gamma^m_{a n} = \frac{1}{2} \gamma^{m s} \left( \partial_a \gamma_{n s} + \partial_n \gamma_{s a} - \partial_s \gamma_{a n} \right);
\end{equation}
(2) the intrinsic curvature of a spacelike hypersurface \( \Sigma_t \) is described by:
\begin{equation}
\left[ D_m, D_n \right] V^a = {}^{(3)}R^a_{b m n} V^b ~~\Leftrightarrow ~~~   {}^{(3)} R^{r}{}_{smn} = \partial_{m} {}^{(3)}\Gamma^{r}{}_{ns} - \partial_{n} {}^{(3)}\Gamma^{r}{}_{ms} 
    + {}^{(3)}\Gamma^{r}{}_{mq} {}^{(3)}\Gamma^{q}{}_{ns} - {}^{(3)}\Gamma^{r}{}_{nl}{}^{(3)} \Gamma^{l}{}_{ms},
\end{equation}
where $V^a$ is arbitrary vector, and $D_m$ is the 3-covariant derivative. The 3-dimensional Ricci tensor can be calculated as
\begin{equation}
    {}^{(3)} R_{a b} =  {}^{(3)} R^k{}_{a k b}.
\end{equation}
 The scalar intrinsic curvature of the hypersurface, \({}^{(3)}R\), can be obtained by contracting the 3-dimensional Ricci tensor with the induced metric:
\begin{equation}\label{R3}
    {}^{(3)} R=
   \gamma^{a b} {}^{(3)} R_{a b}.
\end{equation}
The extrinsic curvature  $K_{ab}$ is 
\begin{equation}\label{Kab}
    K_{mn}=-\gamma^a_m \gamma^b_n \nabla_a n_b,
\end{equation}
where \(\nabla_a\) is the spacetime covariant derivative. The trace of the extrinsic curvature, known as the mean curvature, is:
\begin{equation}\label{K}
    K=g^{mn}K_{mn}=\gamma^{mn}K_{mn}.
\end{equation}
From the Gauss-Codazzi relations,  the three-dimensional Riemann tensor \({}^{(3)}R^m {}_{nab}\), Ricci tensor \({}^{(3)}R_{ab}\), and 3-curvature scalar \({}^{(3)}R\) can be expressed in terms of the full spacetime curvature Riemann tensor \(R^m{}_{nrs}\), Ricci tensor \(R_{ab}\) and curvature scalar $R$ correspondingly as:
\begin{equation}\label{3Riemann}
\gamma^m_a \gamma^n_b \gamma^r_ c \gamma^s_d {} R_{m n r s} = {}^{(3)} R_{a b c d} + K_{a c} K_{b d} - K_{a d} K_{b c},
\end{equation}
\begin{equation}\label{3Ricci}
    {}^{(3)} R_{a b} = \gamma^m_a\gamma^n_b R_{mn} + \gamma_{am} n^n \gamma^r_b n^s R^m {}_{nrs} - K K_{ab} + K_{am} K^m_b ,
\end{equation}
\begin{equation}\label{3rsk}
 R + 2  R_{m n} n^m n^n = {}^{(3)} R + K^2 - K_{m n} K^{m n}.
\end{equation}
The tensors $\gamma^{mn}$, $K_{mn}$, ${}^{(3)} R_{mn}$ are 3-dimensional objects. Hence, their contractions with the normal vector $n_m$ are zero. 

For a general tensor with mixed components $T^{m_1 \dots m_p}_{n_1 \dots n_q}$, we can relate the 
3-derivative and the 3 + 1 -derivative using the induced metric:
\begin{equation}\label{34covder}
    D_{r} T^{a_1 \dots a_p}_{b_1 \dots b_q} =
\gamma^{a_1}_{m_1} \dots \gamma^{a_p}_{m_p} 
\gamma^{n_1}_{b_1} \dots \gamma^{n_q}_{b_q}
\gamma^{s}_{r} \nabla_{s} T^{m_1 \dots m_p}_{n_1 \dots n_q}
\end{equation}
and
\begin{equation}\label{covdmet}
    D_a \gamma_{bc} = 0,~~~~\nabla_a g_{bc}=0.
\end{equation}

The following simple example shows that for a general manifold the foliation is not unique, and is being chosen by hand. Different foliations generate different hypersurfaces with different ${}^{(3)} R$ on them. That is why it is good when the theory admits choosing such a foliation by its dynamics.
Minkowski space-time allows slicing with ${}^{(3)}\!R=0$ and ${}^{(3)}\!R<0$.

\emph{(i) Flat slices.} In standard inertial coordinates $(t,r,\theta,\phi)$,
$ds^2=-dt^2+dr^2+r^2 d\Omega_2^2$, the slices $t=\mathrm{const}$ have ${}^{(3)}\!R=0$.

\emph{(ii) Hyperbolic (Milne) slicing.} Inside the future (or past) light cone, introduce Milne coordinates
$t=\tau\cosh\chi$, $r=\tau\sinh\chi$ ($\tau>0$, $\chi\ge 0$), so
\[
ds^2=-d\tau^2+\tau^2\!\left(d\chi^2+\sinh^2\!\chi\,d\Omega_2^2\right),
\]
and the spatial slices $\tau=\mathrm{const}$ are copies of three-dimensional hyperbolic space $H^3$ with ${}^{(3)}\!R=-6/\tau^2$; see e.g.
\cite{Zenginoglu:2008ScriFixing} for hyperboloidal foliations in GR and \cite{Wald:1984rg} for Milne coordinates as a standard example.

\emph{(iii) On “$k=+1$” slicings.} A global Cauchy foliation of $(3{+}1)$ Minkowski by complete \emph{spatial} $S^3$-slices of constant positive curvature is incompatible with its Cauchy-surface topology. (Minkowski admits smooth Cauchy surfaces diffeomorphic to $\mathbb{R}^3$), \cite{Wald:1984rg, BernalSanchezSplitting}.
Outside the light cone one can write Minkowski in “radial” form with de Sitter level-sets of positive constant \emph{Lorentzian} curvature, e.g., \cite{deBoer:2003MinkowskiSlicing}.
This shows explicitly that ${}^{(3)}\!R$ is slicing-dependent even for the same 4D geometry; any suppression statement must therefore specify the foliation.

\section{Theory with the Lagrange multiplier}\label{sec:lagmult}\setcounter{equation}{0}
Let us start by suppressing the intrinsic curvature using brute force. We consider a theory with a Lagrange multiplier as the simplest mechanism for eliminating the three-dimensional spatial curvature. This approach is motivated by the observed smallness of spatial curvature in cosmology. 
In this section, we distinguish between a global and a local implementation of the
${}^{(3)}R$--constraint.
A \emph{constant} multiplier $\lambda$ enforces an \emph{integral}  constraint
$\int d^4x\,\sqrt{-g}\,{}^{(3)}R=0$,
whereas an \emph{auxiliary field} $\phi(x)$ (with no kinetic term, like in the standard lagrangue multiplier approach) enforces the \emph{local} constraint
${}^{(3)}R(x)=0$ pointwise. 
We analyze both because the global model is technically simple and already captures key solution patterns,
while the local model is the standard field–theoretic implementation.

\subsection{Global (integral) constraint (constant $\lambda$)}

Consider the action
\begin{equation}\label{laglam}
   S= S_{grav} +S_{matter}=\frac{1}{2}\int d^4 x \sqrt{-g} (R-2 \Lambda+ \lambda {}^{(3)} R) + S_{matter},
\end{equation}
where  $R$ is the 4-dimensional Ricci scalar, ${}^{(3)} R$ is intrinsic 3-dimensional curvature scalar on space-like hypersurfaces, $\Lambda$ is a cosmological constant (CC),  and  $\lambda$ is a Lagrange multiplier, and  consider here the case when it is constant. Varying the action by $\lambda$ one has an integral constraint: 
\begin{equation}
   \int  {}^{(3)} R\sqrt{-g}~  d^4 x=0.
\end{equation}
The variation of the action with respect to the metric is \cite{Jha:2022svf},
\bea
\delta S_{grav}  
        =\int d^4 x \left[ \sqrt{-g}(R^{ab}-\frac{1}{2} g^{ab} R +g^{ab} \Lambda) \delta g_{ab} +  \lambda \sqrt{-g}\left(-\frac{1}{2} g^{ab}\, {}^{(3)}R +\left[{}^{(3)}R^{ij}-\frac{1}{N}(D^i D^j N- \gamma^{ij} D_c D^c N ) \right] \frac{\partial \gamma_{ij}}{\partial g_{ab} } \right)\delta g_{ab}  \right].\nonumber
        \eea
Then, the equations of motion are
\begin{equation}
\begin{array}{cccc}
(R^{ab}-\frac{1}{2} g^{ab} R +g^{ab} \Lambda)   +  \lambda \left(-\frac{1}{2} g^{ab}\, {}^{(3)}R +\left[{}^{(3)}R^{ij}-\frac{1}{N}(D^i D^j N- \gamma^{ij} D_c D^c N ) \right] \frac{\partial \gamma_{ij}}{\partial g_{ab} } \right) =  T^{ab}.
\end{array}
\end{equation}
Using (\ref{dergamma}) and the fact that $\gamma^{00}=0$ and $D^0=0$, these equations of motion become
\begin{equation}\label{fieldeq}
\begin{array}{cccc}
       (R^{ab}-\frac{1}{2} g^{ab} R +g^{ab} \Lambda)  +  \lambda \left({}^{(3)} R^{ab}-\frac{1}{2} g^{ab} {}^{(3)} R-\frac{1}{N} D^a D^b N+ \frac{1}{N}\gamma^{ab} D_c D^c N \right) = T^{ab}.
\end{array}
\end{equation}
Not surprisingly, we reproduce the GR result in the case of $\lambda=0$. We have therefore derived a theory of gravity, where the integral of the intrinsic curvature scalar must vanish. We shall show that this theory admits many desired physical solutions such as cosmology and black holes.
 
\subsubsection{Theory with $g^{mn}{}^{(3)}R_{mn}$ term}
The intrinsic curvature scalar is defined as  ${}^{(3)}R =\gamma^{mn}{}^{(3)}R_{mn}$. Since ${}^{(3)}R_{mn}n^m n^n=0$, we can add this zero term to the definition of ${}^{(3)}R $, and due to $\gamma^{mn}=g^{mn}+n^m n^n$, we can write ${}^{(3)}R =g^{mn}{}^{(3)}R_{mn}$.
Then the action is
\begin{equation}\label{lagReq}
    S=\frac{1}{2}\int d^4 x \sqrt{-g} (R -2 \Lambda+ \lambda {}^{(3)} R_{mn}g^{mn}) + S_{matter}.
\end{equation}
In  Appendix \ref{gR}, we have verified that this modification does not affect the theory, and that the equations of motion (\ref{laglam}) --- (\ref{fieldeq}) remain unchanged.

\subsection{Auxiliary field}
Let us assume that we have an auxiliary field $\phi(x)$ without a kinetic term instead of the Lagrange multiplier $\lambda$. The corresponding action is
\begin{equation}\label{auxscflag}
   S=\frac{1}{2}\int d^4 x \sqrt{-g} (R -2\Lambda + \phi {}^{(3)} R) + S_{matter}.
\end{equation}
Varying the action by $\phi$ one has a local constraint
\begin{equation}
     {}^{(3)} R  =0.
\end{equation}
From  \cite{Jha:2022svf}, the resulting equations of motion are similar to \eqref{fieldeq}, but we put $\phi N$ instead of $N$ in brackets after the covariant derivative.
Thus, we got the field equations
\begin{equation}\label{fieldeqau}
\begin{array}{cccc}
          \left(R^{ab}-\frac{1}{2} g^{ab} R +g^{ab} \Lambda\right)  +  \phi \left({}^{(3)} R^{ab}-\frac{1}{2} g^{ab} {}^{(3)} R\right) -\frac{1}{N} D^a D^b (\phi N)+ \frac{1}{N}\gamma^{ab} D_c D^c (\phi N)  =   T^{ab}.
\end{array}
\end{equation}

\subsection{Synchronous gauge}
Of specific interest is the synchronous gauge, with $g_{00}=-1$, $g_{0i}=g_{i0}=0$, $N=1$. The equations of motion (\ref{fieldeq}) then look like a sum of $4$-dimensional and $3$-dimensional Einstein's equations.
\begin{equation}\label{syneq}
    R^{ab} -\frac{1}{2} g^{ab} R +g^{ab} \Lambda + \lambda \left({}^{(3)}R^{ab} -\frac{1}{2} g^{ab} {}^{(3)}R \right) =   T^{ab}.
\end{equation}
Rewriting the $00$ and $0j$ parts where $j=1,2,3$, we get
 
        \bea
    R^{00} &+&\frac{1}{2}  R- \Lambda +\frac{1}{2}\lambda{}^{(3)}R=   T^{00},\cr
    R^{0j}  &=&   T^{0j}\cr
        R^{ij} &-&\frac{1}{2} g^{ij} R +g^{ij} \Lambda + \lambda \left({}^{(3)}R^{ij} -\frac{1}{2} g^{ij} {}^{(3)}R \right) =   T^{ij}.
        \eea
Taking the trace of (\ref{syneq}) we get
\begin{equation}\label{tracesyneq}
    -R-\lambda {}^{(3)}R +4\Lambda =  T.
\end{equation}
Then, one can rewrite (\ref{syneq}) using  (\ref{tracesyneq})
\begin{equation}
        R^{ab}  + \lambda {}^{(3)}R^{ab} =   \left(T^{ab} -\frac{1}{2} g^{ab} T\right)+\Lambda g^{ab}.
\end{equation}
Similarly, with non-vanishing CC, and using the trace the equations can be written as:
\begin{equation}\label{synctrace}
    R^{ab} -\frac{1}{4} g^{ab} R + \lambda \left({}^{(3)}R^{ab} -\frac{1}{4} g^{ab} {}^{(3)}R \right) =   \left(T^{ab} -\frac{1}{4} g^{ab} T\right),
\end{equation}
similar to the GR case.

Let us discuss the energy-momentum tensor conservation. Taking the covariant divergence of both sides of (\ref{syneq}) yields:
\begin{equation}
\nabla_a \left[ R^{ab} - \frac{1}{2} g^{ab} R \right] 
+ \nabla_a (g^{ab} \Lambda) 
+ \lambda \nabla_a \left( {}^{(3)}R^{ab} - \frac{1}{2} g^{ab} {}^{(3)}R \right)
=  \nabla_a T^{ab}.
\end{equation}
Using the contracted Bianchi identity,
\begin{equation}
\nabla_a \left( R^{ab} - \frac{1}{2} g^{ab} R \right) = 0
\end{equation}
and the compatibility of the covariant derivative with the metric,
$\nabla_a g^{ab} = 0$,
we obtain the modified conservation equation:
\begin{equation}
\nabla_a T^{ab} = \lambda \nabla_a \left( {}^{(3)}R^{ab} - \frac{1}{2} g^{ab} {}^{(3)}R \right).
\end{equation}
In the case of ${}^{(3)}R=0$, which is slightly less general than $\int d^4 x \sqrt{-g}{}^{(3)}R=0$, we have
\begin{equation}
\nabla_a T^{ab} =\lambda \nabla_a{}^{(3)}R^{ab}.
\end{equation}
In the synchronous gauge we have $\nabla_i=D_i$ for $i=1,2,3$. Then, because ${}^{(3)}R^{00}={}^{(3)}R^{i0}=0$ the equation simplifies to $\nabla_a T^{ab} =\lambda D_a{}^{(3)}R^{ab}$. Using the contracted Bianchi identities in 3 dimensions $ D_a{}^{(3)}R^{ab}=\frac{1}{2}  D_b{}^{(3)}R=0$ we get $\nabla_a T^{ab} =0$ - the standard energy-momentum tensor conservation.

\subsection{Covariant formalism}\m{covform}
Dependence of physics on foliation or coordinate system can be rather detrimental to any theory aspiring to be realized in Nature. Hence, by using the covariant formalism developed in \cite{Gasperini:2009mu} we show that our theory is invariant under coordinate transformations, as a physical theory should be. We start from the action similar to (\ref{laglam})
\begin{equation}
    S=\frac{1}{2}\int d^4 x \sqrt{-g} (R-2 \Lambda+ \lambda {}^{(3)} R) + S_{matter},
\end{equation}
where $\lambda$ is the constant Lagrange multiplier, but the normal vector is defined as in  \cite{Gasperini:2009mu}\footnote{For the non-constant Lagrange multiplier  $\phi(x)$, we refer the reader to section \ref{note_covariant} where we demonstrate that the covariant formalism works for a more general case of a dynamical field $\phi$ which is equivalent to $\lambda=\xi \phi^2+\eta \phi $. Hence, the non-constant Lagrange multiplier case, is a private case of $\xi=0,\eta=1$ and a vanishing kinetic term.}:

\begin{equation}\label{ncov}
    n_a=-\frac{\nabla_a A}{\sqrt{-X}},
\qquad
X=g^{cd}\nabla_cA\nabla_dA<0,
\end{equation}
where $A$ is a scalar field and $X$ is its kinetic term. Because the right hand side of \eqref{ncov} transforms as a vector under general coordinate transformations, 3-dimensional tensors and scalars like intrinsic or extrinsic curvature are fully covariant now. The foliation is defined by the hypersurfaces of constant $A$. This is a generalization of the ADM formalism in which $A=A(t)$.
Because in \eqref{ncov} we expressed the normal vector in terms of derivatives of $A$,
$A$ is now a dynamical scalar field, and we should vary the action with respect to $g_{ab}$ and $A.$ First, we write the variation of the Lagrangian 
\begin{equation}\label{varcov}
\begin{array}{cccc}
      \delta   (\sqrt{-g} R  -2\sqrt{-g}  \Lambda+ \lambda \sqrt{-g} {}^{(3)}R) = (G^{mn}+g^{mn} \Lambda) \delta g_{mn} +  \lambda \delta   (\sqrt{-g}) {}^{(3)}R +   \lambda \sqrt{-g} \delta \gamma_{mn} {}^{(3)}R^{mn }+   \lambda \sqrt{-g}  \gamma_{mn}  \delta {}^{(3)}R^{mn } \\ = (G^{mn}+g^{mn} \Lambda)\sqrt{-g}  \delta g_{mn} -\frac{ \lambda }{2} \sqrt{-g} g^{mn} {}^{(3)}R   \delta g_{mn} +    \lambda \sqrt{-g} {}^{(3)}R^{mn }\delta \gamma_{mn}  -  \lambda \frac{\sqrt{-g}}{N} (D^m D^n N -\gamma^{mn} D_c D^c N) \delta \gamma_{mn},
\end{array}
\end{equation}
where $N$ can be calculated as $\sqrt{\frac{-g}{\gamma}}$ and we use the formula for $\sqrt{-g}  \gamma_{mn}  \delta {}^{(3)}R^{mn } $ from \cite{Jha:2022svf}. Then, because
\begin{equation}
    \gamma_{mn} = g_{mn} + n_m n_n =  g_{mn}+\frac{\partial_m A \partial_n A }{ - \partial_a A\partial_b A g^{ab}},
\end{equation}
the variation of $\gamma_{mn}$ with respect to $A$ is
\begin{equation}\label{delAgamma}
    \delta_A \gamma_{mn} =\frac{-2 \partial_n A \partial_m \delta A}{\partial_a A\partial_b A g^{ab}}+ \frac{2 \partial_m A \partial_n A \partial_b A \partial_a \delta A g^{ab}}{(\partial_a A\partial_b A g^{ab})^2}.
\end{equation}
The variation of the Lagrangian with respect to $A$ is 
\begin{equation}\label{varg3R}
   \lambda  \delta_A   (\sqrt{-g} {}^{(3)}R)  =\lambda \left(   \sqrt{-g} {}^{(3)}R^{mn }  - \frac{\sqrt{-g}}{N} (D^m D^n N -\gamma^{mn} D_c D^c N)\right) \left(\frac{-2 \partial_n A \partial_m \delta A}{\partial_a A\partial_b A g^{ab}}+ \frac{2 \partial_m A \partial_n A \partial_b A \partial_a \delta A g^{ab}}{(\partial_a A\partial_b A g^{ab})^2}\right).
\end{equation}
 Integrating this expression by parts to get rid of the derivatives of $\delta A$, we get the field equation for $A$:
\begin{equation}
   \partial_a [ \sqrt{-g} B^{an}\frac{-2 \partial_n A}{\partial_a A\partial_b A g^{ab}}]+  \partial_a[ \sqrt{-g}B^{mn} \frac{2 \partial_m A \partial_n A \partial_b A g^{ab}}{(\partial_a A\partial_b A g^{ab})^2}] =0,
\end{equation}
which gives a conserved current 
\begin{equation}\label{eqmA}
    B^{an}\frac{-2 \partial_n A}{\partial_a A\partial_b A g^{ab}}+  B^{mn} \frac{2 \partial_m A \partial_n A \partial_b A g^{ab}}{(\partial_a A\partial_b A g^{ab})^2}=J^a,
\end{equation}
 and\footnote{In the case of the $\phi(x)$, $ B^{mn} = \phi{}^{(3)}R^{mn}-\frac{1}{N} D^m D^n (\phi N) +\frac{1}{N}\gamma^{mn} D_c D^c (\phi N).$}
\begin{equation}\label{defB}
    B^{mn} = {}^{(3)}R^{mn}-\frac{1}{N} D^m D^n N +\frac{1}{N}\gamma^{mn} D_c D^c N.
\end{equation}

Using equations (\ref{Kab}), (\ref{3Ricci}), and (\ref{34covder}) with the property (\ref{covdmet}) one concludes that first and second terms of  $B^{mn}$ in (\ref{defB}) are proportional to $\gamma^m_a\gamma^n_b$, and the third term is proportional to $\gamma^{mn}$. $\partial_n A$ is proportional to the normal vector $n_n$. From this we conclude that all contractions like $B^{mn} \partial_n A$ or $B^{mn} \partial_m A$ are zero, which gives
\begin{equation}\label{covb0}
    B^{an}\frac{-2 \partial_n A}{\partial_a A\partial_b A g^{ab}}+  B^{mn} \frac{2 \partial_m A \partial_n A \partial_b A g^{ab }}{(\partial_a A\partial_b A g^{ab})^2}\equiv0.
\end{equation}
 The equation for $A$ is an equality that is satisfied for all $A$.
 The variation of the induced metric with respect to the space-time metric is
\begin{equation}
    \delta_g \gamma_{mn} = \delta g_{mn}-\frac{\partial_a A\partial_b A\partial_m A\partial_n A g^{aa'}g^{bb'} \delta g_{a'b'}}{(\partial_a A\partial_b A g^{ab})^2}.
\end{equation}
Putting this variation into (\ref{varcov}) we get the equation of motion
\begin{eqnarray}\label{coveqmet}
\begin{array}{cccc}
      (R^{a'b'}-\frac{1}{2} g^{a'b'} R g^{a'b'} \Lambda)  +  \lambda \left({}^{(3)} R^{a'b'}-\frac{1}{2} g^{a'b'} {}^{(3)} R-\frac{1}{N} D^{a'} D^{b'} N+ \frac{1}{N}\gamma^{a'b'} D_c D^c N \right) + \\ \lambda B^{mn}\frac{\partial_a A\partial_b A\partial_m A\partial_n A g^{aa'}g^{bb'}}{(\partial_a A\partial_b A g^{ab})^2}=  T^{a'b'}.
\end{array}
\end{eqnarray}
Last term of this equation is trivial because $B^{mn}$ is ``perpendicular'' to normal vector as we discussed above, and we get
\begin{eqnarray}\label{coveqm}
\begin{array}{cccc}
      (R^{mn}-\frac{1}{2} g^{mn} R +g^{mn} \Lambda)  +  \lambda \left({}^{(3)} R^{mn}-\frac{1}{2} g^{mn} {}^{(3)} R-\frac{1}{N} D^a D^b N+ \frac{1}{N}\gamma^{mn} D_c D^c N \right) = T^{mn},
\end{array}
\end{eqnarray}
i.e. exactly (\ref{fieldeq}). Hence, we find that in the covariant formalism the equations of motion for the metric are the same as in the usual ADM case. This means that this formalism allows us to choose any foliation, and that this formalism is therefore invariant under coordinate transformations. The fact that eq. (\ref{covb0}) is trivial demonstrates that. Hence, we can always find a foliation and impose the vanishing of ${}^{(3)} R$ in a gauge independent and coordinate invariant consistent manner. However, because of the triviality of the equation for $A$, different foliations can give different results. Therefore, we need a dynamical equation for $A$. In   ~\ref{subsec:dynamical-foliation} we show how to \emph{dynamically} select the field $A$ determining foliation by adding to (\ref{laglam}) another terms with Lagrange multipliers so that $n^a$ becomes dynamically fixed, thereby eliminating foliation arbitrariness at the level of dynamics. As a result, strictly speaking we have constructed a fully covariant scalar-tensor theory of gravity with suppressed intrinsic curvature.

\paragraph{Eliminating explicit $3$D objects in favor of $(g_{ab},A)$.}
Once the slicing is encoded by the scalar $A$, all “spatial” quantities are covariant projections of $4$D tensors with the unit normal
$
n_a=-\frac{\nabla_aA}{\sqrt{-X}}
$
and the projector
$
\gamma_{ab}=g_{ab}+n_a n_b
$.
The extrinsic curvature and the intrinsic (projected) covariant derivative are then
\[
K_{ab} =-\,\gamma_a{}^{c}\gamma_b{}^{d}\,\nabla_c n_d,\qquad
D_a T^{b_1\!\dots}_{\ \ \ \ c_1\!\dots}
 =\gamma_a{}^{m}\gamma^{b_1}{}_{n_1}\!\dots\,\gamma_{c_1}{}^{p_1}\!\dots\,\nabla_m T^{n_1\!\dots}_{\ \ \ \ p_1\!\dots}.
\]
All intrinsic curvatures are determined algebraically from the $4$D Riemann tensor and $K_{ab}$ by Gauss–Codazzi:
\begin{align}
\gamma_a{}^{m}\gamma_b{}^{n}\gamma_c{}^{p}\gamma_d{}^{q}R_{mnpq}
&={}^{(3)}\!R_{abcd}+K_{ac}K_{bd}-K_{ad}K_{bc},\label{eq:Gauss}\\
\gamma_a{}^{m}\gamma_b{}^{n}n^{p}R_{mnpq}\gamma^{q}{}_{c}
&=D_a K_{bc}-D_b K_{ac}\qquad\text{(Codazzi)},\label{eq:Codazzi}
\end{align}
which imply, upon contraction,
\begin{align}
{}^{(3)}\!R_{ab}
&=\gamma_a{}^{m}\gamma_b{}^{n}R_{mn}
+\gamma^{mn}R_{m a n b}
+K\,K_{ab}-K_{a}{}^{c}K_{cb},\label{eq:3Ricci}\\
{}^{(3)}\!R
&=\gamma^{ac}\gamma^{bd}R_{abcd}+K^2-K_{ab}K^{ab}
=R+2\,R_{ab}n^a n^b+K^2-K_{ab}K^{ab}.\label{eq:3Rscalar}
\end{align}
Equations \eqref{eq:Gauss}–\eqref{eq:3Rscalar} show that every appearance of ${}^{(3)}\!R$, ${}^{(3)}\!R_{ab}$, $D_a$ (and the induced metric itself) can be written \emph{purely} in terms of $g_{ab}$, $A$ and their $4$D covariant derivatives. In particular, any functional that was originally expressed using “$3$D data” on the slices can be recast as
\[
S_{\rm base}[g;\text{``$3$D$($foliation$)$''}]\;\equiv\;S_{\rm base}[\,g,A\,],
\]
with the dependence on the slicing entering only through the \emph{scalar} $A$. Thus, the full theory is $4$D gravity \emph{plus one scalar}; its equations are manifestly covariant and contain no independent $3$D fields. 
Directly calculating $S=\frac{1}{2}\int d^4 x \sqrt{-g} (R+ \lambda {}^{(3)} R)$ using the expression for the intrinsic curvature \eqref{eq:3Rscalar}, then the extrinsic curvature \eqref{Kab}, then the projectors \eqref{gammagen} and finally the normal vector \eqref{ncov} we get 
 \begin{align}
S
&=\int d^4x\,\sqrt{-g}\Bigg[
\Big(\tfrac12+\lambda\Big)R
+\frac{\lambda}{-X}\Big(\big(\nabla_m\nabla^m A\big)^2
-\nabla_m\nabla_n A\,\nabla^m\nabla^n A-\frac{1}{2(-X)}(\nabla_m X)(\nabla^m X)\cr
&+\frac{1}{{-X}}\,\big(\nabla_m\nabla^m A\big)\,(\nabla^a A\nabla_a X)\Big)
-\frac{2}{X}\,(\nabla_a\lambda)\Big(\nabla^b A\,\nabla^a\nabla_b A-\nabla^a A\,\Box A\Big)\Bigg]+\text{boundary}.\label{pure-scalartensor}
\end{align}
 
In \eqref{pure-scalartensor} the ``kinetic term'' $X\equiv g^{\mu\nu}\nabla_\mu A\nabla_\nu A$ appears in the denominator, so the theory is only defined on the branch $X\neq 0$ (typically $X<0$), where $A$ has a timelike gradient and can be interpreted as a foliation field defining a unit normal $n_\mu=-\nabla_\mu A/\sqrt{-X}$; the inverse powers $1/X$ and $1/X^2$ are thus tied to the normalization of $n_\mu$ and are expected when rewriting Gauss--Codazzi combinations in terms of $A$. 

This is one of most important results of the paper: we started from a theory highly dependent on the foliation and got a scalar-tensor theory, which does not use any foliations at all.
 This structure, however, raises potential stability concerns: the higher-derivative operators built from $\nabla_\mu\nabla_\nu A$ and $\Box A$ must satisfy a degeneracy condition (as in (D)HOST-like constructions) to avoid an extra Ostrogradsky ghost, while the inverse powers of $X$ can drive strong coupling or even singular behavior when $|X|$ becomes small, and the scalar sector may exhibit gradient instabilities depending on the background and on how the $\lambda$-constraint is realized. We defer the discussion of instability to future work.

\subsection{Making the foliation field $A$ dynamical}
\label{subsec:dynamical-foliation}

\subsubsection{General strategy and possible dynamical prescriptions.}
In the covariant formulation, the foliation is encoded by a scalar field $A$,
whose level sets define the hypersurfaces of interest. In \eqref{ncov} the normal is automatically unit timelike,
$n^a n_a=-1$,
and hypersurface orthogonal. Indeed, in differential-form notation the
normal one-form is
\begin{equation}
        n := n_a dx^a = f\,dA,
        \qquad
        f := -(-X)^{-1/2}.
\label{eq:n_form_f_def}
\end{equation}
Therefore
\begin{equation}
        dn = df\wedge dA,
        \qquad
        n\wedge dn
        =
        f\,dA\wedge df\wedge dA
        =
        0 .
\label{eq:Frobenius_A}
\end{equation}
Equivalently, the twist, or vorticity, of the normal congruence,
\begin{equation}
        \omega_{ab}
        :=
        \gamma_a{}^c\gamma_b{}^d\nabla_{[c}n_{d]},
\label{eq:twist_def}
\end{equation}
vanishes identically:
\begin{equation}
        \omega_{ab}=0 .
\end{equation}
Therefore, the scalar $A$ already provides a covariant foliation. What is
still missing is a \emph{dynamical prescription} selecting the physically
preferred foliation among all possible level-set slicing.
This can be done by adding Lagrange-multiplier sectors to a baseline theory
$S_{\rm base}$. We assume that the baseline theory is one of the covariant
formulations discussed above, for which the $A$-equation is either trivial
or does not by itself fix the foliation. The examples include the
$\lambda\,{}^{(3)}\!R$ toy model \eqref{laglam} of Sec.~3 with the covariant
formalism described in Sec.~\ref{covform}, and the non-minimal
intrinsic-curvature couplings of Sec.~4, see \eqref{anothercouplag} and
Sec.~\ref{note_covariant}.

There are several natural ways to make the slicing dynamical:
\begin{itemize}
\item \textbf{Geodesic foliation.}  
One can require the normal congruence to be geodesic, introducing the acceleration
\begin{equation}
a_b:=n^a\nabla_a n_b
\label{accel}
\end{equation}
we claim
\begin{equation}
a_b=0 .
\label{eq:geodesic-condition-list}
\end{equation}
This is imposed by a vector Lagrange multiplier $\Lambda^b$ through
\begin{equation}
S_{\rm geo}
=
\int_{\mathcal M} d^4x\,\sqrt{-g}\;\Lambda^b a_b .
\label{eq:Sgeo-list}
\end{equation}
This choice is analyzed explicitly below and in Appendix~\ref{appSgeo}.

\item \textbf{CMC or maximal slicing.}
One can prescribe the mean curvature of the leaves. It is convenient to
denote the expansion of the normal congruence by
\begin{equation}
        \Theta := \nabla_a n^a .
\label{eq:Theta_prescription_list}
\end{equation}
With the convention
\begin{equation}
        K_{ab}:=-\gamma_a{}^c\gamma_b{}^d\nabla_c n_d,
        \qquad
        K:=\gamma^{ab}K_{ab},
\end{equation}
one has \(K=-\Theta\). Hence fixing the expansion as
\(\Theta=K_0(A)\) is equivalently fixing the mean curvature as
\(K=-K_0(A)\).
Here \(K_0(A)\) is a prescribed, non-dynamical function of the slicing
label \(A\). It specifies the desired value of the expansion on each leaf
\(A=\mathrm{const}\). Since \(A\) is constant along a given leaf,
\(K_0(A)\) is constant on that leaf, which is precisely the
constant-mean-curvature condition. The special case
\(K_0(A)=K_\star=\mathrm{const}\) fixes the same mean curvature for all
leaves, and \(K_\star=0\) gives maximal slicing.
We impose the condition
\begin{equation}
        \Theta=K_0(A)
\label{eq:CMC_condition_list}
\end{equation}
by adding to the action
\begin{equation}
        S_{\rm CMC}
        =
        \int d^4x\,\sqrt{-g}\,
        \chi(x)\left(\nabla_a n^a-K_0(A)\right),
\label{eq:SCMC_list}
\end{equation}
where \(\chi\) is a scalar Lagrange multiplier.
The function \(K_0\) is fixed as part of the prescription and is not varied
in the action.

\item \textbf{Shear-free slicing.}  
The shear of the congruence is
\begin{equation}
\sigma_{ab}
:=
\gamma_a{}^c\gamma_b{}^d\nabla_{(c}n_{d)}
-\frac13\gamma_{ab}\Theta ,
\qquad
\sigma_{ab}=\sigma_{ba},
\qquad
\sigma_{ab}n^b=0,
\qquad
\sigma^a{}_a=0 .
\label{eq:shear-def}
\end{equation}
A shear-free foliation can be imposed by
\begin{equation}
S_{\rm shear}
=
\int d^4x\,\sqrt{-g}\;\Xi^{ab}\sigma_{ab},
\label{eq:Sshear-list}
\end{equation}
where $\Xi^{ab}$ is symmetric, spatial and trace-free.

\item \textbf{Alignment with a timelike Killing direction.}  
If a timelike Killing vector $V^a$ exists in the region considered, one may
dynamically align the foliation normal with it:
\begin{equation}
S_{\rm Kill}
=
\int d^4x\,\sqrt{-g}\;\lambda^a\gamma_a{}^bV_b .
\label{kill-lag}
\end{equation}
The multiplier equation enforces
\begin{equation}
\gamma_a{}^bV_b=0,
\end{equation}
or equivalently $n_a\propto V_a$. This prescription is only meaningful in
regions where such a timelike direction exists.
\end{itemize}

Each multiplier sector contributes three kinds of equations. Variation with
respect to the multiplier imposes the desired slicing condition; variation
with respect to $A$ gives a covariant conservation law; and variation with
respect to the metric gives an additional stress tensor in the metric field
equations. In the following we spell this out explicitly for the geodesic
sector and for the CMC sector.

\subsubsection{Geodesic foliation sector.}
To remove the arbitrariness of the slicing, one may require the scalar-defined
normal $n_a$ to generate geodesic integral curves. We therefore consider
\begin{equation}\label{geo-lag}
S_{\rm tot}[g,A,\Lambda;\Psi]
=
S_{\rm base}[g;\Psi]+S_{\rm geo}[g,A,\Lambda],
\qquad
S_{\rm geo}
=
\int_{\mathcal M} d^4x\,\sqrt{-g}\;\Lambda^{\,b}a_b,
\qquad
a_b:=n^a\nabla_a n_b .
\end{equation}
Here $\Lambda^b$ is a vector Lagrange multiplier and $\Psi$ denotes any matter
fields.
All variations of $S_{\rm geo}$ are derived in Appendix~\ref{appSgeo}. In
brief, variation with respect to the Lagrange multiplier gives
\begin{equation}\label{geo-cond}
\delta_{\Lambda}S_{\rm geo}:\qquad
a_b\equiv n^a\nabla_a n_b=0
\quad
\text{(cf.\ \eqref{eq:EL-lambda})}.
\end{equation}
Thus the normal congruence is geodesic.
Variation with respect to the scalar $A$ gives a covariant conservation law,
\begin{equation}
\delta_A S_{\rm geo}:\qquad
\nabla_c J^c_{(A)}=0,
\end{equation}
with
\begin{equation}
J^c_{(A)}
=
\frac{\Lambda^{\,b}}{\sqrt{-X}}
\Big(
\nabla^c n_b+n^c a_b-\gamma^c{}_b\nabla_a n^a
\Big),
\qquad
\gamma^a{}_b=\delta^a{}_b+n^a n_b .
\label{eq:Jgeo-main}
\end{equation}
This is the explicit geodesic-sector current; see also
\eqref{eq:JcA} and \eqref{eq:deltaA-structure}.
Finally, variation with respect to the metric gives a symmetric stress tensor:
\begin{equation}
\delta_g S_{\rm geo}
=
\frac12
\int_{\mathcal M}\sqrt{-g}\;
T^{\rm(geo)}_{mn}\,\delta g^{mn}
+
\int_{\partial\mathcal M} d\Sigma_a\,\mathcal B^a_{(g)} .
\label{eq:delta-g-Sgeo-main}
\end{equation}
Using \eqref{eq:Tmn-compact-article-again}, this stress tensor can be written
as
\begin{align}
T^{\rm(geo)}_{mn}
&=
-\,g_{mn}\Lambda^{\,b}a_b
+\frac12\Lambda^{\,b}
\Big[
n_m\nabla_n n_b
+n_n\nabla_m n_b
-\nabla_m(n_n n_b)
-\nabla_n(n_m n_b)
\Big]
\cr
&\quad
+\nabla_a
\Bigg\{
\frac12\Lambda^{\,b}
\Big[
n^a n_m\gamma_{nb}
+n^a n_n\gamma_{mb}
-n_m\gamma_n{}^a n_b
-n_n\gamma_m{}^a n_b
\Big]
\Bigg\}
+\frac12\Lambda^{\,b}a_b n_m n_n
\cr
&\quad
+\text{(terms proportional to $a_b$ that vanish on-shell by
\eqref{eq:EL-lambda})}.
\label{geo-EMT}
\end{align}
Therefore the total metric equation takes the form
\begin{equation}
\frac{2}{\sqrt{-g}}\frac{\delta S_{\rm base}}{\delta g^{mn}}
+
T^{\rm(geo)}_{mn}
=
T^{\rm(matt)}_{mn},
\label{eq:metric-eq-total}
\end{equation}
where
\begin{equation}
T^{\rm(matt)}_{mn}
:=
-\frac{2}{\sqrt{-g}}\frac{\delta S_{\rm matt}}{\delta g^{mn}} .
\end{equation}
The key point is that the geodesic condition is no longer a gauge choice.
It follows from the multiplier equation. The scalar $A$ still defines the
leaves, but the allowed leaves are now restricted dynamically by
$a_b=0$. For nonzero $\Lambda^b$, the additional stress
$T^{\rm(geo)}_{mn}$ also feeds back into the metric equations.

\noindent\textbf{Trivial multiplier branch.}
The conservation law $\nabla_cJ^c_{(A)}=0$ always admits the trivial solution
$\Lambda^b\equiv0 .$
On this branch $J^c_{(A)}=0$ identically, $S_{\rm geo}$ contributes no stress
tensor, and the $A$-equation becomes vacuous. Hence the full theory reduces
to the baseline theory $S_{\rm base}$. In this sense the geodesic sector
extends the original setup rather than replacing it.

\noindent\textbf{Existence of $A$: local versus global.}
Since
$n_a=-\frac{\nabla_aA}{\sqrt{-X}}$,
the one-form $n_a$ is proportional to $dA$. Therefore the congruence is
twist-free and admits the slicing $A=\mathrm{const}$ wherever $X<0$.
The multiplier equation
$a_b=n^a\nabla_a n_b=0$
then imposes that the integral curves of $n^a$ are geodesics. Equivalently,
after substituting $n_a(\nabla A)$ into $a_b=0$, one obtains a quasilinear
second-order PDE for $A$; see Appendix~\ref{appSgeo}.
Local existence is guaranteed in the usual Gaussian-normal construction:
around any point and timelike direction one may shoot geodesics orthogonal
to a spacelike hypersurface and take $A$ to be the proper time along them.
Then $X<0$, $\omega_{ab}=0$, and $a_b=0$ in a sufficiently small
neighborhood. Global obstructions may nevertheless occur, for example due to
caustics of the geodesic congruence, regions where $X\to0$, or topological
obstructions. In such cases the construction should be understood patchwise,
with overlapping charts and monotone redefinitions $A\mapsto f(A)$.

\subsubsection{CMC constraint sector.}

We impose the CMC condition by adding to the baseline theory the multiplier sector
\begin{equation}
        S_{\rm CMC}[g,A,\chi]
        =
        \int d^4x\,\sqrt{-g}\;
        \chi(x)\left(\nabla_a n^a-K_0(A)\right),
\label{eq:SCMC_main}
\end{equation}
where \(\chi(x)\) is a scalar Lagrange multiplier. Up to a boundary term,
this action can be written as
\begin{equation}
        S_{\rm CMC}
        =
        -\int d^4x\,\sqrt{-g}\;
        \left(n^a\nabla_a\chi+\chi K_0(A)\right)+\text{boundary}
\label{eq:SCMC_ibp_main}
\end{equation}
This form is useful because the covariant derivative acts only on the
scalar \(\chi\), so no explicit variation of the Levi--Civita connection
is needed.
The multiplier equation gives the CMC constraint itself,
\begin{equation}
        \frac{\delta S_{\rm CMC}}{\delta\chi}=0
        \qquad\Longrightarrow\qquad
        \nabla_a n^a=K_0(A).
\label{eq:CMC_constraint_main}
\end{equation}
The scalar-field equation $A$ takes the current form
\begin{equation}
        \nabla_a J^a_{\rm CMC}
        +
        \chi K_0'(A)
        =
        0,
        \qquad
        J^a_{\rm CMC}
        :=
        \frac{1}{\sqrt{-X}}\,
        \gamma^{ab}\nabla_b\chi .
\label{eq:CMC_A_equation_main}
\end{equation}
Thus, for a constant prescribed mean curvature, \(K_0(A)=K_\star\), the
equation becomes a genuine conservation law,
\begin{equation}
        \nabla_a J^a_{\rm CMC}=0 .
\end{equation}
For a general function \(K_0(A)\), the explicit \(A\)-dependence of the
prescription acts as a source term.
Finally, the metric variation gives an additional symmetric stress tensor
defined by
\begin{equation}
        T^{\rm(CMC)}_{ab}
        :=
        -\frac{2}{\sqrt{-g}}\,
        \frac{\delta S_{\rm CMC}}{\delta g^{ab}} 
        =
        -g_{ab}\left(n^c\nabla_c\chi+\chi K_0\right)
        +
        2n_{(a}\nabla_{b)}\chi
        +
        \left(n^c\nabla_c\chi\right)n_a n_b .
\label{eq:T_CMC_main}
\end{equation}
Therefore, for a gravitational baseline \(S_{\rm base}[g,A;\Psi]\) and a
matter action \(S_{\rm m}\), the metric equation can be written schematically
as
\begin{equation}
        \frac{2}{\sqrt{-g}}
        \frac{\delta S_{\rm base}}{\delta g^{ab}}
        =
        T^{\rm(matt)}_{ab}
        +
        T^{\rm(CMC)}_{ab},
        \qquad
        T^{\rm(matt)}_{ab}
        :=
        -\frac{2}{\sqrt{-g}}
        \frac{\delta S_{\rm m}}{\delta g^{ab}} .
\label{eq:CMC_total_metric_equation}
\end{equation}
The explicit derivation of \eqref{eq:CMC_A_equation_main} and
\eqref{eq:T_CMC_main} is given in
Appendix~\ref{app:CMC-variations}. The important point is that the CMC
condition is not imposed as a coordinate gauge choice: it follows from the
multiplier equation and therefore restricts the scalar-defined foliation
dynamically.


\subsection{Auxiliary-vector covariant completion and mimetic reduction} \label{subsec:auxiliary_vector_mimetic}
\noindent
The construction developed in this subsection was motivated by the suggestion
of Alexander Vikman.  He pointed out that for a non-constant multiplier
\(\lambda\), the intrinsic-curvature modification can be rewritten, up to a
boundary term, in a form closely related to mimetic gravity.  More precisely,
the ADM-type term \(\lambda\,{}^{(3)}\!R\) can be traded for covariant higher order scalar-tensor theory, namely variation of mimetic gravity.  We are grateful to
Alexander Vikman for this observation. In what follows we develop this idea
in detail for the constant-\(\lambda\) model and show how it leads first to
an auxiliary-vector covariant completion and then to its mimetic
specialization.

\paragraph{From the intrinsic-curvature action to a Ricci coupling.}
We start from the ADM-type action\footnote{
In this subsection we omit the cosmological constant only for notational
simplicity. It can be restored in the action by the replacement
\[
        R \;\longrightarrow\; R-2\Lambda ,
\]
or, in the auxiliary-vector form, by writing
\[
        S_n
        =
        \frac12
        \int d^4x\sqrt{-g}
        \left[
        (1+\lambda)R
        +\lambda R_{ab}n^a n^b
        -2\Lambda
        +\rho(n_a n^a+1)
        \right]
        +S_m .
\]
Equivalently, one may keep the geometrical equations without an explicit
$\Lambda$ term and include it as an effective energy-momentum tensor
\[
        T^{(\Lambda)}_{ab}=-\Lambda g_{ab}.
\]
Then the matter source is replaced by
\[
        T^{\rm(matt)}_{ab}
        \;\longrightarrow\;
        T^{\rm(matt)}_{ab}+T^{(\Lambda)}_{ab}
        =
        T^{\rm(matt)}_{ab}-\Lambda g_{ab}.
\]
With the signature $(-,+,+,+)$ this corresponds, on an FLRW background, to an
effective perfect fluid with
\[
        \varrho_\Lambda=\Lambda,
        \qquad
        p_\Lambda=-\Lambda .
\]
}
\begin{equation}
S_{\rm ADM}
=
\frac12
\int d^4x\,\sqrt{-g}
\left(
R+\lambda\,{}^{(3)}\!R
\right)
+S_{\rm m},
\label{eq:ADM_intrinsic_action}
\end{equation}
where \(\lambda\) is taken to be a spacetime constant.  At this stage
\({}^{(3)}\!R\) is the intrinsic curvature of a foliation whose unit
timelike normal is \(n^a\), satisfying
\begin{equation}
n_a n^a=-1,
\qquad
\gamma_{ab}=g_{ab}+n_a n_b .
\label{eq:unit_normal_projector_aux_start}
\end{equation}
Thus, in the derivation below, \(n^a\) is still assumed to be a
hypersurface normal.  Only after the boundary-term reduction is obtained
will we use the resulting four-dimensional bulk expression as the starting
point for an auxiliary-vector covariant completion.
{
With the convention \eqref{Kab}, \eqref{K}, \eqref{accel}, the useful decomposition of \(\nabla_a n_b\), derived in
Appendix~\ref{app:normal-decomposition}, is
\begin{equation}
        \nabla_a n_b=-K_{ab}-n_a a_b,
        \qquad
        \nabla_a n^a=-K .
\label{eq:nabla-n-decomposition}
\end{equation}
The contracted Gauss relation gives
\begin{equation}
R+2R_{mn}n^m n^n
=
{}^{(3)}\!R
+
K^2-K_{mn}K^{mn}.
\label{eq:contracted_Gauss_relation}
\end{equation}
Equivalently,
\begin{equation}
{}^{(3)}\!R
=
R
+
2R_{mn}n^m n^n
-
\left(
K^2-K_{mn}K^{mn}
\right).
\label{eq:R3_from_Gauss}
\end{equation}
We now derive explicitly how the extrinsic-curvature combination is related
to the normal-normal Ricci projection.  Using the commutator identity
\begin{equation}
R_{mn}n^m n^n
=
n_a[\nabla_b,\nabla^a]n^b ,
\label{eq:Rnn_commutator}
\end{equation}
one obtains, by applying the product rule,
\begin{align}
R_{mn}n^m n^n
={}&
\nabla_a
\left(
n^b\nabla_b n^a
-
n^a\nabla_b n^b
\right)
-
(\nabla_a n^b)(\nabla_b n^a)
+
(\nabla_a n^a)^2 .
\label{eq:Rnn_product_rule}
\end{align}
This is the same type of identity used in Appendix E of
\cite{Jha:2022svf}, where the normal-normal Ricci projection is rewritten
through a commutator of covariant derivatives and the total divergence is
then separated from the ADM bulk terms.
Using \eqref{eq:nabla-n-decomposition}, we have
\begin{equation}
n^b\nabla_b n^a=a^a,
\qquad
\nabla_a n^a=-K,
\label{eq:acceleration_and_trace_K}
\end{equation}
and
\begin{equation}
(\nabla_a n^b)(\nabla_b n^a)
=
K_{mn}K^{mn},
\qquad
(\nabla_a n^a)^2=K^2 .
\label{eq:grad_n_contractions}
\end{equation}
Therefore \eqref{eq:Rnn_product_rule} becomes
\begin{equation}
R_{mn}n^m n^n
=
\nabla_m\left(a^m+K n^m\right)
+
K^2-K_{mn}K^{mn}.
\label{eq:Rnn_K_identity}
\end{equation}
Equivalently,
\begin{equation}
K^2-K_{mn}K^{mn}
=
R_{mn}n^m n^n
-
\nabla_m\left(a^m+K n^m\right).
\label{eq:Kcombo_Rnn_boundary}
\end{equation}
Notice that this relation has not been assumed in advance; it follows from
the commutator identity \eqref{eq:Rnn_commutator} and the decomposition
\eqref{eq:nabla-n-decomposition}.
Substituting \eqref{eq:Kcombo_Rnn_boundary} into
\eqref{eq:R3_from_Gauss}, we find
\begin{align}
{}^{(3)}\!R
=R+2R_{mn}n^m n^n-\left[R_{mn}n^m n^n-\nabla_m\left(a^m+K n^m\right)\right]
=
R
+
R_{mn}n^m n^n
+
\nabla_m\left(a^m+K n^m\right).
\label{eq:R3_Ricci_boundary}
\end{align}
Thus the intrinsic-curvature scalar differs from the four-dimensional
Ricci-coupled scalar $R+R_{mn}n^m n^n\equiv\gamma^{mn}R_{mn}$ only by a total divergence.
Substituting this expression into \eqref{eq:ADM_intrinsic_action} gives
\begin{align}
S_{\rm ADM}
&=\frac12\int d^4x\,\sqrt{-g} \left[R+\lambda\,{}^{(3)}\!R\right]+S_{\rm m}
\nonumber\\
&=
\frac12
\int d^4x\,\sqrt{-g}
\left[
(1+\lambda)R
+
\lambda R_{mn}n^m n^n
\right]
+
\frac{1}{2}
\int d^4x\,\sqrt{-g}\,\lambda \, \nabla_m\left(a^m+K n^m\right)
+S_{\rm m}.
\label{eq:SADM_with_boundary_explicit}
\end{align}
}
If \(\lambda\) is constant, the last integral is a boundary contribution 
and the action can be written as
\begin{equation}
        S_{\rm ADM}
        =
        \frac12
        \int d^4x\,\sqrt{-g}
        \left[
        (1+\lambda)R
        +
        \lambda R_{mn}n^m n^n
        \right]
        +S_{\rm m}
        +\text{boundary}.
\label{eq:ADM_Ricci_form_general_n}
\end{equation}
The important point is that the equivalence
\eqref{eq:R3_Ricci_boundary} was derived assuming that \(n^a\)
is a unit hypersurface normal, because only then \({}^{(3)}\!R\) is the
intrinsic curvature of the leaves.  After the reduction, however, the bulk
term
$R_{mn}n^m n^n$
is a four-dimensional covariant scalar.  This motivates a covariant
completion in which \(n_a\) is first promoted to an independent auxiliary
timelike covector, with its normalization imposed by a Lagrange multiplier.

\paragraph{Auxiliary-vector action.}
We now take $n_a$ to be an independent auxiliary covector field.  Its
normalization is imposed by a local Lagrange multiplier $\rho(x)$:
\begin{equation}
S_n
=
\frac12
\int d^4x\,\sqrt{-g}
\left[
(1+\lambda)R
+\lambda R^{mn}n_m n_n
+\rho\left(n_m n^m+1\right)
\right]
+S_{\rm m}
+\text{boundary}.
\label{eq:auxiliary_vector_action}
\end{equation}
Equivalently, the Ricci-coupling term can be written as
$R^{mn}n_m n_n=R_{mn}n^m n^n$.  In the metric variation below we keep the
covariant components $n_a$ fixed; all metric dependence of $n^a$ comes from
raising the index with $g^{ab}$.
Variation with respect to $\rho$ gives the unit constraint
\begin{equation}
n_m n^m=-1.
\label{eq:n_unit_constraint}
\end{equation}
Variation with respect to the auxiliary covector $n_a$ gives
\begin{equation}
\delta_n S_n
=
\int d^4x\,\sqrt{-g}\,
\left(
\lambda R^{ab}n_b+\rho n^a
\right)
\delta n_a ,
\end{equation}
and therefore
\begin{equation}
\lambda R^{ab}n_b+\rho n^a=0.
\label{eq:n_variation_upper}
\end{equation}
Contracting this equation with $n_a$ and using $n_a n^a=-1$, we obtain
\begin{equation}
\rho
=
\lambda R_{mn}n^m n^n .
\label{eq:rho_from_n_equation}
\end{equation}
Thus the independent auxiliary vector is not arbitrary on shell: it must be
an eigenvector of the Ricci tensor. Contracting \eqref{eq:n_variation_upper} with the induced metric $\gamma^{ac}$ we obtain 
\begin{equation}
\gamma^{ac} R_{cb}n^b=0 .
\label{eq:projected_n_equation}
\end{equation}

\paragraph{Metric variation.}
The metric variation of the action \eqref{eq:auxiliary_vector_action} gives
\begin{equation}
(1+\lambda)G_{ab}
-\rho\,n_a n_b
+\lambda\,\mathcal{H}_{ab}[n]
=
T_{ab},
\label{eq:metric_equation_auxiliary_n}
\end{equation}
with
\begin{align}
\mathcal{H}_{ab}[n]
={}&
-\frac12 g_{ab}\,n^{qc}R_{qc}
+n_a{}^q R_{bq}
+n_b{}^q R_{aq}
+\frac12 g_{ab}\nabla_c\nabla_q n^{qc}
-\frac12\nabla_q\nabla_a n_b{}^q
-\frac12\nabla_q\nabla_b n_a{}^q
+\frac12 \Box n_{ab}.
\label{eq:H_ab_auxiliary_n}
\end{align}
where we have introduced the shorthand notation
\begin{equation}
n_{ab}:=n_a n_b,
\qquad
n_a{}^b:=n_a n^b,
\qquad
n^{ab}:=n^a n^b,
\qquad
\Box:=g^{mn}\nabla_m\nabla_n .
\end{equation}
The form of  \(\mathcal H_{ab}[n]\) arises from the standard identity for the integrated variation of \(\int d^4x\sqrt{-g}\,n^{ab}R_{ab}\), equivalently from the formal adjoint of the linearized Ricci, or Lichnerowicz, operator acting on the symmetric tensor \(n^{ab}\); see e.g. Refs.~\cite{Lichnerowicz1961,DeWitt1965,Wald1984}.
Equation \eqref{eq:metric_equation_auxiliary_n} is the covariant
metric equation of the auxiliary-vector formulation.
The variation with respect to the global multiplier $\lambda$ gives
\begin{equation}
0
=
\frac{\partial S_n}{\partial\lambda}
=
\frac12
\int d^4x\,\sqrt{-g}
\left(
R+R_{mn}n^m n^n
\right).
\label{eq:lambda_global_constraint_auxiliary}
\end{equation}
When $n_a$ is hypersurface orthogonal and the boundary terms are discarded,
this is equivalent to the global intrinsic-curvature constraint
\begin{equation}
\int d^4x\,\sqrt{-g}\;{}^{(3)}\!R=0 .
\label{eq:global_R3_constraint_auxiliary}
\end{equation}
Substituting $\rho $ from \eqref{eq:rho_from_n_equation} we have the EFE (Einstein field equations)
\begin{equation}\label{EFE-n-aux}
     (1+\lambda)G_{ab}
- \lambda R_{mn}n^m n^n n_a n_b
+\lambda\,\mathcal{H}_{ab}
=
T_{ab},
\end{equation}
\paragraph{Remark on hypersurface orthogonality.}
At this stage $n_a$ is a general unit timelike auxiliary covector.  A generic
unit timelike covector is not necessarily the normal to a foliation.  To
interpret it literally as a hypersurface normal one must either impose the
Frobenius condition
\begin{equation}
n_{[a}\nabla_b n_{c]}=0,
\label{eq:Frobenius_condition}
\end{equation}
or choose a scalar potential $A$ such that $n_a$ is proportional to
$\nabla_a A$.  The mimetic construction below implements the second option
directly.

\paragraph{Mimetic specialization: $n_a=-\nabla_a A$.}
We now specialize the auxiliary covector to the mimetic, hypersurface-normal
form
\begin{equation}
n_a=-\nabla_a A .
\label{eq:mimetic_normal}
\end{equation}
The unit constraint \eqref{eq:n_unit_constraint} becomes the usual mimetic
constraint
\begin{equation}
\nabla_m A\nabla^m A=-1 .
\label{eq:mimetic_constraint}
\end{equation}
Substituting \eqref{eq:mimetic_normal} into
\eqref{eq:auxiliary_vector_action}, we obtain
\begin{equation}
S_A
=
\frac12
\int d^4x\,\sqrt{-g}
\left[
(1+\lambda)R
+\lambda R_{mn}\nabla^m A\nabla^n A
+\rho\left(\nabla_m A\nabla^m A+1\right)
\right]
+S_{\rm m}
+\text{boundary}.
\label{eq:final_simplified_action}
\end{equation}
This is the second-order, boundary-term equivalent form of the mimetic
completion.
For constant $\lambda$, the Ricci-coupling term is equivalent, up to a
boundary term, to the higher-derivative scalar combination
\begin{equation}
R_{mn}\nabla^m A\nabla^n A
=
(\Box A)^2
-
\nabla_m\nabla_n A\,\nabla^m\nabla^n A
+\nabla_m(\cdots).
\label{eq:Ricci_coupling_to_HD_scalar}
\end{equation}
Therefore, for constant \(\lambda\), the action can be written as
\begin{equation}
        S_A
        =
        \frac12
        \int d^4x\,\sqrt{-g}
        \left[
        (1+\lambda)R
        +
        \lambda
        \left(
        (\Box A)^2
        -
        \nabla_m\nabla_n A\,\nabla^m\nabla^n A
        \right)
        +
        \rho\left(\nabla_m A\nabla^m A+1\right)
        \right]
        +S_{\rm m}
        +\text{boundary}.
\label{eq:mimetic_HD_form_boundary}
\end{equation}

\paragraph{Variation with respect to $A$: conserved current.}
In the mimetic theory, $n_a$ is no longer an independent auxiliary field.
Therefore one should not impose the algebraic equation
\eqref{eq:n_variation_upper}.  Instead,
\begin{equation}
\delta n_a=-\nabla_a\delta A .
\end{equation}
Using the auxiliary-vector variation
\eqref{eq:n_variation_upper}, the $A$-variation becomes
\begin{align}
\delta_A S_A
=
-\int d^4x\,\sqrt{-g}\,
\left(
\lambda R^{ab}n_b+\rho n^a
\right)
\nabla_a\delta A
=
\int d^4x\,\sqrt{-g}\,
\nabla_a
\left(
\lambda R^{ab}n_b+\rho n^a
\right)
\delta A .
\end{align}
Hence the scalar equation is
\begin{equation}
\nabla_a
\left(
\lambda R^{ab}n_b+\rho n^a
\right)=0 .
\label{eq:A_equation_n_form}
\end{equation}
Using $n_a=-\nabla_a A$, this can be written as the conserved-current
equation
\begin{equation}
\nabla_a J^a=0,
\qquad
J^a
:=
2\left(
\lambda R^{ab}+\rho g^{ab}
\right)
\nabla_b A .
\label{eq:A_eom_current}
\end{equation}
The overall factor of $2$ is conventional.
Thus, in the free auxiliary-vector theory the $n_a$-variation gives the
algebraic Ricci-eigenvector condition \eqref{eq:n_variation_upper}, whereas
in the mimetic theory the same structure is converted into the differential
conservation law \eqref{eq:A_eom_current}.

\paragraph{Metric equations in the mimetic case.}
Since in the mimetic formulation the covariant component
$n_a=-\nabla_a A$ is metric independent, the metric equation is obtained
from \eqref{eq:metric_equation_auxiliary_n} by the substitution
$n_a=-\nabla_a A$:
\begin{equation}
(1+\lambda)G_{ab}
-\rho\,n_a n_b
+\lambda\,\mathcal{H}_{ab}[n]
=
T_{ab},
\qquad
n_a=-\nabla_a A .
\label{eq:metric_final_compact}
\end{equation}
Explicitly,
\begin{align}
(1+\lambda)G_{ab}
-\rho\,n_a n_b
+\lambda\Big[
&-\frac12 g_{ab}\,n^{qc}R_{qc}
+n_a{}^q R_{bq}
+n_b{}^q R_{aq}
\nonumber\\
&+\frac12 g_{ab}\nabla_c\nabla_q n^{qc}
-\frac12\nabla_q\nabla_a n_b{}^q
-\frac12\nabla_q\nabla_b n_a{}^q
+\frac12\Box n_{ab}
\Big]
=
T_{ab}.
\label{eq:metric_final_compact_explicit}
\end{align}
Together with the mimetic constraint
\eqref{eq:mimetic_constraint} and the conserved-current equation
\eqref{eq:A_eom_current}, this gives the covariant mimetic completion of the
intrinsic-curvature suppression model. In cosmological applications, the mimetic term behaves like a dust component, as in the basic mimetic gravity theory. We shall analyze this case in section \ref{sec:applications}, when we discuss applications.

\subsubsection{Relation to Einstein-aether and khronometric theories.}
It is useful to note that the auxiliary-vector completion is closely related
to a particular version of Einstein-aether theory
\begin{equation}
S_{\ae}
=
\frac{1}{2}
\int d^4x\,\sqrt{-g}\,
\left[
R
-
K^{ab}{}_{mn}\nabla_a u^m \nabla_b u^n
+
\ell\left(u^a u_a+1\right)
\right]
+S_{\rm m},
\label{eq:Einstein_aether_action}
\end{equation}
where $u^a$ is the aether vector field and $\ell$ is a Lagrange multiplier
enforcing the unit timelike constraint
\begin{equation}
u^a u_a=-1 .
\end{equation}
The tensor $K^{ab}{}_{mn}$ is given by
\begin{equation}
K^{ab}{}_{mn}
=
c_1 g^{ab}g_{mn}
+
c_2 \delta^a_m\delta^b_n
+
c_3 \delta^a_n\delta^b_m
-
c_4 u^a u^b g_{mn},
\label{eq:Einstein_aether_K_tensor}
\end{equation}
where $c_1,c_2,c_3,c_4$ are dimensionless coupling constants.
It is a generally covariant
gravity theory coupled to a dynamical unit timelike vector field
\cite{Jacobson:2000xp,Eling:2004dk,Jacobson:2008aj}. Indeed, after discarding
the boundary term, the Ricci coupling can be written as
\begin{equation}
        R_{ab}n^a n^b
        =
        (\nabla_a n^a)^2
        -
        \nabla_a n_b \nabla^b n^a + \text{div}.
\end{equation}
Therefore, if $\lambda$ is regarded as a fixed coupling constant rather than
as a Lagrange multiplier, the action \eqref{eq:auxiliary_vector_action}
is equivalent, up to a boundary term, to a one-parameter Einstein-aether-type
theory with a unit timelike vector $n^a$. In the standard Einstein-aether
notation this corresponds to the special choice (where the coefficient before $R$ is normalized to 1)
\begin{equation}
        c_1=0,
        \qquad
        c_2=-\frac{\lambda}{1+\lambda},
        \qquad
        c_3=\frac{\lambda}{1+\lambda},
        \qquad
        c_4=0,
\end{equation}
up to convention-dependent signs.

If, in addition, the aether vector is restricted to be hypersurface
orthogonal, either by imposing the Frobenius condition or by writing
$n_a\propto\nabla_a A$, the theory becomes closely related to the
khronometric, or Einstein-khronon, formulation. This is precisely the
hypersurface-orthogonal sector of Einstein-aether theory and is also the
covariant low-energy description of non-projectable Hořava gravity
\cite{Horava:2009uw,Jacobson:2010mx,Blas:2010hb}. In this language the scalar
field $A$ plays the role of the khronon, namely the field that defines the
preferred time foliation. The mimetic specialization $n_a=-\nabla_a A$,
together with the constraint $\nabla_aA\nabla^aA=-1$, is therefore a more
restrictive mimetic/khronon realization of the same auxiliary-vector
construction, similar in spirit to mimetic gravity
\cite{Chamseddine:2013kea,Chamseddine:2014vna}. If $\lambda$ is varied as a
global multiplier, however, the theory should be viewed not as ordinary
Einstein-aether theory, but as an Einstein-aether-like constrained theory
supplemented by the global intrinsic-curvature condition.

\section{Theories with non-minimal coupling to gravity}\label{sec:nonmincoupl}\setcounter{equation}{0}     
We shall now try to construct a dynamical cancellation of ${}^{(3)}R$. For this purpose, we introduce a dynamical scalar field $\phi$, and also couple it to ${}^{(3)}R$. We will mostly use the ADM formalism, and then we will discuss the covariant formulation.  
\subsection{$4-D$ non-minimal coupling}\m{4dcoupl}
Consider a scalar field non-minimally coupled to gravity,
\begin{equation}\label{4dconformallag}
  S= \frac{1}{2}  \int d^4 x \sqrt{-g} \left[R-2\Lambda-(\nabla^a\phi\nabla_a\phi+V(\phi)+\xi R \phi^2 + \eta R \phi)\right] 
\end{equation}
with the metric part of the action $  S_g=  \int [\frac{1}{2} R-2\Lambda]\sqrt{-g}d^4x$ and the scalar field part $  S_{\phi}=  \int -\frac{1}{2}(\nabla^a\phi\nabla_a\phi+V(\phi)+\xi R \phi^2 + \eta R \phi) \sqrt{-g}d^4x$. 
The field energy-momentum is defined as
\begin{equation}
    T_{ab} = -\frac{2}{\sqrt{-g}} \frac{\delta S_{\phi}}{\delta g^{ab}}.
\end{equation}
The equations of motion are 
(see detailed derivation in the Appendix \ref{emt4d}):
\bea \label{eq4dmet}
    G_{ab}+g_{ab} \Lambda &=& \nabla_a\phi \nabla_b\phi-\frac{1}{2}g_{ab} [\nabla^c\phi\nabla_c\phi+V(\phi)]+(g_{ab}\nabla_c\nabla^c- \nabla_a  \nabla_b )(\xi \phi^2+\eta\phi) +(\xi \phi^2+\eta\phi) G_{ab}, \\
\label{eq4dphi}
    \nabla_a \nabla^a \phi &-&\frac{1}{2}V'(\phi) -\xi R \phi -\frac{1}{2} \eta R =0.
\eea
Taking the trace of (\ref{eq4dmet})
we get
\begin{equation}\label{traceeq4met}
    -R+4\Lambda=-\nabla_a\phi \nabla^a\phi-2V(\phi)+3 \nabla_c \nabla^c(\xi \phi^2+\eta\phi)-(\xi \phi^2+\eta\phi)R.
\end{equation}
Using in (\ref{traceeq4met}) that 
\begin{equation}
    \nabla_a \nabla^a \phi^2 = 2 \nabla_a \phi \nabla^a \phi + 2 \phi \nabla_a \nabla^a \phi
\end{equation}
and then replacing $\nabla_a \nabla^a \phi$ by (\ref{eq4dphi}) in (\ref{traceeq4met}) we get
\be
-R + 4\Lambda =
(6\xi - 1) \nabla^a \phi \nabla_a \phi - 2V(\phi)
+ \left(3\xi \phi + \tfrac{3}{2} \eta \right) V'(\phi)
+ R \left[ (6\xi - 1)\xi \phi^2 + (6\xi  - 1) \eta\phi +  \tfrac{3}{2} \eta^2  \right].
\ee
For $\eta=0, \xi=1/6$, and $V=0$ we reproduce the standard conformal coupling of a scalar field to gravity including the vanishing of the trace of the energy momentum tensor. Another interesting situation is that with the conformal coupling $\xi=1/6$, and $V=0$ the trace of the energy momentum tensor deviates from zero only at order $\eta^2$. Hence, one can have a soft breaking of the conformal symmetry by controlling the size of $\eta$, making the breaking technically natural.
Since we will be interested in removing $^{(3)}R$ from the equations, or forcing it to vanish, the conformal coupling will not suffice, and we will need to consider constant solutions instead.
One can see that for $\phi=const.,\, V=0, \Lambda=0$,
\begin{equation}
    R \left[(6 \xi-1)\xi \phi^2 + \eta (6\xi-1) \phi  + \frac{3}{2}\eta^2+1\right]=0.
\end{equation}
Solutions of the quadratic equation for constant $\phi$ in square brackets are then solutions of the traced equation for any value of $R$.
However, there is a different solution that makes the curvature drop out from the action and from the field equations. We first consider a constant solution for \eqref{eq4dphi} which occurs for
 $   \phi =- \frac{\eta}{2\xi} $. Hence, integrating out $\phi$ modifies the strength of curvature in the action and field equations.  Substituting the solution into the action, the Ricci scalar drops out from the action and field equations if $\xi=-\frac{1}{4} \eta^2$. 
This will be the more relevant solution for our case, since we plan to drop $^{(3)}R$ from the equations using this method.
The question then arises, if the Ricci scalar (3 or 4 dimensional) drops from the action, does it appear in the field equations. The answer to this question is provided by Caroll gravity.

\subsection{Carroll gravity}\m{carroll}
Carroll gravity can be described as a power series in a vanishing speed of light limit of GR and compare it to the theories we study.
According to \cite{Hansen:2021fxi,deBoer:2023fnj}, we can rewrite the Einstein--Hilbert action in the form:
\begin{equation}
S_{{EH}} = c^2 S_{{LO}} + c^4 S_{{NLO}} + \mathcal{O}(c^6),
\end{equation}
where LO is leading order term and NLO is next to the leading order term.
The LO term of the Einstein--Hilbert action is
\begin{equation}
S_{{LO}} = \frac{1}{2 G} \int_M \left( K^{mn}K_{mn} - K^2 \right) e \, \mathrm{d}^{d+1}x,
\end{equation}
where \begin{equation}e = \det(\tau_m, e^a_m)\end{equation} is the tetrad determinant. $\tau_m$ is a normal vector to the hypersurfaces of the foliation and $ e^a_m$ are the tangent vectors. This action depends only on the extrinsic curvature invariants and not on the Ricci scalar. The intrinsic curvature is canceled from the leading order action as in the theory \eqref{laglam} of $\lambda=-1$. ${}^{(3)} R$ appears only at next to leading order, so it is very small. Thus, the Carroll gravity expansion of GR is a way to consistently make ${}^{(3)} R$ small or vanishing.

The major point we wish to emphasize here is what are the terms that appear in the equations of motion order by order. As mentioned, the leading order action 
does not include 3- or 4-dimensional Ricci tensors or scalars \cite{deBoer:2023fnj}. As a result, these tensors do not appear in the equations of motion as well. This is important for our further findings. We will show how  ${}^{(3)} R$ is dropped out from the action, which will imply that it also disappears from the field equations.

\subsection{Non-minimal coupling to the intrinsic curvature I}\m{3dconfcoupl}
In this subsection, we will study the case of the intrinsic curvature coupled to a scalar field. We will first consider the following action:
\begin{equation}\label{confR3action}
S = \frac{1}{2}\int d^4x \sqrt{-g}\left[R-2\Lambda - \left( \nabla_a \phi \nabla^a \phi + V(\phi) + \xi {}^{(3)} R \phi^2 \right)\right]
\end{equation}
with gravitational part $S_g = \int \frac{\sqrt{-g}}{2} (R-2\Lambda )d^4 x$ and matter part  $S_m = \int- \frac{1}{2} \left( \nabla_a \phi \nabla^a \phi + V(\phi) + \xi {}^{(3)} R \phi^2 \right) \sqrt{-g} \, d^4 x$. 
The field equations are:
\bea\label{eqphi}
    \Box \phi - \xi {}^{(3)} R \phi -\frac{1}{2} V'=0,\\
\label{eomemtensor}
          R_{ab}-\frac{1}{2} g_{ab} R+g_{ab} \Lambda=  T_{ab},
\eea
where the matter energy-momentum defined as
 $   T_{ab} = -\frac{2}{\sqrt{-g}} \frac{\delta S_m}{\delta g^{ab}}$ is:
\begin{equation}\label{emtensor}
T_{ab} = \nabla_a \phi \nabla_b \phi - \frac{1}{2} g_{ab} \nabla_c \phi \nabla^c \phi - \frac{1}{2} g_{ab} V(\phi) + \frac{\xi}{N}(\gamma_{a b} D_c D^c -D_{a} D_{b})(\phi^2 N) + \xi {}^{(3)} R_{ab} \phi^2   - \frac{1}{2} {}^{(3)} R g_{ab} \xi \phi^2.
\end{equation}
The derivation can be found in Appendix \ref{emt3d}. 
The trace equation takes the form,
\begin{equation}\label{traceeom}
     - R +4\Lambda=- \nabla_a \phi \nabla^a \phi   - 2 V(\phi)
- \xi {}^{(3)} R \phi^2 +\frac{\xi}{N}(2 D_c D^c)(\phi^2 N) .
\end{equation}
Considering the ADM formalism, the following replacements can be performed consistently:
\begin{equation}
    D_i \phi = \gamma_i^a \nabla_a \phi,~~ \gamma_i^a=\delta_i^a + n_i n^a, ~~~n^i=-g^{0i} N,~~n_i = -\delta^0_i N.
    \end{equation}
To simplify the analysis, we will choose the synchronous gauge $N=1$. Using the field equation for $\phi$, and for the conformal coupling in 3 dimensions $\xi=1/4$, and $V=0$, the trace
    equation simplifies considerably,
\be
-R+4\Lambda=\frac{1}{2}\partial_0^2(\phi^2).
\ee
This is the spatial version of conformal coupling of a scalar field to gravity. Thus, as expected in 3D conformal coupling, we find that due to conformal coupling the trace almost vanishes, and we are left with a single term with time derivatives only.

To remove ${}^{(3)} R$ from the equations of motion, or the action, we can consider constant solutions of $\phi$, for example $\phi=\xi^{-1/2}$. However, such a solution fulfills the field equation only for constant ${}^{(3)} R$. So this solution is again a Lagrange multiplier solution in disguise.

\subsection{Non-minimal coupling to the intrinsic curvature II}\m{anothercoupl}
Consider the action,
\begin{equation}\label{anothercouplag}
S = \frac{1}{2}\int  d^4 x \sqrt{-g}\left[R-2\Lambda -  \left( \nabla_a \phi \nabla^a \phi + V(\phi) + \xi {}^{(3)} R \phi^2+ \eta {}^{(3)} R \phi \right)\right] .
\end{equation}
The equations of motion are:
\bea 
\label{eqphi2}
    \Box \phi &-& \xi {}^{(3)} R \phi - \frac{1}{2}\eta {}^{(3)} R -\frac{1}{2} V' =0, \\ 
\label{eqmet}
    G_{ab}+ g_{ab} \Lambda&=& \nabla_a \phi \nabla_b \phi - \frac{1}{2} g_{ab} \nabla_c \phi \nabla^c \phi - \frac{1}{2} g_{ab} V(\phi) + \frac{\xi}{N}(\gamma_{a b} D_c D^c -D_{a} D_{b})(\phi^2 N) + \xi {}^{(3)} R_{ab} \phi^2   - \frac{1}{2} {}^{(3)} R g_{ab} \xi \phi^2  \cr
     &+& \frac{\eta}{N}(\gamma_{a b} D_c D^c -D_{a} D_{b})(\phi N) + \eta {}^{(3)} R_{ab} \phi   - \frac{1}{2} {}^{(3)} R g_{ab} \eta \phi,  
\eea
where $G_{ab}$ is the Einstein tensor.
To remove the intrinsic curvature from all field
equations, we again consider constant solutions of $\phi$, such as
$\Box \phi=V=0$. 
  $  \phi = -\frac{\eta}{2 \xi}$ is a solution of \eqref{eqphi2} for any ${}^{(3)} R$.
Substituting this solution back into the action, we now have 
\begin{equation} \label{eq:integratedout}
    S= \int d^4x\frac{\sqrt{-g}}{2}  \left(K_{ji} K^{ij} - K^2+{}^{(3)} R\left(1+\frac{\eta^2}{ 4\xi}\right)\right).
\end{equation} 
Hence, we control the coupling strength of the intrinsic curvature, and it is no longer similar to the extrinsic curvature. Specifically, it will be completely removed from the action and equations for $\xi=-\eta^2/4$.
The gravity part of Einstein's field equations \eqref{eqmet} will be exactly as (\ref{fieldeq}), only now $\lambda=\frac{\eta^2}{ 4\xi}$, and it is no longer a Lagrange multiplier, but simply a coupling constant.  This is perhaps the most intriguing result of the paper. 
We have constructed a theory where the intrinsic curvature in the field equation enters with different coefficient from the one in the geodesic equation (see in \ref{subs-frlw}). A phenomenological approach of this idea in a formulation of two different curvatures, has recently been used in cosmological analysis \cite{Shimon:2024mbm}.  

\subsection{A note on the covariant formalism}
\label{note_covariant}

The covariant formalism is also applicable to the theory with action
\eqref{anothercouplag}.  In this subsection we only explain how the
argument of subsection~\ref{covform} is modified in the presence of the
non-minimal intrinsic-curvature coupling.  We assume \eqref{ncov}, so that
the unit normal is written in terms of the scalar field \(A\).  The variation
with respect to \(\phi\) does not involve the normal vector and therefore
gives the same scalar-field equation as before, namely \eqref{eqphi2}.
It remains to discuss the variations with respect to \(A\) and \(g_{ab}\).
As in subsection~\ref{covform}, the only object multiplying the variation of
the induced metric is a purely spatial tensor.  In the present case it is
\begin{equation}
    B^{mn}
    =
    {}^{(3)}R^{mn}(\xi \phi^2+\eta\phi)
    -
    \frac{1}{N}
    \left(
    D^mD^n-\gamma^{mn}D_cD^c
    \right)
    (N\xi\phi^2+N\eta\phi),
\label{eq:Bmn_nonminimal_cov}
\end{equation}
where
\begin{equation}
    N=\sqrt{\frac{-g}{\gamma}}.
\end{equation}
The tensor \(B^{mn}\) has the same property as in subsection~\ref{covform}:
it is three-dimensional, or spatial, and symmetric in both indices. Therefore, its
contraction with the normal vanish,
\begin{equation}
    B^{mn}n_n=0 .
\label{eq:Bmn_spatial_nonminimal}
\end{equation}
Since the scalar-defined normal satisfies
\begin{equation}
    \partial_m A=-\sqrt{-X}\,n_m.
\end{equation}
All terms in the \(A\)-variation are proportional to contractions of
\(B^{mn}\) with \(\partial_mA\) or \(\partial_nA\).  Hence, by the same
argument as in subsection~\ref{covform}, the \(A\)-equation is identically
trivial:
\begin{equation}
    B^{an}
    \frac{-2 \partial_n A}{\partial_a A\partial_b A g^{ab}}
    +
    B^{mn}
    \frac{
    2 \partial_m A \partial_n A \partial_b A g^{ab}
    }{
    (\partial_a A\partial_b A g^{ab})^2
    }
    \equiv 0 .
\label{covb01}
\end{equation}
Thus, the covariant rewriting by itself does not generate an independent
equation selecting the foliation.
The metric variation is simplified in the same way.  The additional term
which appears from the \(A\)-dependence of the induced metric is again
proportional to contractions of \(B^{mn}\) with the normal, and therefore
vanishes by \eqref{eq:Bmn_spatial_nonminimal}.  Thus, we do not need to write
the intermediate expression explicitly.  The final metric equations are
\begin{eqnarray}
     R^{a'b'}
     -\frac{1}{2} g^{a'b'} R
     +g^{a'b'} \Lambda
     &=&
     \nabla^{a'} \phi \nabla^{b'} \phi
     - \frac{1}{2} g^{a'b'} \nabla_c \phi \nabla^c \phi
     - \frac{1}{2} g^{a'b'} V(\phi)
     +  (\xi \phi^2+ \eta \phi)
     \left(
     {}^{(3)} R^{a'b'}
     -\frac{1}{2} g^{a'b'} {}^{(3)} R
     \right)
     \cr
     \displaystyle
     &-&\frac{1}{N}
     \left(
     D^{a'} D^{b'}
     + \gamma^{a'b'} D_c D^c
     \right)
     (N\xi \phi^2+N\eta \phi).
\label{eq:cov_metric_nonminimal_final}
\end{eqnarray}

The result is therefore the following.  In the covariant formalism the
metric equations are the same as in the usual ADM formulation, but the
foliation is represented covariantly by the scalar field \(A\).  Hence, the
equations are invariant under coordinate transformations, while the
triviality of \eqref{covb01} shows that the covariant rewriting alone does
not dynamically select a preferred foliation.

Finally, all explicit three-dimensional geometrical quantities can be
eliminated by rewriting \eqref{anothercouplag} in the spirit of
\eqref{pure-scalartensor}.  Using \(X=g^{ab}\nabla_aA\nabla_bA\), one obtains
\begin{align}
S_{\rm bare}
&=
\int d^4x\,\sqrt{-g}\Bigg[
\left(\frac12-(\xi\phi^2+\eta\phi)\right)R
-\frac{(\xi\phi^2+\eta\phi)}{-X}
\Bigg(
\big(\nabla_\mu\nabla^\mu A\big)^2
-\nabla_\mu\nabla_\nu A\,\nabla^\mu\nabla^\nu A
\nonumber\\
&\hspace{3.2cm}
-\frac{1}{2(-X)}
(\nabla_\mu X)(\nabla^\mu X)
+
\frac{1}{-X}
\big(\nabla_\mu\nabla^\mu A\big)
(\nabla^\rho A\nabla_\rho X)
\Bigg)
\nonumber\\
&\hspace{1.2cm}
+\frac{2}{X}
\big(\nabla_\alpha(\xi\phi^2+\eta\phi)\big)
\left(
\nabla^\beta A\,\nabla^\alpha\nabla_\beta A
-
\nabla^\alpha A\,\Box A
\right)
-\nabla_a\phi\nabla^a\phi
-V(\phi)
\Bigg]
+\text{boundary}.
\label{eq:bare_nonminimal_scalar_tensor}
\end{align}
Thus the theory can be written as an exotic scalar-tensor theory without
explicit reference to three-dimensional geometrical objects.  One can then
either make the foliation dynamical by adding the multiplier sectors of
subsection~\ref{subsec:dynamical-foliation}, with \(S_{\rm base}\) taken to
be \eqref{anothercouplag}, or construct the corresponding mimetic version of
\eqref{anothercouplag}.

\section{Applications}\label{sec:applications}\setcounter{equation}{0}
After constructing the theories with vanishing or suppressed intrinsic curvature, let us apply these models in several settings such as Cosmology, Black Holes, Gravitational Waves and the Lense-Thirring effect. Each example is evaluated on a foliation defined by $A$ which is chosen dynamically,  and we take convenient coordinates to produce the ADM foliation characterized by the hypersurfaces of constant time so that $A=f(t)$. This implies that we can make the transformation to any other coordinates because of the existence of the covariant formalism and field reformulation of foliation.
\subsection{FLRW universe}\label{subs-frlw}

We consider the FLRW (Friedmann–Lemaître–Robertson–Walker) metric in the following form:
\begin{equation}\label{metfrid}
    ds^2=-dt^2+a^2(t)\gamma_{ij}dx^i dx^j=-dt^2+a^2(t)\l(\frac{1}{1-kr^2}dr^2+r^2 \l(d\theta^2+ \sin ^2\theta  d\phi^2 \r)\r),
\end{equation}
where $k=+1$ for a positively curved space, $k=0$ for a flat space and $k=-1$ for a negatively curved space.
In all the covariant completions described above, by solving the equations of motion, one can pick the normal vector to be the usual 
\begin{equation}
    n_m=\{-1,0,0,0\}.
\end{equation}
The induced metric (\ref{gamma}) is
\begin{equation}
    \gamma_{mn}=\left(
\begin{array}{cccc}
 0 & 0 & 0 & 0 \\
 0 & \frac{a(t)^2}{1-k r^2} & 0 & 0 \\
 0 & 0 & r^2 a(t)^2 & 0 \\
 0 & 0 & 0 & r^2 a(t)^2 \sin ^2 \theta  \\
\end{array}
\right).
\end{equation}
The extrinsic curvature (\ref{Kab}) is
\begin{equation}
    K_{11} = -\frac{a(t) a'(t)}{1-k r^2},~~~~  K_{22} =-r^2 a(t) a'(t),~~~~   K_{33} =-r^2 a(t) \sin ^2 \theta  a'(t) .
\end{equation}
The intrinsic curvature scalar calculated by (\ref{3Ricci}) and (\ref{R3}) is
\begin{equation}
    {}^{(3)} R=6\frac{k}{a^2}.
\end{equation}

\textbf{Theory with the Lagrange multiplier.}
For the theories with the Lagrange multiplier (\ref{laglam}) and (\ref{lagReq}) we impose the integral constraint $ \int  {}^{(3)} R\sqrt{-g}~  d^4 x=0$. This can be achieved by $k=0$. All non-zero components of $ {}^{(3)} R_{ab} $ are proportional to $k$, thus,
 $   {}^{(3)} R_{ab} =0.$
$N$ is a scalar and $N=1$, then $D_i N =0$.
So, all the terms in brackets in (\ref{fieldeq}) with $\lambda$ are zero, 
so the equations reduce to the usual Friedmann equations for a flat universe 
\bea 
   \left( \frac{\dot{a}}{a} \right)^2   &=& \frac{1}{3} \rho + \frac{\Lambda}{3},\cr
    \frac{\ddot{a}}{a}\quad &=& -\frac{1}{6} \left( \rho + 3p \right) + \frac{\Lambda}{3},
\eea
where $\rho$ is total energy density, and $p$  is the total pressure of all energy-momentums (matter + \eqref{eq:T_CMC_main} or \eqref{geo-EMT}) in the R-H-S of \eqref{fieldeq}.

{
\textbf{Theory with non-minimal coupling to gravity.}
The theory with non-minimal coupling to gravity studied in the subsection \ref{anothercoupl} modifies the coupling of the intrinsic curvature, $ {}^{(3)} R \rightarrow \left(1+\frac{\eta^2}{4\xi}\right)  {}^{(3)} R$. As a result, the Friedmann equations are modified to: 
\bea \label{eq:etaFriedmann}
 \left( \frac{\dot{a}}{a} \right)^2   &=& \frac{1}{3} \rho + \frac{\Lambda}{3} - \frac{k}{a^2}\left(1+\frac{\eta^2}{4\xi}\right),\cr
     \frac{\ddot{a}}{a} &=& -\frac{1}{6} \left( \rho + 3p \right) + \frac{\Lambda}{3}.
\eea

Of particular interest is the case where ${}^{(3)} R$ completely disappears from the equations. {First, as a curiosity,} consider a vacuum solution with no CC. In that case, space can be curved with $k\neq 0$, and in the usual GR case $k=-1$ gives a coasting universe with $a(t)\propto t$. In our case, this term will not appear in the equations, and the Friedmann equations reduce to 
\begin{equation}
   \dot{a}=0,~~\dot{a}^2+2a\ddot{a}=0,
\end{equation}
where the only solution is Minkowski space.

\subsubsection{Geodesic equation vs.\ effective curvature in FLRW}\label{sec:FLRW-geo-clarify}

Geodesics are curves \(x^\mu(\lambda)\) satisfying
\begin{equation}
\ddot{x}^\mu+\Gamma^\mu_{\alpha\beta}\dot{x}^\alpha\dot{x}^\beta=0,
\qquad
\epsilon:=g_{\mu\nu}\dot x^\mu\dot x^\nu=
\begin{cases}
-1 & \text{timelike (massive),}\\
0 & \text{null (lightlike).}
\end{cases}
\label{eq:FLRW-geod-master}
\end{equation}
Here overdot is \(d/d\lambda\), and \(\epsilon\) encodes the normalization.
For \eqref{metfrid} the nonzero components of the Christoffel symbols are
\begin{equation}
\Gamma^t_{ij}=a\dot a\,\gamma_{ij},\qquad
\Gamma^i_{tj}=\Gamma^i_{jt}=\frac{\dot a}{a}\,\delta^i{}_j,\qquad
\Gamma^i_{jk}={}^{(3)}\Gamma^i{}_{jk}(\gamma),\qquad
\Gamma^t_{tt}=\Gamma^t_{ti}=0.
\label{eq:FLRW-Christoffels}
\end{equation}
The spatial Christoffels \({}^{(3)}\Gamma^i{}_{jk}(\gamma)\) are built from the induced metric \(\gamma_{ij}\), whose constant sectional curvature is
\(
{}^{(3)}\!R = 6k/a^2
\).
Thus the same curvature parameter \(k\) that appears in the line element also (and only) enters geodesics via \(\gamma_{ij}\) (equivalently via \({}^{(3)}\Gamma^i{}_{jk}\)).
If we consider the Lagrange multiplier case, the only consistent solution is $k=0$, so the only allowed solution is the flat FLRW solution, its corresponding geodesics. By contrast, one can consider the non-minimal coupling case. There the geodesic equation will have solutions for $k=0,\pm 1$, but the strength of curvature in the EFE will be modified:
\[
\left(\frac{\dot a}{a}\right)^2=\frac{\rho}{3}+\frac{\Lambda}{3}\;-\;\frac{k_{\rm eff}}{a^2},
\qquad 
k_{\rm eff}:=(1+\frac{\eta^2}{4\xi})\,k, 
\]
cf.\ \eqref{eq:etaFriedmann}. Thus, comparing to GR, \emph{the same} intrinsic curvature enters with an \emph{effective coupling} 
\(1+\frac{\eta^2}{4\xi}\) in the Friedmann equation, while geodesics feel the unscaled \(k\) through the metric.

\paragraph{FLRW background in the mimetic theory: current and modified Friedmann equations.}
We choose again the mimetic clock field as
\begin{equation}
A=t,
\qquad
n_a=-\nabla_a A = (-1,0,0,0),
\end{equation}
which satisfies the mimetic constraint
\begin{equation}
\nabla_aA\nabla^aA=-1.
\end{equation}

To avoid confusion, we denote by $\rho_{\rm mim}(t)$ the \emph{mimetic Lagrange multiplier}
entering the current equation \eqref{eq:A_eom_current}, while by $\varrho(t)$ and $p(t)$
we denote the usual matter energy density and pressure in the perfect-fluid stress tensor
\begin{equation}
T_{ab}^{\rm (matt)}
=
\mathrm{diag}\!\left(
\varrho,\;
p\,a^2,\;
p\,a^2r^2,\;
p\,a^2r^2\sin^2\theta
\right).
\end{equation}

From the conserved current equation \eqref{eq:A_eom_current},
\begin{equation}
\nabla_n J^n=0,
\qquad
J^n=2\left(\lambda R^n{}_{m}+\rho_{\rm mim}\,\delta^n{}_m\right)\nabla^m A,
\end{equation}
one finds for the FLRW ansatz
\begin{equation}
J_a=
\left\{
\,2\!\left(\rho_{\rm mim}(t)+3\lambda\frac{\ddot a}{a}\right),\,0,\,0,\,0
\right\},
\qquad
J^a=
\left\{
-2\!\left(\rho_{\rm mim}(t)+3\lambda\frac{\ddot a}{a}\right),\,0,\,0,\,0
\right\}.
\end{equation}
Therefore
\begin{equation}
\nabla_aJ^a=0
\;\Longrightarrow\;
\partial_t\!\left[
a^3\!\left(\rho_{\rm mim}+3\lambda\frac{\ddot a}{a}\right)
\right]=0,
\end{equation}
and hence
\begin{equation}
\rho_{\rm mim}(t)+3\lambda\frac{\ddot a}{a}
=
\frac{C_1}{a^3},
\label{eq:mimetic_current_FLRW_integral}
\end{equation}
where $C_1$ is an integration constant.
The first Friedmann equation is further modified due to the additional $R_{mn}n^mn^n$ term in the Lagrangian.
\begin{align}
\mathcal{E}_{00}
&=
-\rho_{\rm mim}
+\frac{3\big(k(1+\lambda)+\dot a^{\,2}\big)}{a^{2}}
+3\lambda\frac{\ddot a}{a}=\rho,
\label{eq:E00_FRW_correct}
\end{align}
Thus, the integration constant $C_{1}$ behaves as an additional dust-like contribution, similar to the usual mimetic construction.
However, unlike the previous version, the parameter $\lambda$ now affects the background dynamics
not only through the spatial-curvature term $k(1+\lambda)/a^{2}$, but also through the extra acceleration term $6\lambda\,\ddot a/a$ in the first Friedmann equation.
Imposing $^{(3)}R=k=0$, we get the modified Friedmann equations.
\footnote{
The difference between the two FLRW results is an off-shell issue rather than
a contradiction. In the original ADM-type formulation the normal is the
normalized, metric-dependent hypersurface normal, see \eqref{normal}. Hence
the metric variation of the action contains the chain-rule contribution
\[
\left.\frac{\delta S}{\delta g^{ab}}\right|_{n=n(g)}
=
\left.\frac{\delta S}{\delta g^{ab}}\right|_{n}
+
\frac{\delta S}{\delta n_c}\,
\frac{\partial n_c}{\partial g^{ab}} .
\]
Calculating it for the action \eqref{eq:ADM_Ricci_form_general_n} we obtain \begin{equation}
 \frac{\delta S_{\rm ADM}}{\delta n_c}\,
\frac{\partial n_c}{\partial g^{ab}}
=
\int d^4x\,
\frac{\delta S_{\rm ADM}}{\delta n_c}
\frac{\partial n_c}{\partial g^{ab}}\,
\delta g^{ab}
=
\frac{\lambda}{2}
\int d^4x\,\sqrt{-g}\,
N^2
\left(R_{mn}n^m n^n\right)
\delta^0_a\delta^0_b\,
\delta g^{ab}.
\label{eq:normal_chain_rule_addition_variation}
\end{equation}

Equivalently, this contribution is encoded in the variation of the induced
metric, cf. \eqref{gamma} and \eqref{dergamma}. In the original
$\lambda\,{}^{(3)}R$ formulation this extra normal-dependent term is
multiplied by the spatial tensor \eqref{defB} and vanishes by the
transversality relation \eqref{covb0}, so that one recovers
\eqref{coveqm}, i.e. the same equations as \eqref{fieldeq}.

In the auxiliary-vector/mimetic completion the off-shell variational principle
is different: the normal is first treated as an auxiliary covector, and in the
mimetic specialization the covariant component $n_a=-\nabla_a A$ is kept
metric-independent, with the unit constraint imposed by the multiplier
$\rho$. Therefore the chain-rule term present for the metric-dependent
normalized normal is not included in the same way. The resulting difference is
precisely a normal-sector contribution proportional to $\lambda$. On an FLRW
background this contribution appears in the normal-normal, equivalently
$00$, component, because the chosen normal has only a non-trivial time
component, as in \eqref{normal}. This is why the $\lambda$-terms disappear in
the original flat FLRW equations but survive in the mimetic Friedmann
equations.
}

\begin{eqnarray}
       3 H^2
        -\frac{C_1}{a^3}+6\lambda \frac{\ddot{a}}{a}=\sum_i\rho_i+\Lambda
\label{eq:mimetic_Friedmann_standard_H}\cr
        2\frac{\ddot a}{a}
        +H^2=-\sum_i p_i+\Lambda,
\label{eq:mimetic_Friedmann_standard_acc}
\end{eqnarray}
where $i$ denotes different energy components. Assuming perfect fluids $p_i=w_i \rho_i$, and using the continuity equation, we can manipulate the equations to finally lead to:
\be
H^2=\frac{1}{3}\sum_i\frac{1+3\lambda w_i}{1+\lambda}\frac{\rho_{i0}}{a^{3(1+w_i)}},
\ee
 where the CC is simply the $w_i=-1$ case and the integration constant is shifted to $C_1\rightarrow C_1/3(1+\lambda)$. Hence, the effect of the lagrange multiplier $\lambda$ is simply some rescaling the initial densities $\rho_{i0}$ but with no actual physical effect.

\subsection{Schwarzschild metric}
Consider a general spherically symmetric metric  in the form
\begin{equation} \label{BHmet}
ds^{2} =-A(r)dt^{2} +B(r) dr^{2} +r^{2} \left(d\theta ^{2} +\sin ^{2} \theta d\phi ^{2} \right),
\end{equation}
for example, CMC foliation chosen by \eqref{eq:SCMC_main} produces static slices ($A=t$). Or, equivalently, choose a timelike symmetry slices by \eqref{kill-lag}  that leads to the same result for this solution.
Then the normal vector obtained this way (\ref{normal}) is
\begin{equation}\label{normalSc}
   n_m= \left\{-\sqrt{A(r)},0,0,0\right\}.
\end{equation}
Let's call this foliation as static foliation.
The induced metric (\ref{gamma}) is
\begin{equation}
    \gamma_{mn} =\left(
\begin{array}{cccc}
 0 & 0 & 0 & 0 \\
 0 & B(r) & 0 & 0 \\
 0 & 0 & r^2 & 0 \\
 0 & 0 & 0 & r^2 \sin ^2 \theta  \\
\end{array}
\right).
\end{equation}
The extrinsic curvature (\ref{Kab}) $K_{mn}$ is zero.
The 3-curvature (\ref{R3}) calculated with (\ref{3Ricci}) is
\begin{equation}\label{BHR3}
    {}^{(3)}R=\frac{2 \left(r B'(r)+B(r)^2-B(r)\right)}{r^2 B(r)^2}.
\end{equation}
\textbf{Theory with the Lagrange multiplier.}
To make integral constraint $ \int  {}^{(3)} R\sqrt{-g}~  d^4 x=0$ we solve 
\begin{equation}
   \int d^4 x \frac{2 \sin \theta  \sqrt{A(r) B(r)} \left(r B'(r)+B(r)^2-B(r)\right)}{B(r)^2}=0.
\end{equation}
 After integration by $0<\theta<\pi$ and  $0<\phi<2\pi$, imposing $A(r)=1/B(r)$, one has
\begin{equation}
   \int_0^\infty dr \frac{  \sqrt{A(r)B(r)} \left(r B'(r)+B(r)^2-B(r)\right)}{B(r)^2}= \int_0^\infty dr \left( \frac{ \left(r B'(r)-B(r)\right)}{B(r)^2}+1\right) =0.
\end{equation}
Integrating it by parts we have
\begin{equation}
    \frac{(B(r)-1)r}{B(r)}|^{\infty}_{0}=0.
\end{equation}
For the Schwarzschild solution we get that 
\begin{equation}
   \frac{(B(r)-1)r}{B(r)}\equiv 2M ,
\end{equation}
so as expected the integral vanishes. For more general solutions, we can expand the asymptotic series near zero and infinity:
\begin{eqnarray}
    B(r\gg 1)=\sum_{n=0}^{\infty} \frac{a_n}{r^n},\cr
    B(r\ll 1)=\sum_{n=0}^{\infty} c_n r^n. \cr
\end{eqnarray}
For a finite integral we must have $a_0=1$, i.e. asymptotically flat metric. Then there are two possibilities:
\begin{eqnarray}\label{bhconstraints}
    a_1\neq 0 \Rightarrow c_0=0,\quad a_1=-\frac{1}{c_1},\cr
    a_1=0 \Rightarrow c_0,c_1 \quad \textit{unconstrained}.
\end{eqnarray}
The first possibility is the Schwarzschild solution. This result is also verified for the case of a CC. The integral diverges at $r\rightarrow \infty$ in the presence of the CC, because the metric is no longer asymptotically flat.
However,  solving the field equations (\ref{fieldeq}) with no matter, zero \eqref{eq:T_CMC_main} (obtained with vanishing Lagrange multiplier $\chi$), $\Lambda=0$ for the metric (\ref{BHmet}), we obtain that only the Schwarzschild solution satisfies it for every $\lambda$. Hence, we have $A(r)={1-2M/r}, B(r)=\frac{1}{1-2M/r}$ as the only solution that satisfies both the field equations \eqref{fieldeq} and the constraint $ \int  {}^{(3)} R\sqrt{-g}~  d^4 x=0$.

For the local constraint (${}^{(3)}R=0$) we have
\begin{equation}
    \frac{2 \left(r B'(r)+B(r)^2-B(r)\right)}{r^2 B(r)^2}=0,
\end{equation}
that gives
\begin{equation}
    B(r)=\frac{1}{1+const/r}=\frac{1}{1-2M/r},
\end{equation}
i.e.  Schwarzschild solution, which is of course a solution of the field equations \eqref{fieldeq} too.

\textbf{Theory with non-minimal coupling:}
We do not have any intrinsic curvature constraints for the theory with non-minimal coupling to the intrinsic curvature. The field equations (\ref{fieldeq}) with $\Lambda=0$ (equivalently, \eqref{eqmet} and \eqref{eqphi2} with $\Lambda=0$, $\phi=-\frac{\eta}{2\xi}$) are satisfied by the Schwarzschild solution for any value of $\lambda=\eta^2/4\xi$.

\subsection{Lemaitre Metric}
Black holes are not just limited to the Schwarzschild case. Let us show that our procedure works and generates viable black hole solutions also in the presence of a CC or charged black holes. For that purpose, we perform a coordinate transformation from the Schwarzschild static coordinates $(t,r,\theta, \phi)$ to Lemaitre freely falling coordinates $(\tau,\rho,\theta, \phi)$
\begin{equation}\label{SchtoLemTransf}
\begin{array}{cccc}
    d\rho =dt+ \frac{dr}{f\sqrt{1-f}}\,,  \\
d\tau =dt+\frac{dr}{f}\sqrt{1-f},
\end{array}
\end{equation}%
with $f(r)=A(r)=1/B(r)$, the  Schwarzschild metric (\ref{BHmet}) transforms to Lemaitre metric:
\begin{equation} \label{metriclem}
ds^2=-d \tau^2+(1-f)d \rho^2+r^2 d\theta^2 + r^2 \sin^2 \theta d\phi^2 .
\end{equation}
However, $f(r)$ can be more general and depend on the spherically symmetric energy-momentum of matter (including \eqref{geo-EMT}).
This is the synchronous frame.
Let's build a new foliation. Using \eqref{geo-cond} we tie $n^{a}$ to free-fall geodesics from rest at infinity ($A=\tau$).
In this new foliation, in the Lemaitre coordinates, the normal vector (\ref{normal}) is
\begin{equation}
     n_m=   \{-1,0,0,0\}.
\end{equation}
Let's call this foliation as Lemaitre foliation.
The induced metric (\ref{gamma}) is
\begin{equation}
\gamma_{mn}=\left(
\begin{array}{cccc}
 0 & 0 & 0 & 0 \\
 0 & 1-f(r) & 0 & 0 \\
 0 & 0 & r^2 & 0 \\
 0 & 0 & 0 & r^2 \sin ^2 \theta  \\
\end{array}
\right).
\end{equation}
In the field equations (\ref{fieldeq}), ${}^{(3)}R_{ij}$ and ${}^{(3)}R$ are zero, all terms with $N$ are zero because it is constant $N=1$. Thus, all the terms with $\lambda$ are zero,  
so only GR part of the field equations remain, thus, $f=1-2M/r$ for the vacuum case with $\Lambda=0$ \eqref{geo-EMT}.

Let us consider a static charged black hole solution. Both in the Schwarzschild coordinate system and in the Lemaitre coordinate system, the matter energy-momentum tensor is:
\begin{equation}\label{chargerhs}
    T^m{}_n = \left(\begin{array}{cccc}
        -Q^2/r^4 &  0&0&0\\
          0 & - Q^2/r^4&0&0\\
            0 &  0&Q^2/r^4&0\\
              0 &  0&0&Q^2/r^4\\
    \end{array} \right).
\end{equation}
Deriving the field equations with this metric and energy-momentum tensor, all terms with $\lambda$ in \eqref{fieldeq} vanish. Thus, the theory \eqref{laglam} reproduces GR. When we consider the theories with  dynamical foliation choice, namely geodesic, we solve \eqref{fieldeq} with the RHS which includes \eqref{geo-EMT}. Thus,  if \eqref{geo-EMT} is zero, we still get the GR charged black hole solution $f=1-2M/r+Q^2/r^2$.
Similarly, if one considers the vacuum field equations (\ref{fieldeq}) with a CC, $\Lambda\neq 0$ one gets Schwarzschild-(Anti) de-Sitter solution $f={1-2M/r-\Lambda r^2/3  }$.

To summarize, both for charged black holes and the Schwarzschild-(Anti) de Sitter solution, all terms with  $\lambda$  vanish, 
 and these solutions are allowed both for theories with the Lagrange multiplier and the theories with non-minimal coupling 
in the Lemaitre foliation.

\subsection{Weak gravitational waves with two polarizations}
Consider the metric for a weak gravitational wave with two polarizations \cite{Landau_Lifshitz_1975}:
\begin{equation}
 g_{mn}  = \left(
\begin{array}{cccc}
 -1 & 0 & 0 & 0 \\
 0 & 1+{h_+}(t,z) & {h_x}(t,z) & 0 \\
 0 & {h_x}(t,z) & 1-{h_+}(t,z) & 0 \\
 0 & 0 & 0 & 1 \\
\end{array}
\right),
\end{equation}
where ${h_+}(t,z),~{h_x}(t,z)$ are assumed to be small, and we adopt synchronous geodesic slicing with \eqref{geo-lag} generated by a local free-fall congruence ($A=t$ along $u^{a}$).
Then, the normal vector (\ref{normal}) and the induced metric are:
\be 
     n_m=\{-1,0,0,0\},\quad
  \gamma_{mn} = \left(
\begin{array}{cccc}
 0 & 0 & 0 & 0 \\
 0 & {h_+}(t,z)+1 & {h_x}(t,z) & 0 \\
 0 & {h_x}(t,z) & 1-{h_+}(t,z) & 0 \\
 0 & 0 & 0 & 1 \\
\end{array}
\right).
\ee 
The intrinsic curvature scalar (\ref{R3}) is of the second order in $h_{ij}$:
\begin{equation}
   {}^{(3)}R = 4 \left({h_+}(t,z) {h_+}^{(0,2)}(t,z)+{h_x}(t,z) {h_x}^{(0,2)}(t,z)\right)+3 {h_+}^{(0,1)}(t,z)^2+3 {h_x}^{(0,1)}(t,z)^2 + O(h^4_{ij}).
\end{equation}
It satisfies the local constraint at the 1-st order. 
The vacuum field equations \eqref{fieldeq} 
are of the 1st order  
\begin{equation}\label{waveeom}
    \left(
\begin{array}{cccc}
 0 & 0 & 0 & 0 \\
 0 & \frac{1}{2}  \left(-\lambda  {h_+}^{(0,2)}(t,z)-{h_+}^{(0,2)}(t,z)+{h_+}^{(2,0)}(t,z)\right) & \frac{1}{2}  \left(-\lambda 
   {h_x}^{(0,2)}(t,z)-{h_x}^{(0,2)}(t,z)+{h_x}^{(2,0)}(t,z)\right) & 0 \\
 0 & \frac{1}{2}  \left(-\lambda  {h_x}^{(0,2)}(t,z)-{h_x}^{(0,2)}(t,z)+{h_x}^{(2,0)}(t,z)\right) & \frac{1}{2} \left(\lambda 
   {h_+}^{(0,2)}(t,z)+{h_+}^{(0,2)}(t,z)-{h_+}^{(2,0)}(t,z)\right) & 0 \\
 0 & 0 & 0 & 0 \\
\end{array}
\right)=0.
\end{equation}
These are well-known wave equations with the known solutions,
\bea
    {h_+}(t,z)= f_1(z-\sqrt{\lambda +1} t)+ f_2(z+\sqrt{\lambda +1} t),\cr
    {h_x}(t,z) = f_3 (z-\sqrt{\lambda +1} t)+ f_4(z+\sqrt{\lambda +1} t),
\eea 
where $f_1,f_2,f_3,f_4$ are free functions. 
Thus, the wave is propagating with the speed $\sqrt{\lambda +1}$. 
To avoid superluminality, we then require $\lambda\leq 0$. This can potentially be tested by experiments. Current constraints on propagation of gravitational waves imply $|\lambda|\ll 1$. 
Specifically, if the intrinsic curvature drops out of the equations completely, then $\lambda=-1$ in \eqref{laglam} or $\phi=-\frac{\eta}{2\xi}$ and $\frac{\eta^2}{4\xi}=-1$ in \eqref{anothercouplag}, and gravitational waves do not propagate, reducing the theory to lowest order Caroll Gravity.

\paragraph{Weak gravitational waves: vanishing current in vacuum and modified propagation speed.}
We choose again the synchronous mimetic slicing
\begin{equation}
A=t,
\qquad n_a=(-1,0,0,0).
\end{equation}
From the 00-component of the equations for metric we get the vanishing mimetic multiplier,
\begin{equation}
\rho_{\rm mim}=0,
\end{equation}
the current \eqref{eq:A_eom_current} vanishes at the linearized level,
\begin{equation}
J^a=0,
\end{equation}
so the scalar-current equation is automatically satisfied.
Linearizing the metric field equations \eqref{eq:metric_final_compact} (and omitting the overall common prefactor), we get the same \eqref{waveeom} and the same propagation speed.

\subsection{Lense-Thirring metric}
 The Lense-Thirring metric is an approximation of the Kerr metric in the weak-field and slow-rotation limit. We take it in the form \cite{Gray:2021roq}:
\begin{equation} \label{LTmet}
ds^2 = -f \, dt^2 + \frac{dr^2}{f} + 2a(f - 1)\sin^2\theta \, dt \, d\phi + r^2(\sin^2\theta \, d\phi^2 + d\theta^2),
\end{equation}
where $f=1-2M/r$. We propose the foliation by the hypersurfaces of constant time --- static foliation.
Then the normal vector (\ref{normal}) is
\begin{equation}\label{normalLT}
   n_m= \left\{-\frac{{\sqrt{4 a^2 M^2 \sin ^2 \theta +r^3 (r-2 M)}}}{r^2},0,0,0\right\}.
\end{equation}
The induced metric (\ref{gamma}) is
\begin{equation}
    \gamma_{mn} =\left(
\begin{array}{cccc}
 \frac{2 M}{r}+\frac{(r-2 M) r^3+4 a^2 M^2 \sin ^2 \theta }{r^4}-1 & 0 & 0 & -\frac{2 a M \sin ^2 \theta }{r} \\
 0 & \frac{1}{1-\frac{2 M}{r}} & 0 & 0 \\
 0 & 0 & r^2 & 0 \\
 -\frac{2 a M \sin ^2 \theta }{r} & 0 & 0 & r^2 \sin ^2 \theta  \\
\end{array}
\right).
\end{equation}
The extrinsic curvature (\ref{Kab}) $K_{mn}$ is
\begin{equation}
    \begin{array}{cccc}
      K_{0 1} =K_{1 0} = -\frac{6 a^2 M^2 \sin ^2 \theta  }{r^3\sqrt{{4 a^2 M^2 \sin ^2 \theta -2 M r^3+r^4}}};

\\

K_{1 3} = K_{3 1} =  \frac{3 a M \sin ^2 \theta }{\sqrt{4 a^2 M^2 \sin ^2 \theta -2 M r^3+r^4}}.
    \end{array}
\end{equation}
We check that this slicing can be picked with the CMC Lagrangian \eqref{eq:SCMC_main} or Killing field \eqref{kill-lag}  with the corresponding vanishing Lagrange multipliers.
The 3-curvature (\ref{R3}) calculated with (\ref{3Ricci}) is zero.
The vacuum field equations (\ref{fieldeq}) are satisfied in the linear order of $M$ and $a$ for every $\lambda$ including the case $\lambda=-1$ in \eqref{laglam} or $\phi=-\frac{\eta}{2\xi}$ and $\frac{\eta^2}{4\xi}=-1$ in \eqref{anothercouplag}. Hence, the Lense-Thirring solution is a solution for the theories (\ref{laglam}), (\ref{lagReq}) and (\ref{anothercouplag}) in a weak field limit. 
The normal vector \eqref{normalLT} with the Lense-Thirring metric satisfy the  auxiliary normal vector theory equations \eqref{EFE-n-aux}.
It cannot be expressed as divergence of any field, as may be expected by the mimetic version.
\section{Newman-Janis algorithm }\label{sec:njalg}\setcounter{equation}{0}
In this section, we apply the modified Newman-Janis algorithm developed in \cite{Chaturvedi:2023ctn} to obtain the rotating black hole for theories (\ref{laglam}) and (\ref{anothercouplag}).
\subsection{Obtaining the metric}
According to \cite{Chaturvedi:2023ctn,Drake:1998gf}, the algorithm contains five steps. First, writing the Schwarzschild
metric (\ref{BHmet}), considered as a static spherically symmetric seed metric, in advanced
Eddington-Finkelstein coordinates $(u,r,\theta,\phi)$ (i.e., the $g_{rr}$ component is eliminated by a change
of coordinates and a cross-term is introduced):
\begin{equation} \label{seedmet}
    ds^2=-(1-\frac{2M}{r})du^2 - 2dudr + r^2 (d\theta^2 + \sin^2 \theta d\phi^2),
\end{equation}
where 
\begin{equation}
    du = dt - \frac{dr}{1-2M/r}.
\end{equation}
The second step of the algorithm involves expressing
the contravariant form of the seed metric (\ref{seedmet}) in terms of a
null tetrad, $e^m_a=(l^m, n^m, m^m, \bar{m}^m)$
 as:
\begin{equation}\label{defmetnj}
    g_{mn}=-l^m n^n-l^n n^m+m^m \bar{m}^n+m^n \bar{m}^m,
\end{equation}
where
\begin{equation}
    l_m l^m =  n_m n^m= m_m m^m = l_m m^m= n_m m^m=0,~~~ l_m n^m =- m_m \bar{m}^m=-1
\end{equation}
with $ \bar{m}_m$ being the complex conjugate of the $m_m$.
For the Schwarzschild spacetime (\ref{seedmet}), the \textit{null tetrad vectors} \( (l^{m}, n^{m}, m^{m}, \bar{m}^{m}) \) are
\begin{equation}
    \begin{array}{cccc}
         l^{m} = \delta^{m}_{1}, \\
         n^{m} = \delta^{m}_{0} - \frac{1}{2} \left( 1 - \frac{2M}{r} \right) \delta^{m}_{1}, \\
       m^{m} = \frac{1}{\sqrt{2} r} \left( \delta^{m}_{2} + \frac{i}{\sin \vartheta} \delta^{m}_{3} \right),  \\
    \bar{m}^{m} = \frac{1}{\sqrt{2} r} \left( \delta^{m}_{2} - \frac{i}{\sin \vartheta} \delta^{m}_{3} \right).
    \end{array}
\end{equation}
The third step, let the coordinate \( r \) take complex values so the complex conjugate of \( r \) appears
\begin{equation}
    \begin{array}{cccc}
        l^{m} = \delta^{m}_{1} , \\
        n^{m} = \delta^{m}_{0} - \frac{1}{2} \left( 1 - M \left[ \frac{1}{r} + \frac{1}{\bar{r}} \right] \right) \delta^{m}_{1},\\
        m^{m} = \frac{1}{\sqrt{2} r} \left( \delta^{m}_{2} + \frac{i}{\sin \vartheta} \delta^{m}_{3} \right),\\
        \bar{m}^{m} = \frac{1}{\sqrt{2} \bar{r}} \left( \delta^{m}_{2} - \frac{i}{\sin \vartheta} \delta^{m}_{3} \right).
    \end{array}
\end{equation}
Then for  the fourth step we perform the following complex coordinate transformation on
the null vectors
\begin{equation}\begin{array}{cccc}
 r'=r+i a R(\theta ),\\
    \theta'=\theta+i a P(\theta ),\\
     \phi'=\phi+i a Q(\theta ),\\
      \theta'=\theta.
      \end{array}
\end{equation}
After this transformation  the null vectors become
\begin{equation}\label{tetradaftertrans}
    \begin{array}{cccc}
        l^m=\{0,1,0,0\},\\
       n^m=\left\{1,\frac{M {r'}}{a^2 R(\theta )^2+{r'}^2}-\frac{1}{2},0,0\right\},\\
       m^m=\left\{\frac{a P'(\theta )}{\sqrt{2} (a R(\theta )-i {r'})},\frac{a R'(\theta )}{\sqrt{2} (a
   R(\theta )-i {r'})},\frac{1}{\sqrt{2} ({r'}+i a R(\theta ))},\frac{i \left(a Q'(\theta
   )+\csc \theta \right)}{\sqrt{2} ({r'}+i a R(\theta ))}\right\},\\
       \bar{m}^m=\left\{\frac{a P'(\theta )}{\sqrt{2} (a R(\theta )+i {r'})},\frac{a R'(\theta )}{\sqrt{2} (a
   R(\theta )+i {r'})},\frac{1}{\sqrt{2} ({r'}-i a R(\theta ))},-\frac{i \left(a Q'(\theta
   )+\csc \theta \right)}{\sqrt{2} ({r'}-i a R(\theta ))}\right\}.
    \end{array}
\end{equation}
Constructing the metric from (\ref{defmetnj}),(\ref{tetradaftertrans}),  and inverting the obtained metric, the nonzero components are:
\begin{equation}
\begin{array}{cccc}
   g_{00} =-\frac{a^2 R ^2+{r'} ({r'}-2 M)}{a^2 R ^2+{r'}^2} , \\ g_{03} =g_{30} =\frac{a
   \left(P'  \left(a^2 R ^2+{r'} ({r'}-2 M)\right)+R'  \left(a^2
   R ^2+{r'}^2\right)\right)}{\left(a^2 R ^2+{r'}^2\right) \left(a Q'(\theta
   )+\csc \theta \right)}, \\
g_{10} = g_{01} =-1   , \\ g_{13} = g_{31} = \frac{a P' }{a Q' +\csc \theta }, \\
  g_{22} =a^2 R ^2+{r'}^2  ,\\
 g_{33} = -\frac{-a^4 R^4+{r'} \left(a^2 ({r'}-2 M) P'(\theta)^2+2 a^2 {r'} P'(\theta) R' -{r'}^3\right)+R^2 \left(2 a^4 P' R' +a^4
  P'(\theta )^{2}-2 a^2 {r'}^2\right)}{\left(a^2 R^2+{r'}^2\right) \left(a Q' +\csc(\theta )\right)^2}.
\end{array}
\end{equation}
The fifth step is to get the  Boyer–Lindquist (BL) coordinates. For this we perform the following coordinate transformation:
\begin{equation}
    dt=du+\frac{-a^2 P'(\theta ) R'(\theta )+a^2 R(\theta )^2+{r'}^2}{a^2 R'(\theta )^2+a^2 R(\theta )^2-2 M {r'}+{r'}^2},~~
    d \phi = d \phi' -\frac{a R'(\theta
   ) \left(a Q'(\theta )+\csc \theta \right)}{a^2 R'(\theta )^2+a^2 R(\theta )^2-2 M{r'}+{r'}^2}
\end{equation}
and get the BL metric in the form
\begin{equation}\label{metnj}
\begin{array}{cccc}
 g_{00}=-\frac{a^2 R ^2+{r'} ({r'}-2 M)}{a^2 R ^2+{r'}^2}, \\ g_{03}= g_{30}=\frac{a \left(P'  \left(a^2 R ^2+{r'} ({r'}-2 M)\right)+R' 
   \left(a^2 R ^2+{r'}^2\right)\right)}{\left(a^2 R ^2+{r'}^2\right) \left(a
   Q' +\csc \theta \right)}, \\
  g_{11}= \frac{a^2 R^2+{r'^{2}}}{a^2 R^{'2}+a^2 R ^2+{r'} ({r'}-2 M)},
     \\
  g_{22}= a^2 R ^2+r'^{2},  \\
  g_{33}= -\frac{-a^4 R^4+{r'} \left(a^2
   ({r'}-2 M) P'^{2}+2 a^2 {r'} P'  R' -r'^{3}\right)+R(\theta
   )^2 \left(2 a^4 P'  R' +a^4 P'^2-2 a^2 r'^{2}\right)}{\left(a^2
   R^2+r'^{2}\right) \left(a Q' +\csc \theta \right)^2} .\\
\end{array}
\end{equation}
All other components vanish.
\subsection{Solving the field equations}
In this subsection we will solve the field equations (\ref{fieldeq}) with $\Lambda=0$ and zero matter and effective energy-momentum tensors using the metric (\ref{metnj}) with $r' \goto r$.
As usual, we take the foliation defined by the hypersurfaces of constant time.
The normal vector (\ref{normal}) is
\begin{equation}
    n_m=\left\{-{\sqrt{\frac{\left(a^2 R(\theta )^2+r^2\right) \left(a^2
   R'(\theta )^2+a^2 R(\theta )^2+r (r-2 M)\right)}{a^4 R(\theta )^4+r \left(a^2 (2 M-r) P'(\theta )^2-2 a^2 r P'(\theta ) R'(\theta )+r^3\right)-R(\theta )^2 \left(2 a^4 P'(\theta ) R'(\theta )+a^4 P'(\theta )^2-2 a^2 r^2\right)}}},0,0,0\right\}.
\end{equation}
The nonzero components of the induced metric (\ref{gamma}) are
\begin{equation}
\begin{array}{cccc}

\gamma_{0 0} = \frac{\left(a^2 R(\theta )^2+r^2\right) \left(a^2 R'(\theta )^2+a^2 R(\theta )^2+r (r-2 M)\right)}{a^4 R(\theta )^4+r \left(a^2 (2 M-r) P'(\theta )^2-2 a^2 r P'(\theta ) R'(\theta )+r^3\right)-R(\theta )^2 \left(2 a^4 P'(\theta ) R'(\theta )+a^4 P'(\theta )^2-2 a^2 r^2\right)}-\frac{a^2 R(\theta )^2+r (r-2 M)}{a^2 R(\theta )^2+r^2};

\\

\gamma_{0 3} = \gamma_{3 0} = \frac{a \left(P'(\theta ) \left(a^2 R(\theta )^2+r (r-2 M)\right)+R'(\theta ) \left(a^2 R(\theta )^2+r^2\right)\right)}{\left(a^2 R(\theta )^2+r^2\right) \left(a Q'(\theta )+\csc \theta \right)};

\\

\gamma_{1 1} = \frac{a^2 R(\theta )^2+r^2}{a^2 R'(\theta )^2+a^2 R(\theta )^2+r (r-2 M)};

\\

\gamma_{2 2} = a^2 R(\theta )^2+r^2;

\\

\gamma_{3 3} = \frac{a^4 R(\theta )^4+r \left(a^2 (2 M-r) P'(\theta )^2-2 a^2 r P'(\theta ) R'(\theta )+r^3\right)-R(\theta )^2 \left(2 a^4 P'(\theta ) R'(\theta )+a^4 P'(\theta )^2-2 a^2 r^2\right)}{\left(a^2 R(\theta )^2+r^2\right) \left(a Q'(\theta )+\csc \theta \right)^2}.

\\
\end{array}
\end{equation}
We take the slowly rotating limit where $a$ is small and then get the intrinsic curvature (\ref{R3}),(\ref{3Ricci}) to first order in $a$
\begin{equation}
 {}^{(3)} R =  \frac{2 a \sin \theta  \left(Q^{(3)}(\theta )+4 \cot \theta  Q''(\theta )+Q'(\theta ) \left(3 \cot ^2 \theta -\csc ^2 \theta \right)\right)}{r^2}+O(a^2)
\end{equation}
and the equations of motion (\ref{fieldeq}) are
\begin{equation}\label{fieldeqnj}
\begin{array}{cccc}
\sqrt{-g} E^{00}= 
\frac{(1+\lambda) r^3 \sin \theta}{2(r-2M)} {}^{(3)} R, \\ 
\sqrt{-g}  E^{11}=\frac{(1+\lambda)(2M-r)r \sin \theta}{2}{}^{(3)} R\\
\sqrt{-g} E^{03}= \frac{a \left(-2 M P^{(3)}(\theta )-2 M \cot \theta  P''(\theta )+(2 M-r) \csc ^2 \theta  P'(\theta )+R'(\theta ) \left(4 M-r \csc ^2 \theta \right)+r P^{(3)}(\theta )+r \cot \theta  P''(\theta )+r R^{(3)}(\theta )+r \cot \theta  R''(\theta
   )\right)}{2 r^2 (r-2 M)}.
\end{array}
\end{equation}
These field equations are equivalent to those of GR except  the presence of the multiplier $1+\lambda$ in $\sqrt{-g} E^{00}=0$ and $\sqrt{-g} E^{11}=0$. We see that $\sqrt{-g} E^{00}=0$ and $\sqrt{-g} E^{11}=0$ are equivalent to ${}^{(3)}R=0$. Except the case $\lambda=-1$ their solution is
\begin{equation}
    Q(\theta )= \left(-c_1+\frac{i c_2}{2}\right) \csc \theta+i \left(c_1+\frac{i c_2}{2}\right) \log \left(\cot  \frac{\theta }{2} \right)+c_3,
\end{equation}
where $c_1,c_2,c_3$ are integration constants. To have real solutions only we should put $c_2=2i c_1$ that gives
    $Q(\theta )= c_3-2 c_1 \csc (\theta )$, and
we shall consider the simplest possible solution 
 $  Q(\theta )=c_3$.

In the case $\lambda=-1$ in \eqref{laglam} or  $\phi=-\frac{\eta}{2\xi}$ and $\frac{\eta^2}{4\xi}=-1$ in \eqref{anothercouplag}, $Q(\theta )$ can be an arbitrary function. Then ${}^{(3)}R$ can become non-zero. However, since in these theories ${}^{(3)}R$ is canceled out from the field equations, it can be arbitrary. 

Then we have one equation $\sqrt{-g} E^{03}$ for two functions $ R(\theta ), P(\theta )$. This equation should be satisfied for every $r$, therefore we must equate to zero the sum of the coefficients at the zero and first powers of r of the expression in the numerator of $\sqrt{-g}E_{03}$. This gives us two equations:
\begin{eqnarray}
 P^{(3)}(\theta )+\cot \theta  P''(\theta )+\csc ^2 \theta  \left(-P'(\theta )\right)+R^{(3)}(\theta )+\cot \theta  R''(\theta )-\csc ^2 \theta  R'(\theta )=0,\\
 -2  P^{(3)}(\theta )-2 \cot \theta  P''(\theta )+2 \csc ^2 \theta  P'(\theta )+4 R'(\theta )=0.
\end{eqnarray}
Summing up  these equations we easily get that 
\begin{equation}
   R(\theta )= \cos \theta  \left(b_2 \log \left(\cot  \frac{\theta }{2} \right)+b_1\right)+b_3,
\end{equation}
where $b_1,b_2,b_3$ are integration constants. Unfortunately, the solution for $P(\theta )$ cannot be expressed explicitly. It is the solution of the equation
\begin{eqnarray}\label{solP}
 -2  P^{(3)}(\theta )-2 \cot \theta  P''(\theta )+2 \csc ^2 \theta  P'(\theta )+4 R'(\theta )=0,\\
 R'(\theta )=-b_2 \cot \theta -\sin \theta  \left(b_2 \log \left(\cot  \frac{\theta }{2} \right)+b_1\right).
\end{eqnarray}
For $b_1=b_2=1$ we found numerically the solution for $P(\theta )$ assuming that $P(1)=P'(1 )=0,~~P''(1)=3$. The solution is depicted in figure \ref{P[theta]numP2=3}.
\begin{figure}[h!]
    \centering
    \includegraphics[width=0.8\textwidth]{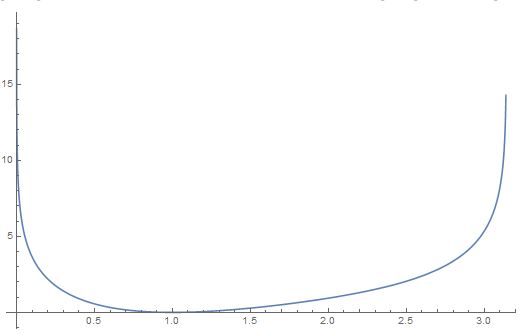} 
    \caption{Numerical solution for $P(\theta )$}
    \label{P[theta]numP2=3}
\end{figure}
Taking the GR case where $R'(\theta )=-\sin \theta $ the solution of (\ref{solP}) is 
\begin{equation}
  P(\theta )=  -d_2 \log (\sin \theta )+d_1 \log \left(\tan  \frac{\theta }{2} \right)+d_3-\cos \theta ,
\end{equation}
where $d_1,d_2,d_3$ are constants of integration.
We see that the GR solution $P(\theta )=-\cos \theta ,~~R(\theta )=\cos \theta ,~~Q(\theta )=0$ satisfies (\ref{fieldeqnj}) and ${}^{(3)}R=0$. This result is expected because we got field equations proportional to those of GR in the slowly rotating limit. Hence, the Kerr solution is a solution of the theory (\ref{laglam}) in the slow rotating limit, and we have constructed a rotating black hole solution for arbitrary mass $M$.

\section{Conclusions}\label{sec:concl}
In this work, we studied several approaches of suppressing the intrinsic curvature \({}^{(3)}R\):  

First, we used \textbf{a Lagrange multiplier.} This method imposes a strict constraint, \(\int {}^{(3)}R \sqrt{-g} \, d^4x = 0\), which effectively eliminates intrinsic curvature from the dynamical equations in some foliation. Within this framework, we obtained cosmological solutions, 
black hole solutions with or without a cosmological constant, as well as charged black holes. We further found rotating black hole solutions in the slow-rotating limit, both in the Lense-Thirring metric and the Newman-Janis algorithm.
       Finally, we also obtained gravitational waves solution with a modified speed of $\sqrt{1+\lambda}$.

Second, we used dynamic cancellation via \textbf{couplings to a scalar field} (\(\phi=-\frac{\eta}{2\xi},~~\lambda =\frac{\eta^2}{4\xi}\)). This approach introduces interactions between the intrinsic curvature and a scalar field, allowing \({}^{(3)}R\) to be dynamically neutralized. We have obtained the same solutions in this framework as well, only now $\lambda$ is a coupling constant, and not a lagrange multiplier.

Third, we constructed a \textbf{covariant auxiliary-vector} completion of the same
intrinsic-curvature mechanism. Using the Gauss--Codazzi relation, the ADM
term $\lambda\,{}^{(3)}R$ can be rewritten, up to a boundary term, as a
four-dimensional Ricci coupling,
$
    R+\lambda\,{}^{(3)}R
    \simeq
    (1+\lambda)R+\lambda R_{ab}n^a n^b .
$
Promoting $n_a$ to an independent unit timelike auxiliary covector, with its
normalization imposed by a Lagrange multiplier, gives a fully covariant
formulation of the same constraint. When $n_a$ is hypersurface orthogonal and
boundary terms are neglected, the variation with respect to $\lambda$ is
equivalent to the original global intrinsic-curvature constraint,
$
    \int d^4x\sqrt{-g}\,{}^{(3)}R=0 .
$

We also showed that the auxiliary-vector construction admits a \textbf{mimetic
specialization}, $n_a=-\nabla_a A$, with the mimetic constraint
$\nabla_a A\nabla^a A=-1$. In this form the preferred foliation is generated
by the scalar field $A$, and the algebraic auxiliary-vector equation is
replaced by a conserved-current equation for the mimetic scalar. Thus the
auxiliary-vector and mimetic formulations provide additional covariant
realizations of the same idea: the intrinsic curvature can be cancelled or
suppressed not only by the original ADM multiplier term, but also by its
Ricci-coupled and mimetic completions.

Finally, we have seen that  \(\lambda = -1\) reduces to Carroll gravity. Hence, our theory provides a different UV completion to the lowest order Carroll gravity action. Better yet, since we have an additional "small parameter" $\lambda$, we will not necessarily be limited to the same expansion in powers of the speed of light $c$. It would be interesting to consider these different UV completions, and their possible connections more thoroughly. 

As a final step we \emph{promoted} the slicing to a physical ingredient by supplementing the baseline action with a family of \emph{alignment} terms (Sec.~3.5): the geodesic term
\(S_{\rm geo}[g,A,\Lambda]\),
the CMC term \(S_{\rm CMC}[g,A,\chi]\), and, when applicable, analogous terms that align the normal with other preferred timelike directions (e.g.\ a timelike Killing field, shear-free slicings, etc.). In the baseline theory the foliation choice is \emph{a priori} arbitrary and the slicing label can be regarded as non-dynamical; by contrast, once these alignment terms are included the would-be slicing label becomes a genuine field: we trade the auxiliary notion of a ``geometric foliation'' for a new scalar degree of freedom \(A\), with \(n_a=-\nabla_aA/\sqrt{-X}\), and we explicitly rewrite all ``3D'' objects fully in terms of \((g_{ab},A)\). The resulting theory is therefore an \emph{exotic scalar--tensor theory} that is fully covariant and manifestly coordinate-invariant: the multiplier variations impose the desired geometric conditions (e.g.\ geodesicity, CMC, Killing alignment), the variation with respect to \(A\) yields a conserved current (cf.\ \eqref{eq:JcA} and its CMC/Killing analogues), and the metric variation produces an additional symmetric stress tensor (cf.\ \eqref{eq:Tmn-compact-article-again} and \eqref{eq:T_CMC_main}). In this way, the dynamics \emph{selects} a physically preferred vorticity-free slicing, and one can ``forget the foliation'' and work instead with the predictive covariant system \(\{g_{ab},A\}\); setting the new multipliers to zero continuously returns the baseline theory.

We then demonstrate that this covariant scalar--tensor theory admits the standard classes of solutions of interest, including FLRW cosmologies, black-hole spacetimes (in particular charged and rotating solutions), and gravitational-wave propagation on cosmological and black-hole backgrounds. 
The observational bounds on the gravitational-wave speed, translate into a stringent constraint on \(\lambda\) the term controlling the intrinsic-curvature: It must be very small.
Nevertheless, except the non-minimal coupling FLRW case, once the $^{(3)}R=0$ is imposed, $\lambda$ only appears if the traceless part of the 3D Ricci tensor is non-vanishing, $^{(3)}R_{ab}\neq 0$. In all our examples, such a situation did not occur. Specifically in FLRW, black holes, and gravitational waves solution, $\lambda$ was unconstrained allowing it fulfill the gravitational waves speed constraints, and still have vanishing or very small intrinsic curvature (spatial curvature).

The presented research deepens our understanding of mechanisms to cancel the intrinsic curvature and their implications for astrophysical and cosmological models. For future progress, one may 
      continue exploring other classes of solutions within these theories, including quantum corrections and
     extend the application of the modified Newman-Janis algorithm to study fast rotating black holes.
    Finally, in the non-minimal coupling approach, we have created a distinction between the curvature appearing in the geodesic equation, and the effective  one in the field equations. It is specifically clear in the context of Cosmology, where $k\neq 0$ in the metric and geodesic equation, becomes $k\left(1+\frac{\eta^2}{4\xi}\right)$ in the field equations. This could have interesting physical implications for current cosmic anomalies. In recent work \cite{Shimon:2024mbm}, a phenomenological approach was adopted to analyze current observations. In that sense, our theory could potentially provide the underlying theory for such an analysis. It will be interesting to perform a similar analysis using our theory and parametrization.
This work opens new perspectives for further investigations in modified gravity theories and their applications in astrophysics and cosmology.

{\bf Acknowledgments:} IBD and EE are supported in part by the “Program of Support of High Energy Physics” Grant by
the Israeli Council for Higher Education. EE has been supported in part by the ministry of absorption, the “Program of Support of High Energy Physics” Grant by Israeli Council for Higher
Education and  by the Israel Science Fund (ISF) grant No. 1698/22.
\appendix

\section{Theory with  ${}^{(3)}R =g^{mn}{}^{(3)}R_{mn}$}\m{gR}
\setcounter{equation}{0}
Considering the action \eqref{lagReq}
\begin{equation}\label{lagReq2}
    S=\frac{1}{2}\int d^4 x \sqrt{-g} (R -2 \Lambda+ \lambda {}^{(3)} R_{mn}g^{mn}) + S_{matter}.
\end{equation}
The variation of the new term gives
\begin{equation}
\begin{array}{cccc}
     \delta(\sqrt{-g} g^{mn} {}^{(3)}R_{mn}) &=&\sqrt{-g} \delta g^{mn} {}^{(3)}R_{mn} -\frac{1}{2}  \sqrt{-g} g_{ab} \delta  g^{ab} g^{mn} {}^{(3)}R_{mn}+\sqrt{-g} g^{mn} \delta({}^{(3)}R_{mn})\\
      &=& \sqrt{-g} \delta g^{mn} {}^{(3)}R_{mn} -\frac{1}{2}  \sqrt{-g} g_{ab} \delta  g^{ab} g^{mn} {}^{(3)}R_{mn}+    \sqrt{-g} (\gamma^{mn}-n^m n^n) \delta({}^{(3)}R_{mn}) .
\end{array}
\end{equation}
So, the only new term appears comparing to (\ref{laglam}) is $- \sqrt{-g} ~n^m n^n \delta{}^{(3)}R_{mn}$.
We will use the following expressions for variations derived in the Appendix \ref{gammadelR}:
\begin{equation}
\delta {}^{(3)}R_{ab} = D_c \delta {}^{(3)}\Gamma^c_{ab} - D_a \delta {}^{(3)}\Gamma^c_{cb},
\end{equation}
for 3-Ricci tensor variation and
\begin{equation}
\delta {}^{(3)}\Gamma^s_{mn} 
= \frac{1}{2} \gamma^{s t} \left( D_m \delta \gamma_{n t} + D_n \delta \gamma_{t m} - D_t \delta \gamma_{m n} \right),
\end{equation}
for the variation of the Christoffel symbols.
\begin{equation}
    \begin{array}{cccc}
- n^a n^b \delta^{(3)} R_{ab} = -n^a n^b \left( D_c \delta^{(3)} \Gamma^c_{ab} - D_a \delta^{(3)} \Gamma^c_{cb} \right)
= - n^a n^b D_c \delta^{(3)} \Gamma^c_{ab} = - n^a n^b \frac{1}{2} \gamma^{cd} D_c \left( D_a \delta \gamma_{bd} + D_b \delta \gamma_{da} - D_d \delta \gamma_{ab} \right)
\\
= n^a n^b \frac{1}{2} D_c D^c \delta \gamma_{ab} = n^a n^b \frac{1}{2} D_c \left( \gamma^{cs} \gamma^{a'}_a \gamma^{b'}_b \nabla_s \delta \gamma_{a' b'} \right)
= n^a n^b\frac{1}{2} \gamma^{c s} \gamma^{a'}_a \gamma^{b'}_b  D_c \nabla_s \delta \gamma_{a' b'} = 0,
    \end{array}
\end{equation}
where we used that $n^a D_a...=0,~~ D_a \gamma^{b}_c=0,~~ D_a\gamma^{c s}=0$.
This means that the equations of motion of this theory are the same as for (\ref{laglam}) --- (\ref{fieldeq}).

\section{Derivation of $\gamma^{cd} \delta {}^{(3)}R_{cd}$ }\m{gammadelR}
\setcounter{equation}{0}
We start with
\begin{equation}\label{2.7}
\int \gamma^{cd} \delta {}^{(3)}R_{cd}\sqrt{-g} , d^4x.
\end{equation}
Going to 3 dimensions we have
\begin{equation}\label{2.8}
{}^{(3)}R_{mn} = \partial_r {}^{(3)}\Gamma^r_{mn} - \partial_m {}^{(3)}\Gamma^r_{rn} + {}^{(3)}\Gamma^s_{mn} {}^{(3)}\Gamma^r_{sr} - {}^{(3)}\Gamma^s_{rn} {}^{(3)}\Gamma^r_{sm}.
\end{equation}
In a locally inertial reference frame, we find
\begin{equation}\label{2.9a}
\delta {}^{(3)}R_{mn} = \partial_r \delta {}^{(3)}\Gamma^r_{mn} - \partial_m \delta {}^{(3)}\Gamma^r_{rn}
= D_r \delta {}^{(3)}\Gamma^r_{mn} - D_m \delta {}^{(3)}\Gamma^r_{rn}.
\end{equation}
Since this equation is tensorial (as $\delta {}^{(3)}\Gamma^r_{mn}$ is a 3D tensor), it holds in all coordinate systems, and we conclude:
\begin{equation}\label{2.10a}
\delta {}^{(3)}R_{ab} = D_c \delta {}^{(3)}\Gamma^c_{ab} - D_a \delta {}^{(3)}\Gamma^c_{cb},
\end{equation}
\begin{equation}\label{2.10b}
\gamma^{ab} \delta {}^{(3)}R_{ab} = \gamma^{ab} D_c \delta {}^{(3)}\Gamma^c_{ab} - \gamma^{ab} D_a \delta {}^{(3)}\Gamma^c_{cb}
= \gamma^{ab} D_c \delta {}^{(3)}\Gamma^c_{ab} - \gamma^{cb} D_c \delta {}^{(3)}\Gamma^a_{ab}
= D_c \left[ \gamma^{ab} \delta {}^{(3)}\Gamma^c_{ab} - \gamma^{cb} \delta {}^{(3)}\Gamma^a_{ab} \right].
\end{equation}
Next, we compute the variation of the Christoffel symbols:
\begin{eqnarray}\label{2.11a}
\delta {}^{(3)}\Gamma^s_{mn} &=& \frac{1}{2} \delta \left( \gamma^{s t} \left( \partial_m \gamma_{n t} + \partial_n \gamma_{t m} - \partial_t \gamma_{m n} \right) \right)
= \frac{1}{2} \delta \gamma^{s t} \left( \partial_m \gamma_{n t} + \partial_n \gamma_{t m} - \partial_t \gamma_{m n} \right) + \frac{1}{2} \gamma^{s t} \left( \partial_m \delta \gamma_{n t} + \partial_n \delta \gamma_{t m} - \partial_t \delta \gamma_{m n} \right)\cr
&=& \frac{1}{2} \gamma^{s t} \left( D_m \delta \gamma_{n t} + D_n \delta \gamma_{t m} - D_t \delta \gamma_{m n} \right),
\end{eqnarray}
where we replaced partial derivatives by covariant 3-derivatives and used that $D_a \gamma_{bc}=0$.
For the relevant 3-Christoffel symbol variation:
\begin{equation}\label{2.11e}
\delta {}^{(3)}\Gamma^c_{ab} = \frac{1}{2} \gamma^{cd} \left( D_a \delta \gamma_{bd} + D_b \delta \gamma_{da} - D_d \delta \gamma_{ab} \right).
\end{equation}
We now compute:
\begin{equation} 
\label{2.12b}
\gamma^{ab} \delta {}^{(3)}\Gamma^c_{ab} = \frac{1}{2} \gamma^{ab} \gamma^{cd} \left( 2 D_a \delta \gamma_{bd} - D_d \delta \gamma_{ab} \right),
\end{equation}
\begin{equation}\label{2.13b}
\gamma^{cb} \delta {}^{(3)}\Gamma^a_{ab} = \frac{1}{2} \gamma^{cb} \gamma^{ad} D_b \delta \gamma_{da}.
\end{equation}
Using these, we find:
\bea
\label{2.14a}
\gamma^{ab} \delta {}^{(3)}R_{ab} &=& D_c 
\left[ \gamma^{ab} \delta {}^{(3)}\Gamma^c_{ab} - \gamma^{cb} \delta {}^{(3)}\Gamma^a_{ab} \right]
= \frac{1}{2} D_c 
\left[ \gamma^{ab} \gamma^{cd} 
\left( 2 D_a \delta \gamma_{bd} - D_d \delta \gamma_{ab} \right) 
- \gamma^{cb} \gamma^{ad} D_b \delta \gamma_{da} \right]\cr
&=& \gamma^{ab} \gamma^{cd} D_c D_a \delta \gamma_{bd} - \frac{1}{2} \gamma^{ab} D_c D^c \delta \gamma_{ab} - \frac{1}{2} \gamma^{ad} D_c D^c \delta \gamma_{ad}
= \gamma^{ab} \gamma^{cd} D_c D_a \delta \gamma_{bd} - \gamma^{ab} D_c D^c \delta \gamma_{ab}. \cr
\eea 
Using the definition of the induced metric (\ref{gammagen}) we have
\begin{equation}
\gamma^{ab} \delta {}^{(3)}R_{ab}= \gamma^{ab} \gamma^{cd} D_c D_a \delta (g_{bd}+n_b n_d) - \gamma^{ab} D_c D^c \delta (g_{ab}+n_a n_b).
\end{equation}
Because $D_b\gamma^{ac}=0$ and all terms like $\gamma^{ab}n_b \delta n_d$ are zero, we have
\begin{equation}
\gamma^{ab} \delta {}^{(3)}R_{ab}= \gamma^{ab} \gamma^{cd} D_c D_a \delta g_{bd} - \gamma^{ab} D_c D^c \delta g_{ab}.
\end{equation}
Recalling now that
$\delta g_{ab} = -g_{ac} g_{bd} \delta g^{cd}$,
and using that $D_bg_{ac}=0$ we see that
\begin{equation}\label{2.15a}
\gamma^{ab} \delta {}^{(3)}R_{ab} = \gamma_{ab} D_c D^c \delta g^{ab} - \gamma^a_{b_1} \gamma^c_{d_1} D_a D_c  \delta g^{b_1 d_1} =  \gamma_{ab} D_c D^c \delta g^{ab} - (\delta^a_{b_1}+n^a n_{b_1} )( \delta^c_{d_1}+n^c n_{d_1}) D_a D_c  \delta g^{b_1 d_1} .
\end{equation}
Because $n^aD_a...=0$, we have
\begin{equation}\label{2.15b}
\gamma^{ab} \delta {}^{(3)}R_{ab}= \left( \gamma_{ab} D_c D^c - D_a D_b \right) \delta g^{ab}.
\end{equation}
\section{Geodesic-foliation term and its variations}\label{appSgeo}
\setcounter{equation}{0}
We supplement the action by a linear Lagrange–multiplier term that enforces geodesicity of the foliation normal:
\begin{equation}
S_{\rm geo}=\int_{\mathcal M}\! d^4x\,\sqrt{-g}\;\Lambda^{\,b}\,a_b,
\qquad
a_b=n^a\nabla_a n_b,
\qquad
n_a=-\frac{\nabla_a A}{\sqrt{-X}},\quad
X=g^{cd}\nabla_cA\,\nabla_dA<0 .
\label{eq:Sgeo-def}
\end{equation}
Throughout, bulk total divergences are converted to surface terms on $\partial\mathcal M$ using Stokes’ theorem,
\(
\int_{\mathcal M}\!\sqrt{-g}\,\nabla_a V^a=\int_{\partial\mathcal M}\! d\Sigma_a\,V^a
\),
with $d\Sigma_a$ the outward-pointing hypersurface element.

\subsection*{Variation with respect to $\Lambda^b$}
The variation with respect to non-constant Lagrange multipliers $\Lambda^b$ gives the geodesic equation:
\begin{equation}
\delta_{\Lambda}S_{\rm geo}=\int_{\mathcal M}\!\sqrt{-g}\ \delta\Lambda^{\,b}\,a_b
\quad\Longrightarrow\quad
{\,a_b\equiv n^a\nabla_a n_b=0\, .}
\label{eq:EL-lambda}
\end{equation}

\subsection*{Variation with respect to $A$ (metric fixed)}
We first need the $A$-variations of $n_a$ and $n^a$. Using $n_a=-\nabla_a A/\sqrt{-X}$ and
\(
\delta X=2\,\nabla^c A\,\nabla_c\delta A
\),
one finds the compact projector forms
\begin{equation}
{
\delta_A n_b=-\frac{1}{\sqrt{-X}}\ \gamma_b{}^{c}\,\nabla_c\delta A,
\qquad
\delta_A n^{a}=-\frac{1}{\sqrt{-X}}\ \gamma^{ac}\,\nabla_c\delta A .
}
\label{eq:delta-n-A}
\end{equation}
Now
\begin{align}
\delta_A S_{\rm geo}
&=\int_{\mathcal M}\!\sqrt{-g}\ \Lambda^{\,b}\,\delta_A\!\big(n^a\nabla_a n_b\big)
=\int_{\mathcal M}\!\sqrt{-g}\ \Lambda^{\,b}\Big[ (\delta_A n^a)\,\nabla_a n_b + n^a\nabla_a(\delta_A n_b)\Big] .
\end{align}
Substituting \eqref{eq:delta-n-A} and collecting the terms proportional to $\nabla_c\delta A$, we obtain
\begin{align}
\delta_A S_{\rm geo}
&=-\int_{\mathcal M}\!\sqrt{-g}\ \Lambda^{\,b}\left[
\frac{1}{\sqrt{-X}}\ \gamma^{ac}\,(\nabla_a n_b)\,\nabla_c\delta A
+ n^a\,\nabla_a\!\left(\frac{1}{\sqrt{-X}}\ \gamma_b{}^{c}\,\nabla_c\delta A\right)
\right].
\end{align}
Integrating by parts once on the first term and twice on the second one (moving derivatives off $\delta A$), and discarding total divergences in the bulk, we can write
\begin{equation}
\delta_A S_{\rm geo}
=\int_{\mathcal M}\!\sqrt{-g}\ \big(\nabla_c J^c_{(A)}\big)\,\delta A
\;+\;\underbrace{\int_{\partial\mathcal M}\! d\Sigma_c\ \mathcal{B}_{(A)}^{\,c}}_{=:B_A},
\label{eq:deltaA-structure}
\end{equation}
with the $A$-current
\begin{equation}
{
J^c_{(A)}=
\Lambda^{\,b}\left[
\frac{1}{\sqrt{-X}}\ \gamma^{ac}\,\nabla_a n_b
+ n^a\,\nabla_a\!\Big(\frac{1}{\sqrt{-X}}\ \gamma_b{}^{c}\Big)
-\nabla_a\!\Big(\frac{1}{\sqrt{-X}}\ n^a\,\gamma_b{}^{c}\Big)\right]=\frac{\Lambda^b}{\sqrt{-X}}(\nabla^c n_b+n^c a_b-\gamma^c{}_b \nabla_a n^a),
}
\label{eq:JcA}
\end{equation}
and boundary density
\begin{equation}
\mathcal{B}_{(A)}^{\,c}
=-\,\Lambda^{\,b}\left[
\frac{1}{\sqrt{-X}}\ \gamma^{ac}\,n_b\,\nabla_a\delta A
+ n^a\,\frac{1}{\sqrt{-X}}\ \gamma_b{}^{c}\,\nabla_a\delta A
- n^c\,\frac{1}{\sqrt{-X}}\ \gamma_b{}^{a}\,\nabla_a\delta A
\right].
\end{equation}
Hence the Euler–Lagrange equation for $A$ that follows from $S_{\rm geo}$ reads
\begin{equation}
{ \ \nabla_c J^c_{(A)}=0 \ } \quad \text{(no assumption on $a_b$ is needed).}
\label{eq:A-EL}
\end{equation}
On the \emph{geodesic shell} $a_b=0$, the expression \eqref{eq:JcA} simplifies because $\nabla_a n_b$ is purely spatial ($\gamma^a{}_c\gamma^b{}_d\nabla_a n_b$), but \eqref{eq:A-EL} remains a nontrivial conservation law unless one chooses the trivial branch $\Lambda^b\equiv 0$, in which case $J^c_{(A)}=0$ and $\delta_A S_{\rm geo}=B_A$ is pure boundary.

\subsection*{Variation with respect to the metric $g_{mn}$}
We now vary \eqref{eq:Sgeo-def} with respect to $g_{mn}$, keeping $A$ and $\Lambda^b$ fixed. We take $\delta g^{mn}$ as the independent variation and use the elementary identities
\begin{equation}
\delta g_{mn}=-g_{mp}g_{nq}\,\delta g^{pq},\qquad
\delta\sqrt{-g}=-\tfrac12\sqrt{-g}\,g_{mn}\,\delta g^{mn},
\label{eq:metric-ids-var}
\end{equation}
as well as the Levi--Civita variations
\begin{equation}
\delta\Gamma^{c}{}_{ab}
=\tfrac12\,g^{cd}\big(\nabla_a\delta g_{bd}+\nabla_b\delta g_{ad}-\nabla_d\delta g_{ab}\big),\qquad
\delta(\nabla_a V_b)=\nabla_a(\delta V_b)-\delta\Gamma^{c}{}_{ab}V_c .
\label{eq:deltaGamma-var}
\end{equation}
For the normalized normal (with $A$ fixed) we need
\begin{equation}
\delta_g n_b=-\tfrac12\,n_b\,n_m n_n\,\delta g^{mn},\qquad
\delta_g n^{a}=\delta g^{am}n_m-\tfrac12\,n^{a}\,n_m n_n\,\delta g^{mn},
\label{eq:dng-var}
\end{equation}
which follow from $n_a=-\nabla_aA/\sqrt{-X}$ and \eqref{eq:metric-ids-var}.

\emph{First split of the variation.} Starting from
\(
S_{\rm geo}=\int_{\mathcal M}\!\sqrt{-g}\,\Lambda^{\,b}\,n^a\nabla_a n_b
\),
we vary each factor:
\[
\delta_g S_{\rm geo}
=\int_{\mathcal M}\!\delta(\sqrt{-g})\;\Lambda^{\,b}a_b
+\int_{\mathcal M}\!\sqrt{-g}\ \Lambda^{\,b}\,\delta_g\!\big(n^a\nabla_a n_b\big).
\]
For the second term we use
\[
\delta_g\!\big(n^a\nabla_a n_b\big)
=(\delta_g n^a)\,\nabla_a n_b
+n^a\,\delta_g(\nabla_a n_b)
=(\delta_g n^a)\,\nabla_a n_b
+n^a\nabla_a(\delta_g n_b)
-\underbrace{n^a\,\delta\Gamma^{c}{}_{ab}\,n_c}_{\text{from }\delta(\nabla n)} .
\]
Inserting \eqref{eq:metric-ids-var}–\eqref{eq:dng-var} we obtain the canonical three-block form
\begin{align}
\delta_g S_{\rm geo}
&=\int_{\mathcal M}\!\sqrt{-g}\Big[
\underbrace{-\tfrac12\,g_{mn}\Lambda^{b}a_b\,\delta g^{mn}}_{\text{Block I: measure}}
+\underbrace{\ \Lambda^{b}\big((\delta_g n^a)\nabla_a n_b+n^a\nabla_a(\delta_g n_b)\big)\ }_{\text{Block II: metric dependence of $n^a,n_b$}}
\underbrace{\ -\Lambda^{b}n^a\,\delta\Gamma^{c}{}_{ab}\,n_c\ }_{\text{Block III: connection variation}}
\Big].
\label{eq:dgSgeo-split-corr}
\end{align}

\emph{Block I (measure).} The first bracket in \eqref{eq:dgSgeo-split-corr} directly gives
\[
\delta_g S_{\rm geo}^{\rm(meas)}=\int\!\sqrt{-g}\Big(-\tfrac12 g_{mn}\Lambda^{b}a_b\Big)\delta g^{mn}
\quad\Rightarrow\quad
T^{(0)}_{mn}=-\,g_{mn}\,\Lambda^{b}a_b .
\]

\emph{Block II (metric dependence of $n^a,n_b$).} Substituting \eqref{eq:dng-var} into the middle bracket of \eqref{eq:dgSgeo-split-corr} and integrating by parts once to remove $\nabla(\delta g)$ from the term $n^a\nabla_a(\delta_g n_b)$, we collect five explicit bulk pieces (plus a boundary term):
\begin{align}
\int\!\sqrt{-g}\;K^{(II)}_{mn}\,\delta g^{mn},\qquad
K^{(II)}_{mn}
&=\Lambda^{b}n_{(m}\nabla_{n)}n_b
-\tfrac12\,\Lambda^{b}a_b\,n_m n_n
-\tfrac12\,\Lambda^{b}a_b\,n_m n_n \nonumber\\
&\quad
-\tfrac12\,\Lambda^{b} n^a n_b\big[(\nabla_a n_m)n_n+n_m(\nabla_a n_n)\big]
+\tfrac12\,\nabla_a\!\big(\Lambda^{b} n^a n_b\,n_m n_n\big).
\label{eq:KII-article-again}
\end{align}
\noindent\textit{Term-by-term origin of \eqref{eq:KII-article-again}.} 
(i) From $(\delta_g n^a)\nabla_a n_b$ and \eqref{eq:dng-var} we get two \emph{algebraic} pieces in $\delta g^{mn}$:
\[
\underbrace{\Lambda^{b}\,(\delta g^{am}n_m)\,\nabla_a n_b}_{\Rightarrow\ \ \Lambda^{b}n_{(m}\nabla_{n)}n_b}
\ ,\qquad
\underbrace{\Lambda^{b}\!\left(-\tfrac12\,n^{a}n_p n_q\,\delta g^{pq}\right)\!\nabla_a n_b}_{\Rightarrow\ \ -\tfrac12\,\Lambda^{b}a_b\,n_m n_n}\,.
\]
(ii) From $n^a\nabla_a(\delta_g n_b)$ and \eqref{eq:dng-var} we first expand $\nabla_a(n_p n_q)$, producing two \emph{algebraic} contributions
\[
\underbrace{-\tfrac12\,\Lambda^{b} n^a(\nabla_a n_b)\,n_p n_q\,\delta g^{pq}}_{\Rightarrow\ \ -\tfrac12\,\Lambda^{b}a_b\,n_m n_n}
\ ,\qquad
\underbrace{-\tfrac12\,\Lambda^{b} n^a n_b\,\nabla_a(n_p n_q)\,\delta g^{pq}}_{\Rightarrow\ \ -\tfrac12\,\Lambda^{b} n^a n_b\big[(\nabla_a n_m)n_n+n_m(\nabla_a n_n)\big]}\!,
\]
and one term with a derivative on the variation which we remove by a \emph{single} integration by parts:
\[
\underbrace{-\tfrac12\,\Lambda^{b} n^a n_b n_p n_q\,\nabla_a(\delta g^{pq})}_{\stackrel{\text{IBP}}{\Longrightarrow}\ \ +\tfrac12\,\nabla_a\!\big(\Lambda^{b} n^a n_b\,n_m n_n\big)\ \ (+\ \text{boundary})}\,.
\]
Collecting these five bulk contributions yields exactly $K^{(II)}_{mn}$ in \eqref{eq:KII-article-again}; the corresponding stress-tensor piece is $T^{(II)}_{mn}=2\,K^{(II)}_{(mn)}$, and the two $a_b$-terms combine to $-\Lambda^{b}a_b\,n_m n_n$.

\emph{Block III (connection variation).} The last bracket in \eqref{eq:dgSgeo-split-corr} uses \eqref{eq:deltaGamma-var}. After removing $\nabla(\delta g)$ by parts and converting $\delta g_{pq}\to\delta g^{mn}$ with \eqref{eq:metric-ids-var}, the bulk integrand is
\begin{align}
\int\!\sqrt{-g}\;C_{mn}\,\delta g^{mn},\qquad
C_{mn}
=-\tfrac12\Big[
\nabla_a(\Lambda^{\,b} n^a n^d)\,g_{bm}g_{dn}
+\nabla_m(\Lambda^{\,b} n^a n^d)\,g_{ba}g_{dn}
-\nabla_n(\Lambda^{\,b} n^a n^d)\,g_{ba}g_{dm}
\Big],
\label{eq:Cmn-article-again}
\end{align}
so the Block~III contribution to the stress tensor is \(T^{(III)}_{mn}=2\,C_{(mn)}\). The three surface pieces produced here are included in the net boundary flux \(\mathcal B^{a}_{(g)}\) below.

\emph{Putting everything together.} Restoring the Hilbert normalization we write
\begin{equation}
\delta_g S_{\rm geo}
=\frac12\int_{\mathcal M}\!\sqrt{-g}\;T^{\rm(geo)}_{mn}\,\delta g^{mn}
\;+\;\int_{\partial\mathcal M}\! d\Sigma_a\,\mathcal B^{a}_{(g)},
\end{equation}
with the fully explicit \emph{intermediate} bulk tensor
\begin{equation}
{
\begin{aligned}
T^{\rm(geo)}_{mn}\big|_{\text{interm}}
&= -\,g_{mn}\Lambda^{b}a_b
+2\,\Lambda^{b}n_{(m}\nabla_{n)}n_b
-\,\Lambda^{b}a_b\,n_m n_n
-\Lambda^{b} n^a n_b\big[(\nabla_a n_m)n_n+n_m(\nabla_a n_n)\big]
\\
&\quad
+\nabla_a\!\big(\Lambda^{b} n^a n_b\,n_m n_n\big)
+\,2\,C_{(mn)}\ .
\end{aligned}}
\label{eq:Tmn-interm-article-again}
\end{equation}
A convenient compact rearrangement is obtained by combining the total divergences (the last term in \eqref{eq:KII-article-again} together with the $2C_{(mn)}$-piece) into a single superpotential. Using the spatial projector $\gamma_{ab}=g_{ab}+n_an_b$ one arrives at
\begin{equation}
\begin{aligned}
T^{\rm(geo)}_{mn}
&=
-\,g_{mn}\,\Lambda^{\,b} a_b
+\frac12\,\Lambda^{\,b}
\Big[
n_m\nabla_n n_b
+n_n\nabla_m n_b
-\nabla_m(n_n n_b)
-\nabla_n(n_m n_b)
\Big]
\\
&\quad
+\nabla_a\!\Bigg\{
\frac12\,\Lambda^{\,b}
\Big[
n^a n_m\gamma_{nb}
+n^a n_n\gamma_{mb}
-n_m\gamma_n{}^{a} n_b
-n_n\gamma_m{}^{a} n_b
\Big]
\Bigg\}
\\
&\quad
+\frac12\,\Lambda^{\,b} a_b\,n_m n_n
+\text{(terms $\propto a_b$ that vanish on-shell by \eqref{eq:EL-lambda})}.
\end{aligned}
\label{eq:Tmn-compact-article-again}
\end{equation}

Here $\mathcal B^{a}_{(g)}$ is the sum of all boundary contributions generated when removing $\nabla(\delta g)$ in Blocks II and III; it is linear in $\delta g^{mn}$ with coefficients built from $\Lambda^b$, $n^a$, and $g_{mn}$.

\paragraph{On-shell simplification.}
Imposing the geodesic constraint $a_b=0$ from \eqref{eq:EL-lambda} removes every $a_b$-proportional piece; $\nabla_a n_b$ is purely spatial in this case, and $T^{\rm(geo)}_{mn}$ reduces to the algebraic transport term $\frac12\,\Lambda^{\,b}
\Big[
n_m\nabla_n n_b
+n_n\nabla_m n_b
-\nabla_m(n_n n_b)
-\nabla_n(n_m n_b)
\Big]$ plus the superpotential in \eqref{eq:Tmn-compact-article-again}. On the special branch $\Lambda^b\equiv 0$ the bulk tensor vanishes identically, leaving only the boundary flux $\mathcal B^{a}_{(g)}$.

}

\section{Variations of the CMC constraint sector}
\label{app:CMC-variations}
\setcounter{equation}{0}
In this appendix we derive the scalar and metric variations of the CMC
sector used in Sec.~\ref{subsec:dynamical-foliation}. We assume that boundary terms vanish
or are cancelled by appropriate boundary conditions. The CMC action is
\begin{equation}
        S_{\rm CMC}[g,A,\chi]
        =
        \int d^4x\,\sqrt{-g}\;
        \chi\left(\nabla_a n^a-K_0(A)\right),
\label{eq:SCMC_app_start}
\end{equation}
where
\begin{equation}
        X=g^{ab}\nabla_a A\nabla_b A<0,
        \qquad
        n_a=-\frac{\nabla_a A}{\sqrt{-X}},
        \qquad
        \gamma_{ab}=g_{ab}+n_a n_b .
\label{eq:CMC_defs_app}
\end{equation}
Using
\begin{equation}
        \nabla_a(\chi n^a)
        =
        n^a\nabla_a\chi+\chi\nabla_a n^a ,
\end{equation}
we rewrite \eqref{eq:SCMC_app_start}, up to a boundary term, as
\begin{equation}
        S_{\rm CMC}
        \doteq
        -\int d^4x\,\sqrt{-g}\;
        \left(n^a\nabla_a\chi+\chi K_0(A)\right).
\label{eq:SCMC_app_ibp}
\end{equation}
This form will be used below.

\subsection{Variation with respect to the multiplier}

Varying \eqref{eq:SCMC_app_start} with respect to \(\chi\) gives
\begin{equation}
        \delta_\chi S_{\rm CMC}
        =
        \int d^4x\,\sqrt{-g}\;
        \left(\nabla_a n^a-K_0(A)\right)\delta\chi .
\end{equation}
Therefore
\begin{equation}
        \frac{\delta S_{\rm CMC}}{\delta\chi}=0
        \qquad\Longrightarrow\qquad
        \nabla_a n^a=K_0(A).
\label{eq:CMC_constraint_app}
\end{equation}
With the convention \(K_{ab}=-\gamma_a{}^c\gamma_b{}^d\nabla_c n_d\), this
is equivalently \(K=-K_0(A)\).

\subsection{Variation with respect to \(A\)}

In the \(A\)-variation we keep \(g_{ab}\) and \(\chi\) fixed. The variation
of \(X\) is
\begin{equation}
        \delta_A X
        =
        \delta_A\left(g^{ab}\nabla_a A\nabla_b A\right)
        =
        2\nabla^a A\,\nabla_a(\delta A).
\label{eq:deltaA_X_app}
\end{equation}
Hence
\begin{equation}
        \delta_A\sqrt{-X}
        =
        -\frac{\delta_A X}{2\sqrt{-X}} .
\label{eq:deltaA_sqrtX_app}
\end{equation}
Directly varying
\[
        n_a=-\frac{\nabla_a A}{\sqrt{-X}}
\]
gives
\begin{align}
        \delta_A n_a
        &=
        -\frac{\nabla_a(\delta A)}{\sqrt{-X}}
        -
        \nabla_a A\,
        \delta_A\left(({-X})^{-1/2}\right)
        \nonumber\\
        &=
        -\frac{\nabla_a(\delta A)}{\sqrt{-X}}
        -
        \frac{\nabla_a A}{2(-X)^{3/2}}\,\delta_A X .
\label{eq:deltaA_n_step_app}
\end{align}
Using
\begin{equation}
        \nabla_a A=-\sqrt{-X}\,n_a,
        \qquad
        \delta_A X
        =
        -2\sqrt{-X}\,n^b\nabla_b(\delta A),
\end{equation}
we find
\begin{equation}
        \delta_A n_a
        =
        -\frac{1}{\sqrt{-X}}
        \left(
        \nabla_a(\delta A)
        +
        n_a n^b\nabla_b(\delta A)
        \right).
\label{eq:deltaA_n_unprojected_app}
\end{equation}
Equivalently,
\begin{equation}
        \delta_A n_a
        =
        -\frac{1}{\sqrt{-X}}\,
        \gamma_a{}^b\nabla_b(\delta A),
        \qquad
        \delta_A n^a
        =
        -\frac{1}{\sqrt{-X}}\,
        \gamma^{ab}\nabla_b(\delta A).
\label{eq:deltaA_n_projected_app}
\end{equation}
This variation is orthogonal to \(n^a\), as required by the fixed
normalization \(n_a n^a=-1\).

Now varying \eqref{eq:SCMC_app_ibp} gives
\begin{align}
        \delta_A S_{\rm CMC}
        &=
        -\int d^4x\,\sqrt{-g}\;
        \left(
        \delta_A n^a\nabla_a\chi
        +
        \chi K_0'(A)\delta A
        \right)
        \nonumber\\
        &=
        \int d^4x\,\sqrt{-g}\;
        \left[
        \frac{1}{\sqrt{-X}}\,
        \gamma^{ab}(\nabla_a\chi)\nabla_b(\delta A)
        -
        \chi K_0'(A)\delta A
        \right].
\label{eq:deltaA_S_CMC_app}
\end{align}
After integrating the first term by parts, we obtain
\begin{equation}
        \delta_A S_{\rm CMC}
        =
        -\int d^4x\,\sqrt{-g}\;
        \left[
        \nabla_b\left(
        \frac{1}{\sqrt{-X}}\,
        \gamma^{ba}\nabla_a\chi
        \right)
        +
        \chi K_0'(A)
        \right]\delta A .
\label{eq:deltaA_S_CMC_final_app}
\end{equation}
Therefore the Euler--Lagrange equation for \(A\) is
\begin{equation}
        \nabla_b\left(
        \frac{1}{\sqrt{-X}}\,
        \gamma^{ba}\nabla_a\chi
        \right)
        +
        \chi K_0'(A)
        =
        0 .
\label{eq:CMC_A_EOM_app}
\end{equation}
Equivalently,
\begin{equation}
        \nabla_a J^a_{\rm CMC}
        +
        \chi K_0'(A)
        =
        0,
        \qquad
        J^a_{\rm CMC}
        :=
        \frac{1}{\sqrt{-X}}\,
        \gamma^{ab}\nabla_b\chi .
\label{eq:CMC_current_app}
\end{equation}
For \(K_0(A)=K_\star=\mathrm{const}\), this reduces to the conservation law
\(\nabla_a J^a_{\rm CMC}=0\).

\subsection{Variation with respect to the metric}

We now vary \eqref{eq:SCMC_app_ibp} with respect to \(g^{ab}\), keeping
\(A\) and \(\chi\) fixed. Since \(\chi\) is a scalar,
\(\nabla_a\chi=\partial_a\chi\) does not vary through the Levi--Civita
connection. Thus only \(\sqrt{-g}\) and \(n^a\) vary.

We use
\begin{equation}
        \delta_g\sqrt{-g}
        =
        -\frac12\sqrt{-g}\,g_{ab}\delta g^{ab}.
\label{eq:delta_sqrtg_CMC_app}
\end{equation}
At fixed \(A\), \(A_a:=\nabla_a A\) is fixed, while
\begin{equation}
        \delta_g X
        =
        A_a A_b\,\delta g^{ab}.
\label{eq:deltag_X_CMC_app}
\end{equation}
Therefore
\begin{equation}
        \delta_g\sqrt{-X}
        =
        -\frac{\delta_g X}{2\sqrt{-X}} .
\label{eq:deltag_sqrtX_CMC_app}
\end{equation}
Since \(n_a=-A_a/\sqrt{-X}\), we find
\begin{align}
        \delta_g n_a
        &=
        -\frac{A_a}{2(-X)^{3/2}}\delta_g X
        \nonumber\\
        &=
        \frac12 n_a n_c n_d\,\delta g^{cd}.
\label{eq:deltag_ncov_CMC_app}
\end{align}
Raising the index gives
\begin{align}
        \delta_g n^a
        &=
        \delta(g^{ab}n_b)
        \nonumber\\
        &=
        \delta g^{ab}n_b
        +
        g^{ab}\delta_g n_b
        \nonumber\\
        &=
        \left(
        \delta^a{}_{(c}n_{d)}
        +
        \frac12 n^a n_c n_d
        \right)\delta g^{cd}.
\label{eq:deltag_ncon_CMC_app}
\end{align}

Varying \eqref{eq:SCMC_app_ibp} gives
\begin{align}
        \delta_g S_{\rm CMC}
        &=
        -\int d^4x\,
        \left[
        \delta_g\sqrt{-g}
        \left(n^c\nabla_c\chi+\chi K_0(A)\right)
        +
        \sqrt{-g}\,\delta_g n^c\nabla_c\chi
        \right]
        \nonumber\\
        &=
        \int d^4x\,\sqrt{-g}\;
        \left\{
        \frac12 g_{ab}
        \left(n^c\nabla_c\chi+\chi K_0(A)\right)\delta g^{ab}
        -
        \delta_g n^c\nabla_c\chi
        \right\}.
\label{eq:deltag_S_CMC_intermediate_app}
\end{align}
Substituting \eqref{eq:deltag_ncon_CMC_app} and using the symmetry of
\(\delta g^{ab}\), we obtain
\begin{align}
        \delta_g S_{\rm CMC}
        =
        \frac12\int d^4x\,\sqrt{-g}\;
        \Big[
        &g_{ab}\left(n^c\nabla_c\chi+\chi K_0\right)
        -
        2n_{(a}\nabla_{b)}\chi
        \nonumber\\
        &-
        \left(n^c\nabla_c\chi\right)n_a n_b
        \Big]\delta g^{ab}.
\label{eq:deltag_S_CMC_final_app}
\end{align}
Hence
\begin{equation}
        \frac{1}{\sqrt{-g}}
        \frac{\delta S_{\rm CMC}}{\delta g^{ab}}
        =
        \frac12
        \Big[
        g_{ab}\left(n^c\nabla_c\chi+\chi K_0\right)
        -
        2n_{(a}\nabla_{b)}\chi
        -
        \left(n^c\nabla_c\chi\right)n_a n_b
        \Big].
\label{eq:CMC_metric_variation_app}
\end{equation}
With the standard definition
\begin{equation}
        T^{\rm(CMC)}_{ab}
        :=
        -\frac{2}{\sqrt{-g}}
        \frac{\delta S_{\rm CMC}}{\delta g^{ab}},
\end{equation}
this gives
\begin{equation}
        T^{\rm(CMC)}_{ab}
        =
        -g_{ab}\left(n^c\nabla_c\chi+\chi K_0\right)
        +
        2n_{(a}\nabla_{b)}\chi
        +
        \left(n^c\nabla_c\chi\right)n_a n_b .
\label{eq:T_CMC_app}
\end{equation}
This is the stress tensor quoted in the main text.

\section{Decomposition of \(\nabla_a n_b\)}
\label{app:normal-decomposition}
\setcounter{equation}{0}
In this appendix we derive the decomposition of \(\nabla_a n_b\) used in
Eq.~\eqref{eq:nabla-n-decomposition}. The unit timelike normal satisfies
\begin{equation}
        n^a n_a=-1 .
\label{eq:app-unit-normal}
\end{equation}
Taking a covariant derivative of this normalization condition and using
metric compatibility, \(\nabla_c g_{ab}=0\), gives
\begin{equation}
        0=\nabla_c(n^a n_a)=2n^a\nabla_c n_a .
\label{eq:app-normalization-derivative}
\end{equation}
Therefore
\begin{equation}
        n^a\nabla_c n_a=0 .
\label{eq:app-normal-orthogonality}
\end{equation}
Thus \(\nabla_c n_a\) is orthogonal to \(n^a\) in its second index. We now
decompose the identity tensor into its spatial and normal parts. We have
\begin{equation}
        \delta_a{}^c=\gamma_a{}^c-n_a n^c .
\label{eq:app-identity-decomposition}
\end{equation}
Using this decomposition in both indices of \(\nabla_a n_b\), one obtains
\begin{align}
\nabla_a n_b
&=
\delta_a{}^c\delta_b{}^d\nabla_c n_d
\nonumber\\
&=
\left(\gamma_a{}^c-n_a n^c\right)
\left(\gamma_b{}^d-n_b n^d\right)
\nabla_c n_d
\nonumber\\
&=
\gamma_a{}^c\gamma_b{}^d\nabla_c n_d
-\gamma_a{}^c n_b n^d\nabla_c n_d
-n_a n^c\gamma_b{}^d\nabla_c n_d
+n_a n_b n^c n^d\nabla_c n_d .
\label{eq:app-nabla-n-expanded}
\end{align}
The second and fourth terms in Eq.~\eqref{eq:app-nabla-n-expanded} vanish
because of Eq.~\eqref{eq:app-normal-orthogonality}. The first term is fixed
by the definition of the extrinsic curvature \eqref{Kab}
\begin{equation}
        \gamma_a{}^c\gamma_b{}^d\nabla_c n_d=-K_{ab}.
\label{eq:app-projected-term}
\end{equation}
For the remaining term we use the acceleration of the normal congruence \eqref{accel}.
After lowering the index and again using \(\nabla_c g_{ab}=0\), this becomes
\begin{equation}
        a_b=n^c\nabla_c n_b .
\label{eq:app-acceleration-lowered}
\end{equation}
Moreover, \(a_b\) is spatial, since
\begin{equation}
        n^b a_b=n^b n^c\nabla_c n_b=0 .
\label{eq:app-acceleration-spatial}
\end{equation}
Consequently the spatial projector does not change \(a_b\):
\begin{equation}
        \gamma_b{}^d a_d=a_b .
\label{eq:app-projector-on-acceleration}
\end{equation}
The third term in Eq.~\eqref{eq:app-nabla-n-expanded} is therefore
\begin{equation}
        -n_a n^c\gamma_b{}^d\nabla_c n_d
        =
        -n_a\gamma_b{}^d a_d
        =
        -n_a a_b .
\label{eq:app-third-term}
\end{equation}
Combining Eqs.~\eqref{eq:app-projected-term} and
\eqref{eq:app-third-term}, we find
\begin{equation}
        \nabla_a n_b=-K_{ab}-n_a a_b .
\label{eq:app-nabla-n-final}
\end{equation}

Taking the trace of Eq.~\eqref{eq:app-nabla-n-final} gives the corresponding
identity for the divergence of the normal:
\begin{align}
\nabla_a n^a
&=
g^{ab}\nabla_a n_b
\nonumber\\
&=
-g^{ab}K_{ab}-g^{ab}n_a a_b .
\label{eq:app-trace-step}
\end{align}
The second term vanishes because \(n^b a_b=0\). Since \(K_{ab}\) is spatial
in both indices,
\begin{equation}
        n^aK_{ab}=0,
        \qquad
        n^bK_{ab}=0,
\label{eq:app-K-spatial}
\end{equation}
the metric \(g^{ab}\) may be replaced by the induced metric \(\gamma^{ab}\)
inside the contraction with \(K_{ab}\):
\begin{equation}
        g^{ab}K_{ab}=\gamma^{ab}K_{ab}=K .
\label{eq:app-K-trace}
\end{equation}
Thus
\begin{equation}
        \nabla_a n^a=-K .
\label{eq:app-divergence-normal-final}
\end{equation}
Equations~\eqref{eq:app-nabla-n-final} and
\eqref{eq:app-divergence-normal-final} reproduce the decomposition quoted in
Eq.~\eqref{eq:nabla-n-decomposition}.

\section{Derivation of the energy-momentum tensor for the $4$-dimensional conformal coupling}\m{emt4d}
\setcounter{equation}{0}
For simplicity, in derivations we will consider the case of $\eta=0$ in \eqref{4dconformallag}, but the result can be easily extended to the case $\eta \neq 0$. We need to compute:
\begin{equation}
\frac{\delta S}{\delta g^{ab}}=    \frac{\delta}{\delta g^{ab}} \left[ \int \left( \nabla_c \phi \nabla^c \phi + V(\phi) + \xi R \phi^2 \right) \sqrt{-g} \, d^4x \right]. \label{eq:variation_total}
\end{equation}
The kinetic term depends on the metric through the contraction of indices and the volume element. Its variation is:
\begin{equation}
    \frac{\delta}{\delta g^{ab}} \left[ g^{cd} \nabla_c \phi \nabla_d \phi \sqrt{-g} \right]
    = \nabla_a \phi \nabla_b \phi \sqrt{-g} - \frac{1}{2} g_{ab} g^{cd} \nabla_c \phi \nabla_d \phi \sqrt{-g},
\end{equation}
where we used:
\begin{equation}
    \frac{\delta \sqrt{-g}}{\delta g^{ab}} = -\frac{1}{2} \sqrt{-g} g_{ab}.
\end{equation}
The variation of the potential term is straightforward:
\begin{equation}
    \frac{\delta}{\delta g^{ab}} \left[ \int V(\phi) \sqrt{-g} \, d^4x \right] = V(\phi) \frac{\delta \sqrt{-g}}{\delta g^{ab}} = -\frac{1}{2} g_{ab} V(\phi) \sqrt{-g}.
\end{equation}
Finally, for the curvature coupling term, we compute:
\begin{equation}
    \frac{\delta}{\delta g^{ab}} \left[ \int \xi R \phi^2 \sqrt{-g} \, d^4x \right]
    = \xi \frac{\delta R_{cd}}{\delta g^{ab}} g^{cd} \phi^2 \sqrt{-g} 
    + \xi R_{cd} \frac{\delta g^{cd}}{\delta g^{ab}} \phi^2 \sqrt{-g} 
    + \xi R \phi^2 \frac{\delta \sqrt{-g}}{\delta g^{ab}}. \label{eq:curvature_coupling}
\end{equation}
Substituting the variation of the volume element and recognizing the Einstein tensor \( G_{ab} = R_{ab} - \frac{1}{2} g_{ab} R \), we write:
\begin{equation}
    \frac{\delta}{\delta g^{ab}} \left[ \int \xi R \phi^2 \sqrt{-g} \, d^4x \right] = \xi \frac{\delta R_{cd}}{\delta g^{ab}}g^{cd} \phi^2 \sqrt{-g} + \xi G_{ab} \phi^2 \sqrt{-g}.
\end{equation}
Thus, so far, the stress-energy tensor becomes:
\begin{equation}
    T_{ab} = \nabla_a \phi \nabla_b \phi - \frac{1}{2} g_{ab} \left( \nabla_c \phi \nabla^c \phi + V(\phi) \right)
    + \xi \phi^2 \left( G_{ab} + \frac{\delta R_{cd}}{\delta g^{ab}}g^{cd} \right). \label{eq:stress_energy_intermediate}
\end{equation}
The remaining term involving the Ricci tensor variation is best computed under an integral:
\begin{equation}
    \int \xi g^{cd} \delta R_{cd} \phi^2 \sqrt{-g} \, d^4x.
\end{equation}
From \cite{Wald:1984rg}, Eq. (3.4.5), we know that the Ricci tensor is expressed in terms of the connection coefficients as:
\begin{equation}
    R_{m n} = \partial_r \Gamma^r_{m n} - \partial_m \Gamma^r_{r n} 
    + \Gamma^s_{m n} \Gamma^r_{s r} - \Gamma^s_{r n} \Gamma^r_{s m}. \label{eq:ricci_tensor}
\end{equation}
When varying the Ricci tensor, we consider its dependence on the Christoffel symbols. In a locally inertial frame, the variation is given by:
\begin{equation}
    \delta R_{m n} = \partial_r \delta \Gamma^r_{m n} - \partial_m \delta \Gamma^r_{r n}, \label{eq:variation_ricci_tensor}
\end{equation}
where \( \delta \Gamma^r_{m n} \) represents the variation of the Christoffel symbols. This expression can be generalized to any coordinate system since \( \delta \Gamma^r_{m n} \) is a tensor (as it represents differences between Christoffel symbols in different metrics).
Rewriting Eq.~(\ref{eq:variation_ricci_tensor}) in terms of covariant derivatives, we have:
\begin{equation}
    \delta R_{m n} = \nabla_r \delta \Gamma^r_{m n} - \nabla_m \delta \Gamma^r_{r n}. \label{eq:variation_ricci_covariant}
\end{equation}
Contracting this with the metric \( g^{ab} \), we find:
\begin{equation}
    g^{ab} \delta R_{ab} = g^{ab} \nabla_c \delta \Gamma^c_{ab} - g^{ab} \nabla_a \delta \Gamma^c_{cb}= \nabla_c \left[ g^{ab} \delta \Gamma^c_{ab} - g^{bc} \delta \Gamma^a_{ab} \right]. \label{eq:ricci_variation_contracted}
\end{equation}

Next, we compute the variation of the Christoffel symbols. The Christoffel symbols are defined as:
\begin{equation}
    \Gamma^s_{m n} = \frac{1}{2} g^{s t} \left( \partial_m g_{n t} + \partial_n g_{t m} - \partial_t g_{m n} \right).
\end{equation}
The variation of \( \Gamma^s_{m n} \) with respect to the metric gives:
\begin{equation}
    \delta \Gamma^s_{m n} = \frac{1}{2} \delta g^{s t} 
    \left( \partial_m g_{n t} + \partial_n g_{t m} - \partial_t g_{m n} \right) 
    + \frac{1}{2} g^{s t} 
    \left( \partial_m \delta g_{n t} + \partial_n \delta g_{t m} - \partial_t \delta g_{m n} \right). \label{eq:variation_christoffel}
\end{equation}
Replacing partial derivatives by covariant and using that covariant derivative of metric is zero we have
\begin{equation}
    \delta \Gamma^s_{m n} = \frac{1}{2} g^{s t} 
    \left( \nabla_m \delta g_{n t} + \nabla_n \delta g_{t m} - \nabla_t \delta g_{m n} \right). \label{eq:variation_christoffel_simplified}
\end{equation}
We start with the variation of the first term:
\begin{equation}
    g^{ab} \delta \Gamma^c_{ab} = \frac{1}{2} g^{ab} g^{cd} \left( \nabla_a \delta g_{bd} + \nabla_b \delta g_{ad} - \nabla_d \delta g_{ab} \right). \label{eq:variation_gGamma}
\end{equation}
Using the symmetry of \( g^{ab} \) and \( \delta g_{ab} \), we combine terms:
\begin{equation}
    g^{ab} \delta \Gamma^c_{ab} = \frac{1}{2} g^{cd} g^{ab} \left( 2 \nabla_a \delta g_{bd} - \nabla_d \delta g_{ab} \right). \label{eq:variation_gGamma_simplified}
\end{equation}
For the second term, we calculate:
\begin{equation}
    g^{bc} \delta \Gamma^a_{ab} = \frac{1}{2} g^{bc} g^{ad} \left( \nabla_a \delta g_{bd} + \nabla_b \delta g_{ad} - \nabla_d \delta g_{ab} \right). \label{eq:variation_gGamma_2}
\end{equation}
This simplifies under the symmetries of \( g^{ab} \) and \( \delta g_{ab} \) to:
\begin{equation}
    g^{bc} \delta \Gamma^a_{ab} = \frac{1}{2} g^{ad} g^{bc}\nabla_b \delta g_{da}. \label{eq:variation_gGamma_2_simplified}
\end{equation}
Substituting the results from Eq.~(\ref{eq:variation_gGamma_simplified}) and Eq.~(\ref{eq:variation_gGamma_2_simplified}) into the contracted Ricci scalar variation, after simple calculations we have:
\begin{equation}
    g^{ab} \delta R_{ab} =  g^{cd}  g^{ab} \nabla_c\nabla_a \delta g_{bd} -g^{ab}\nabla_c\nabla^c \delta g_{ab}. \label{eq:ricci_combined_variation}
\end{equation}
We now simplify using \( \delta g^{ab} = -g^{ac} g^{bd} \delta g_{cd} \), which gives:
\begin{equation}
    g^{ab} \delta R_{ab} = \left( g_{ab} \nabla_c \nabla^c - \nabla_a \nabla_b \right) \delta g^{ab}. \label{eq:ricci_variation_final_form}
\end{equation}
This is the final expression for the variation of the Ricci scalar, expressed in terms of the metric variations \( \delta g_{ab} \).
So, we have the term:
\begin{equation}\label{eq:2.16a}
\begin{array}{cccc}
     \int \xi g^{cd} \delta R_{cd} \phi^2 \sqrt{-g} \, d^4x = 
    \int \xi \phi^2 \left( g_{ab} \nabla_c \nabla^c \delta g^{ab} - \nabla_a \nabla_b \delta g^{ab} \right) \sqrt{-g} \, d^4x \\= \int \xi \delta g^{ab} \left( g_{ab} \nabla_c \nabla^c - \nabla_a \nabla_b \right) \phi^2 \sqrt{-g} \, d^4x+ surface ~~terms. 
\end{array}
\end{equation}
Here, the surface terms arise from integration by parts, and we discard them assuming they vanish at the boundary. The final expression for the stress-energy tensor becomes:
\begin{equation}
    T_{ab} = \nabla_a \phi \nabla_b \phi - \frac{1}{2} g_{ab} \left( \nabla_c \phi \nabla^c \phi + V(\phi) \right) 
    + \xi \phi^2 G_{ab} + \xi \left( g_{ab} \nabla_c \nabla^c - \nabla_a \nabla_b \right) \phi^2. \label{eq:2.18}
\end{equation}
\section{Derivation of the scalar field energy-momentum tensor with coupling to $^{(3)}R$}\m{emt3d}
\setcounter{equation}{0}
Let us derive the field equations for \eqref{anothercouplag}. Taking the variation of $S_m$ with respect to metric  we get
\begin{equation}
    T_{ab} = \nabla_a \phi \nabla_b \phi - \frac{1}{2} g_{ab} \nabla_c \phi \nabla^c \phi - \frac{1}{2} g_{ab} V(\phi) + \frac{\delta}{\sqrt{-g} \delta g^{ab}} \left( \int \xi {}^{(3)} R \phi^2 \sqrt{-g} \, d^4 x \right),
\end{equation}
where the variations of the kinetic and potential terms are derived in the Appendix \ref{emt4d}.
Writing the last term in detail we have
\begin{equation}
\frac{\delta}{\sqrt{-g} \delta g^{ab}} \left( \int \xi {}^{(3)} R \phi^2 \sqrt{-g} \, d^4 x \right) =   \xi \frac{ \delta {}^{(3)} R_{cd}}{\delta g^{ab}}  \, \gamma^{cd} \, \phi^2 + \xi^3 R_{cd} \,\frac{\delta \gamma^{cd}}{\delta g^{ab}} \, \phi^2 - \frac{1}{2} \xi {}^{(3)} R \, g_{ab} \, \phi^2 . \end{equation}
Then, rewriting the first term of the right hand side according to \cite{Jha:2022svf} (see derivation in Appendix \ref{gammadelR}) we get
\begin{equation}
\gamma^{ab} \delta {}^{(3)} R_{ab} = \left( \gamma_{ab} D_c D^c - D_a D_b \right) \delta g^{ab}. \end{equation}
Integrating it by parts twice we get
\begin{equation}
\int \gamma^{ab} \xi  \delta {}^{(3)} R_{ab} \phi^2 \sqrt{-g} \, d^4x = \int \xi \delta g^{ab} \left( \gamma_{ab} D_c D^c - D_a D_b \right) (N \phi^2)  \sqrt{\gamma} \, d^4 x + boundary~~terms .\end{equation}
So we have
\begin{equation}
 \frac{1}{\sqrt{-g}} \xi \frac{\delta^3 R_{cd}}{\delta g^{ab}} \, \gamma^{cd}  = \frac{\xi}{N} \left( \gamma_{ab} D_c D^c - D_a D_b \right) (N \phi^2).
\end{equation}
Rewriting the second term we have
\begin{equation}
    \xi^3 R_{cd} \,\delta \gamma^{cd} \, \phi^2 =  \xi^3 R_{cd} \,\delta (g^{cd}+n^c n^d) \, \phi^2 =  \xi^3 R_{cd} \,\delta g^{cd} \, \phi^2 ,
\end{equation}
where we use that the terms like $R_{cd} n^c$ vanish.
Thus, we get the energy-momentum tensor
\begin{equation}
T_{ab} = \nabla_a \phi \nabla_b \phi - \frac{1}{2} g_{ab} \nabla_c \phi \nabla^c \phi - \frac{1}{2} g_{ab} V(\phi) + \frac{\xi}{N}(\gamma_{a b} D_c D^c -D_{a} D_{b})(\phi^2 N) + \xi {}^{(3)} R_{ab} \phi^2   - \frac{1}{2} {}^{(3)} R g_{ab} \xi \phi^2.
\end{equation}

\bibliography{references}

\bibliographystyle{Style}

\end{document}